\documentclass[11pt,reqno]{amsart}

\usepackage[T1]{fontenc}
\usepackage[utf8]{inputenc}
\usepackage{lmodern}
\usepackage{amsmath,amssymb,amsthm,mathtools}
\usepackage{booktabs,array,multirow}
\usepackage{enumitem}
\usepackage{microtype}
\usepackage[margin=1.12in]{geometry}
\usepackage{xcolor}
\usepackage{tikz}
\usetikzlibrary{arrows.meta,positioning,calc,decorations.pathreplacing,shapes.geometric,fit,backgrounds}
\usepackage{aliascnt}
\usepackage[colorlinks=true,linkcolor=blue!45!black,citecolor=blue!45!black,urlcolor=blue!45!black]{hyperref}
\usepackage[capitalise,nameinlink]{cleveref}

\setlist[enumerate]{label=\textup{(\roman*)},leftmargin=2.2em,itemsep=2pt,topsep=4pt}
\setlist[itemize]{leftmargin=1.6em,itemsep=2pt,topsep=4pt}
\newtheorem{theorem}{Theorem}[section]
\newcommand{\sharedthm}[3][plain]{%
  \theoremstyle{#1}\newaliascnt{#2}{theorem}\newtheorem{#2}[#2]{#3}\aliascntresetthe{#2}%
  \crefname{#2}{#3}{#3s}}
\sharedthm{proposition}{Proposition}
\sharedthm{lemma}{Lemma}
\sharedthm{corollary}{Corollary}\crefname{corollary}{Corollary}{Corollaries}
\sharedthm[definition]{definition}{Definition}
\sharedthm[definition]{certificate}{Certificate}
\sharedthm[remark]{remark}{Remark}
\theoremstyle{plain}
\numberwithin{equation}{section}
\crefname{equation}{}{}
\Crefname{equation}{Equation}{Equations}

\newcommand{\Z}{\mathbb Z}
\newcommand{\R}{\mathbb R}
\newcommand{\T}{\mathbb T}
\newcommand{\one}{\mathbf 1}
\newcommand{\cE}{\mathcal E}
\newcommand{\cO}{\mathcal O}
\newcommand{\cS}{\mathcal S}
\newcommand{\cT}{\mathcal T}
\newcommand{\cR}{\mathcal R}
\newcommand{\Prob}{\mathbb P}
\newcommand{\dd}{\mathrm d}
\newcommand{\rT}{\mathsf T}
\newcommand{\rA}{\mathsf A}
\newcommand{\rD}{\mathsf D}
\newcommand{\rS}{\mathsf S}
\newcommand{\Uord}{U_{\mathrm{ord}}}
\newcommand{\avg}[1]{\langle #1\rangle}

\newcommand{\ceil}[1]{\lceil #1\rceil}
\newcommand{\repo}{\url{https://github.com/whyihaveyou/ising3d-critical-interval}}
\newcommand{\doi}[1]{\href{https://doi.org/#1}{\nolinkurl{doi:#1}}}

\newcommand{\rlowfrac}{\frac{275442362401406827838944754146320499627}{178669509598593172161055245853679500373}}
\newcommand{\rupfrac}{\frac{18874999999514449}{11930000000000000}}

\hypersetup{pdftitle={Upper and lower bounds for the critical point of the three-dimensional Ising model},pdfauthor={Zhongpu Qiu},pdfkeywords={Ising model, critical point, connective constant, reflection positivity, computer-assisted proof}}
\begin{document}

\title[Bounds for the critical point of the 3D Ising model]{Upper and lower bounds for the critical point\\ of the three-dimensional Ising model}

\author{Zhongpu Qiu}
\address{Beijing Normal University, Beijing, China; and Shanghai Artificial Intelligence Laboratory, Shanghai, China}
\email{slszqzp@outlook.com}

\date{November 23, 2026}


\begin{abstract}
We prove that the critical point $K_c$ of the nearest-neighbour Ising ferromagnet on $\Z^3$ satisfies $0.2164203058<K_c<0.2293910295$, equivalently $1.541630483<e^{2K_c}<1.582145851$, with both endpoints given by explicit rational numbers. To our knowledge these are the sharpest rigorous bounds available. They improve the classical bounds of Fisher (1967) and of Fr\"ohlich, Simon and Spencer (1976), narrowing the rigorously known interval by a factor of about $3.5$, and the interval contains the numerical estimate of the critical point. For the lower bound we refine Fisher's self-avoiding-walk argument. A shortest path in the high-temperature expansion forbids all bonds spanned by its vertices, and the resulting gain is estimated through local correlation integrals and summed over paths by a finite-memory transfer bound. As a by-product, the connective constant of $\Z^3$ is shown to be at most $4.73195$. For the upper bound we prove an infrared bound whose coefficient involves the nearest-neighbour correlation. We keep the zero Fourier mode as an unknown and force it to be positive by exact local relations for the spins on a small shell. The finite inequalities entering the proof are verified in exact rational arithmetic, and the passage to the thermodynamic limit is proved analytically.
\end{abstract}

\maketitle

\setcounter{tocdepth}{1}
\tableofcontents

\section{Introduction}\label{sec:intro}

\subsection{The three-dimensional Ising model}

The Ising model is the simplest model of a phase transition. In three dimensions its critical point is the prototype of a universality class that also contains the liquid--vapour critical point and uniaxial magnets~\cite{PV02}. In two dimensions the model is exactly solvable. Kramers and Wannier located the critical point by duality~\cite{KW41}, Onsager computed the free energy~\cite{Ons44}, and Yang computed the spontaneous magnetization~\cite{Yan52}. No comparable solution is known in three dimensions. Nevertheless the critical behaviour has been determined with remarkable precision by high-temperature series~\cite{PV02}, Monte Carlo simulation~\cite{FL91,FXL18}, the Monte Carlo renormalization group~\cite{RBS21} and, more recently, the conformal bootstrap~\cite{EPPRSV12,KPSV16}.

The rigorous theory of the transition has also advanced substantially. In every dimension $d\ge2$ the transition is \emph{sharp}~\cite{ABF87,DCT16}. In the language of physics, there is a single critical point: correlations decay exponentially throughout the high-temperature phase, and the susceptibility diverges exactly where the spontaneous magnetization appears. In three dimensions the spontaneous magnetization moreover vanishes continuously at the critical point~\cite{ADCS15}, so the transition is of second order. In four and more dimensions the critical behaviour is mean-field and the scaling limit is Gaussian~\cite{AD21}. Three dimensions is thus both the physically most relevant case and the one in which the critical behaviour is least understood. These results concern the \emph{nature} of the transition. The present paper is concerned with its \emph{location}: we determine the critical point of the simple-cubic model to within an explicit interval of relative width $2.6\%$.

\subsection{The critical point}

Let $\Lambda\subset\Z^3$ be finite and let $E(\Lambda)$ denote the set of unoriented nearest-neighbour bonds $\{x,y\}$ with $x,y\in\Lambda$, each bond counted once. For a dimensionless coupling $K=J/(k_{\mathrm B}T)>0$, a dimensionless magnetic field $h\in\R$ and a boundary condition $\eta\in\{-1,+1\}^{\Z^3}$, the Ising Gibbs measure on $\{-1,+1\}^\Lambda$ is
\begin{equation}\label{eq:gibbs}
 \mu^{\eta}_{\Lambda,K,h}(\sigma)=\frac{1}{Z^\eta_{\Lambda}(K,h)}
 \exp\Big(K\!\!\sum_{\{x,y\}\in E(\Lambda)}\!\!\sigma_x\sigma_y
 +K\!\!\sum_{\substack{x\in\Lambda,\ y\notin\Lambda\\ |x-y|=1}}\!\!\sigma_x\eta_y
 +h\sum_{x\in\Lambda}\sigma_x\Big).
\end{equation}
We write $\avg{\,\cdot\,}^+_{\Lambda,K,h}$ for the expectation with the plus boundary condition $\eta\equiv+1$ and $B_R=\{-R,\dots,R\}^3$. The plus-state spontaneous magnetization and the critical point are
\begin{equation}\label{eq:Kc}
 m_+(K)=\lim_{R\to\infty}\avg{\sigma_0}^+_{B_R,K,0},\qquad
 K_c=\inf\{K>0:\ m_+(K)>0\}.
\end{equation}
Thus $K_c$ is the critical inverse temperature measured in units of $J$, and we refer to it as the critical point. The limit exists and $m_+$ is nondecreasing in $K$ by the Griffiths inequalities (\cref{sec:prelim}). Throughout we use the two parametrizations
\begin{equation}\label{eq:ru}
 r=e^{2K},\qquad u=\tanh K=\frac{r-1}{r+1},
\end{equation}
and we write $r_c=e^{2K_c}$.

The most precise numerical estimate, obtained from Monte Carlo simulations on lattices of up to $1024^3$ sites, is $K_c=0.221654626(5)$~\cite{FXL18}, i.e.\ $r_c\approx1.557854$; see~\cite{FL91,RBS21,PV02} for earlier and independent estimates. Rigorous bounds on $K_c$ go back to the beginnings of the subject.

\emph{Lower bounds.} Fisher~\cite{Fis67} bounded the two-point function by a sum over self-avoiding walks, $\avg{\sigma_x\sigma_y}\le\sum_{\omega:x\to y}u^{|\omega|}$. Consequently $\tanh K_c\ge1/\mu$, where $\mu$ is the connective constant of $\Z^3$~\cite{Fis67,PS19}. Fisher evaluated this bound with the upper bound $\mu<4.865$ of Fisher and Sykes~\cite{FS59}, which gives $k_{\mathrm B}T_c/J<4.80$, i.e.\ $K_c>0.2085$. The value of $\mu$ is not known exactly, and its rigorous upper bounds have since been improved. The best published bound, $\mu\le4.7387$, was obtained by P\"onitz and Tittmann with finite-memory automata~\cite{PT00}; see~\cite{Alm93} for an earlier eigenvalue method and~\cite{GL19} for a survey of connective constants.\footnote{The bound $\mu<4.7114$ stated in~\cite{MJHMJG00} is in error; see~\cite[Sect.~7.2]{CLS07}.} A different line of work derives bounds from Callen's correlation identities~\cite{SBO83,Vig85}. The bound $k_{\mathrm B}T_c/J\le5.0010$ of~\cite{Vig85}, i.e.\ $K_c\ge0.19996$, is quoted in~\cite{BLU17} as the best rigorous upper bound on the critical temperature of the three-dimensional model. Inserting $\mu\le4.7387$ into Fisher's inequality gives
\[
 r_c>\frac{\mu+1}{\mu-1}>1.5349,\qquad\text{i.e.}\qquad K_c>0.2142,\quad k_{\mathrm B}T_c/J<4.668 .
\]
We have not found this numerical consequence recorded in the literature. The automaton of \cref{sec:counting} improves the bound on the connective constant to $\mu\le4.73195$ and hence gives $r_c>1.5359$ (\cref{prop:fisher}).

\emph{Upper bounds.} Peierls-type contour arguments~\cite{Pei36,Gri64,WGF67} give an ordered phase at large $K$. The infrared bound~\cite{FSS76} gives considerably more. Combined with the sum rule $\avg{\sigma_0^2}=1$, it yields $K_c\le\mathsf W/6\approx0.25273$, i.e.\ $r_c<1.658$, where $\mathsf W=1.516386\ldots$ is Watson's integral~\cite{Wat39}.

Together these bounds confine $r_c$ to an interval of relative width about $8\%$. The bounds of Fisher and of Fr\"ohlich, Simon and Spencer compare the Ising model with a simpler object, the self-avoiding walk on one side and the Gaussian free field on the other. Neither uses the local structure of the Ising interaction. In percolation, rigorous bounds on critical probabilities have long been obtained by comparison arguments combined with finite computations~\cite{vdBE96}. More recently, computer-assisted methods have located critical probabilities within much narrower intervals~\cite{BBW05,RW07}. These intervals hold with a rigorously controlled confidence level, such as $99.9999\%$, rather than with certainty. The bounds of the present paper are deterministic. The aim of the present work is to bring such quantitative control to the Ising model, by combining classical correlation inequalities with finite local computations certified in exact arithmetic.

\subsection{Main result}

\begin{theorem}[Critical interval]\label{thm:main}
For the nearest-neighbour Ising ferromagnet on $\Z^3$,
\begin{equation}\label{eq:main}
 r_-<e^{2K_c}<r_+,\qquad\text{where}\quad
 \begin{aligned}
 r_-&=\rlowfrac,\\[2pt]
 r_+&=\rupfrac .
 \end{aligned}
\end{equation}
In decimal form, $1.541630483120638<r_-$ and $r_+<1.582145850755612$, and
\begin{equation}\label{eq:mainK}
 0.2164203058063<K_c<0.2293910294998 .
\end{equation}
\end{theorem}

\begin{figure}[t]
\centering
\begin{tikzpicture}[x=1.62cm,y=1cm]
\fill[blue!12] (1.1630483,-0.18) rectangle (5.2145851,0.18);
\draw[-{Stealth[length=5pt]}] (0,0) -- (7.35,0) node[right] {$r=e^{2K}$};
\foreach \t/\lab in {0/1.53,1/1.54,2/1.55,3/1.56,4/1.57,5/1.58,6/1.59,7/1.60}
  {\draw (\t,0.07) -- (\t,-0.07) node[below,font=\scriptsize] {$\lab$};}
\foreach \t in {0.5,1.5,2.5,3.5,4.5,5.5,6.5}{\draw (\t,0.04) -- (\t,-0.04);}
\newcommand{\mk}[4]{\draw[#4,thick] (#1,-0.18) -- (#1,#2); \node[#4,above,font=\scriptsize,align=center] at (#1,#2) {#3};}
\draw[black!70,thick] (0.4945302,-0.18) -- (0.4945302,0.55); \node[black!70,above left=0pt and -6pt,font=\scriptsize,align=right] at (0.4945302,0.55) {Fisher with\\$\mu\le4.7387$};
\mk{1.1630483}{0.9}{$r_-$}{blue!70!black}
\draw[black!70,thick] (2.7854029,-0.18) -- (2.7854029,0.55); \node[black!70,above left=0pt and -4pt,font=\scriptsize,align=right] at (2.7854029,0.55) {Monte Carlo~\cite{FXL18}\\$r_c\approx1.55785$};
\mk{5.2145851}{0.9}{$r_+$}{blue!70!black}
\mk{5.6312094}{1.25}{\cref{cor:singlesite}\\$\sqrt{1+\mathsf W}$}{black!70}
\draw[-{Stealth[length=4pt]},black!70,thick] (6.75,0.55) -- (7.25,0.55);
\draw[-{Stealth[length=4pt]},black!70,thick] (-0.25,1.45) -- (-0.85,1.45);
\node[black!70,above,font=\scriptsize,align=center] at (-0.55,1.5) {\cite{Fis67}: $1.5174$\\\cite{Vig85}: $1.4917$};
\node[black!70,above,font=\scriptsize,align=center] at (6.95,0.6) {\cite{FSS76}: $1.6578$};
\end{tikzpicture}
\caption{The interval of \cref{thm:main} (shaded), compared with Fisher's bound combined with the connective-constant bound of~\cite{PT00}, Fisher's original evaluation~\cite{Fis67}, the bound of~\cite{Vig85} quoted in~\cite{BLU17} and the bound of Fr\"ohlich, Simon and Spencer (FSS)~\cite{FSS76} (the last three off scale), the single-site bound of \cref{cor:singlesite}, and the most precise Monte Carlo estimate~\cite{FXL18}. \Cref{prop:fisher} gives $1.53591$, just to the right of Fisher's bound.}
\label{fig:interval}
\end{figure}
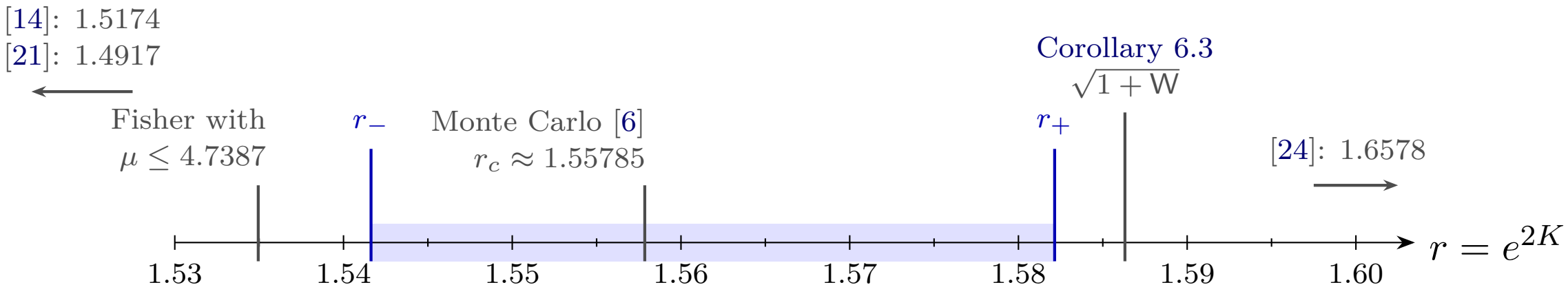

The interval has width $r_+-r_-=0.04051536763497\ldots$. This is about $3.5$ times narrower than the interval $1.5174<r_c<1.6578$ given by the bounds of Fisher~\cite{Fis67} and of Fr\"ohlich, Simon and Spencer~\cite{FSS76}. It contains the most precise Monte Carlo value~\cite{FXL18}. The two inequalities are proved by different arguments, one in each phase. For the lower bound we show (\cref{thm:lower}) that on a right neighbourhood of $r_-$ the zero-field susceptibility of every sufficiently large even torus is bounded uniformly in the box, which forces $m_+=0$. For the upper bound we show (\cref{thm:upper}) that on an explicit interval of length $4.075\cdot10^{-11}$ containing $r_+$ the squared magnetization density of large tori has a strictly positive lower limit, which forces $m_+>0$. By the sharpness of the phase transition~\cite{ABF87,DCT16}, the interval~\eqref{eq:main} therefore also encloses the threshold between finite and infinite susceptibility.

\subsection{Main ideas}

Three ideas lead to \cref{thm:main}.

\emph{The cost of a shortest path.} In Fisher's argument a walk is weighted by $u^{|\omega|}$ alone. If the walk is chosen as a \emph{shortest} connection inside a high-temperature graph, then none of the bonds spanned by its vertices, apart from its own steps, can be present. Removing all these bonds from the normalization lowers the weight of the walk by a factor $e^{-I(P,u)}$ that is exponentially small in its length. We write $I(P,u)$ as a sum of correlation integrals by switching off all removed bonds simultaneously, and we bound each integral from below on a fixed finite graph around the bond.

\emph{Counting paths with chords and shapes.} The local gain must be paid for at the chords of the path. These are summed by an exact exponential moment of the chord number, obtained from a growth argument of Rosenbluth type. Further gains attached to the shapes of consecutive bonds are summed by a transfer operator on a finite automaton with a $16$-step memory. Both estimates are uniform in the size of the torus. Applied to unweighted walks, the same automaton improves the rigorous bound on the connective constant of $\Z^3$.

\emph{An infrared bound that keeps the zero mode.} A representation of the Ising weight by planar unit vectors gives, through reflection positivity, the infrared bound with coefficient $u^{-1}+u-2\avg{\sigma_0\sigma_{e_1}}$ in place of $1/K$. With the sum rule alone this already gives $r_c<\sqrt{1+\mathsf W}<1.58632$ (\cref{cor:singlesite}). We go further by retaining the zero Fourier mode as an unknown. Exact local relations, controlled through a mixture representation of the law of the spins on a small shell, then force it to be positive.

The analytic framework of \cref{sec:paths,sec:counting,sec:infrared,sec:ordered} uses the lattice only through its reflection symmetry and through finitely many local computations. Its application to other nearest-neighbour ferromagnets is work in progress.

\subsection{Outline of the proof}

\subsubsection*{The disordered side}
Work on an even torus $\T_L=(\Z/L\Z)^3$. In the high-temperature expansion, $\avg{\sigma_x\sigma_y}_L$ is the ratio of the weight of bond sets with odd vertices $\{x,y\}$ to the weight $\cE(u)$ of even bond sets. In every such bond set choose a shortest path $P$ from $x$ to $y$. Its chords, namely the lattice bonds joining nonconsecutive vertices of $P$, are absent, for otherwise $P$ would not be shortest. Writing $D(P)$ for the set of all bonds spanned by the vertices of $P$, one obtains
\[
 \avg{\sigma_x\sigma_y}_L\le\sum_{P:x\to y}u^{|P|}e^{-I(P,u)},\qquad
 I(P,u)=\log\frac{\cE(u)}{\cE(u\one_{E\setminus D(P)})}
\]
(\cref{lem:selection}). The quantity $I(P,u)\ge0$ is the free-energy cost of switching off all bonds in $D(P)$. We switch them off simultaneously along a one-parameter interpolation. The cost then becomes a sum, over $e\in D(P)$, of integrals of the correlation across $e$ in the interpolated system (\cref{lem:interp}). Griffiths' second inequality lets us bound each integrand from below on a fixed $28$-site graph surrounding $e$ (the \emph{prism}, \cref{fig:prism}). Only the occupation of ten sites by the path enters. Exact integration of $1024$ rational functions gives a table of local lower bounds.

These local integrals are redistributed to the vertices of the path by a system of half-edge charges (\cref{sec:paths}). The result is a lower bound
\[
 I(P,u)+\theta\,c(P)\ \ge\ \gamma\,(|P|-13)_+ +\sum_i w_{\tau_i},
\]
valid for every path on every sufficiently large torus. Here $c(P)=|D(P)|-|P|$ is the number of chords, $\theta$ is a small price paid per chord, $\gamma$ is a vertex gain guaranteed at every interior vertex, and $w_{\tau_i}\ge0$ are extra gains attached to the shapes $\tau_i$ of consecutive triples of bonds (\cref{fig:shapes}). The chord price is necessary. A Hamiltonian path through the whole box spans every bond, so no gain proportional to $|P|$ can be uniform without it (\cref{rem:hamilton}).

The weights are then summed over paths (\cref{sec:counting}). Two inputs suffice. The first is a transfer bound for the shape-weighted number of self-avoiding walks, obtained from a finite automaton that remembers a $16$-step suffix and has $12\,994\,574$ states. The second is an exact bound for the exponential moment $\sum_P(5/4)^{c(P)}$ of the chord number. It follows from a Rosenbluth-type growth argument. H\"older's inequality combines the two. The outcome is a geometric bound for the susceptibility, uniform in $L$, whenever a rational inequality holds. A field-tilt argument converts this bound into $m_+=0$ (\cref{prop:disorder}).

\subsubsection*{The ordered side}
On the same tori let $q_L$ be the second moment of the magnetization density, so that $\widehat G_L(0)=L^3q_L$ for the Fourier transform of the two-point function. We prove the infrared bound $\lambda(k)\widehat G_L(k)\le u^{-1}+u-2g_L$ for all $k\ne0$, where $g_L$ is the nearest-neighbour correlation (\cref{prop:infrared}). Integrated against five nonnegative trigonometric test polynomials, it gives linear inequalities between finitely many local correlations and the unknown limit $q$ of $q_L$.

Further linear relations hold in every torus. These are translation identities, reflection-positivity squares, and comparisons with a graph from which a small region has been deleted. Each relation can be written as the mean of an explicit function of the $18$ spins on the shell $\{\|x\|_1=2\}$, integrated against their true marginal law $\mu_L$. A suitable linear combination $R$ of all $120$ relations satisfies $0\le\lim_L\mu_L(R)+Qq$ with an explicit $Q>0$. We then prove an \emph{upper} bound $\mu_L(R)\le\Uord<0$ (\cref{sec:ordered}). The unknown marginal $\mu_L$ is represented exactly as a mixture of $14\cdot2^{18}$ explicit finite measures. Finitely many inequalities restrict the mixture weights. The two bounds give $q\ge-\Uord/Q>0$, and a second field-tilt argument converts this into $m_+^2\ge q$ (\cref{prop:order}).

\begin{figure}[t]
\centering
\begin{tikzpicture}[
  node distance=4.2mm and 5mm,
  ana/.style={draw,rounded corners=2pt,text width=6.3cm,align=center,font=\footnotesize,inner sep=3pt,fill=white},
  cer/.style={draw,rounded corners=2pt,text width=6.3cm,align=center,font=\footnotesize,inner sep=3pt,fill=blue!8},
  res/.style={draw,thick,rounded corners=2pt,text width=6.3cm,align=center,font=\footnotesize\bfseries,inner sep=3pt,fill=black!6},
  arr/.style={-{Stealth[length=4pt]},semithick}]
\node[ana] (l1) {Shortest-path selection, \cref{lem:selection}\\[-1pt] $G_L\le\sum_Pu^{|P|}e^{-I(P,u)}$};
\node[ana,below=of l1] (l2) {Interpolation identity, \cref{lem:interp}\\[-1pt] $I(P,u)=\sum_{e\in D(P)}\int_0^u\frac{C_e}{1+sC_e}$};
\node[cer,below=of l2] (l3) {Prism integrals, \cref{cert:prism}\\[-1pt] $1024$ patterns, $64$ panels};
\node[cer,below=of l3] (l4) {Allocation, \cref{cert:allocation} $\Rightarrow$ \cref{prop:cost}\\[-1pt] $I+\theta c\ge\gamma(n-13)_++\sum w_\tau$};
\node[cer,below=of l4] (l5) {Path sums, \cref{cert:transfer,lem:rosenbluth,lem:holder}\\[-1pt] $\sum_P\Omega(P)e^{\theta_2c(P)}\le C_2(U_2e^{\theta_2/4})^n$};
\node[ana,below=of l5] (l6) {Uniform susceptibility bound, \cref{thm:lower}\\[-1pt] and disorder criterion, \cref{prop:disorder}};
\node[res,below=of l6] (l7) {$r_c>r_-$};
\node[ana,right=of l1] (r1) {Infrared bound, \cref{prop:infrared}\\[-1pt] $\lambda(k)\widehat G_L(k)\le u^{-1}+u-2g_L$};
\node[cer,below=of r1] (r2) {Spectral rows, \cref{lem:spectral}\\[-1pt] Green enclosures, \cref{cert:green}};
\node[ana,below=of r2] (r3) {Translation, reflection and deletion rows\\[-1pt] \cref{lem:translation,lem:reflection,lem:deletion}};
\node[cer,below=of r3] (r4) {Residual $R_y$ on $18$ shell spins\\[-1pt] \cref{prop:lowerrow}:\  $0\le\lim_L\mu_L(R_y)+Qq$};
\node[cer,below=of r4] (r5) {Mixture cap, $\mu_L(R_y)\le\Uord<0$\\[-1pt] \cref{prop:cap}, \cref{cert:fields}};
\node[cer,below=of r5] (r6) {$q\ge-\Uord/Q$ on an interval\\[-1pt] \cref{cert:interval}, \cref{prop:order}};
\node[res,below=of r6] (r7) {$r_c<r_+$};
\foreach \a/\b in {l1/l2,l2/l3,l3/l4,l4/l5,l5/l6,l6/l7,r1/r2,r3/r4,r4/r5,r5/r6,r6/r7}
  \draw[arr] (\a) -- (\b);
\draw[arr] (r2.east) -- ++(3mm,0) |- (r4.east);
\begin{scope}[on background layer]
\node[fit=(l1)(l7),draw=black!25,dashed,rounded corners=4pt,inner sep=5pt,label={[font=\small]above:Disordered side (\S\S\ref{sec:paths}--\ref{sec:lower})}] {};
\node[fit=(r1)(r7),draw=black!25,dashed,rounded corners=4pt,inner sep=5pt,label={[font=\small]above:Ordered side (\S\S\ref{sec:infrared}--\ref{sec:ordered})}] {};
\end{scope}
\end{tikzpicture}
\caption{Structure of the proof of \cref{thm:main}. Shaded boxes contain a finite certificate together with the argument that reduces the corresponding analytic statement to it. White boxes are purely analytic.}
\label{fig:flow}
\end{figure}
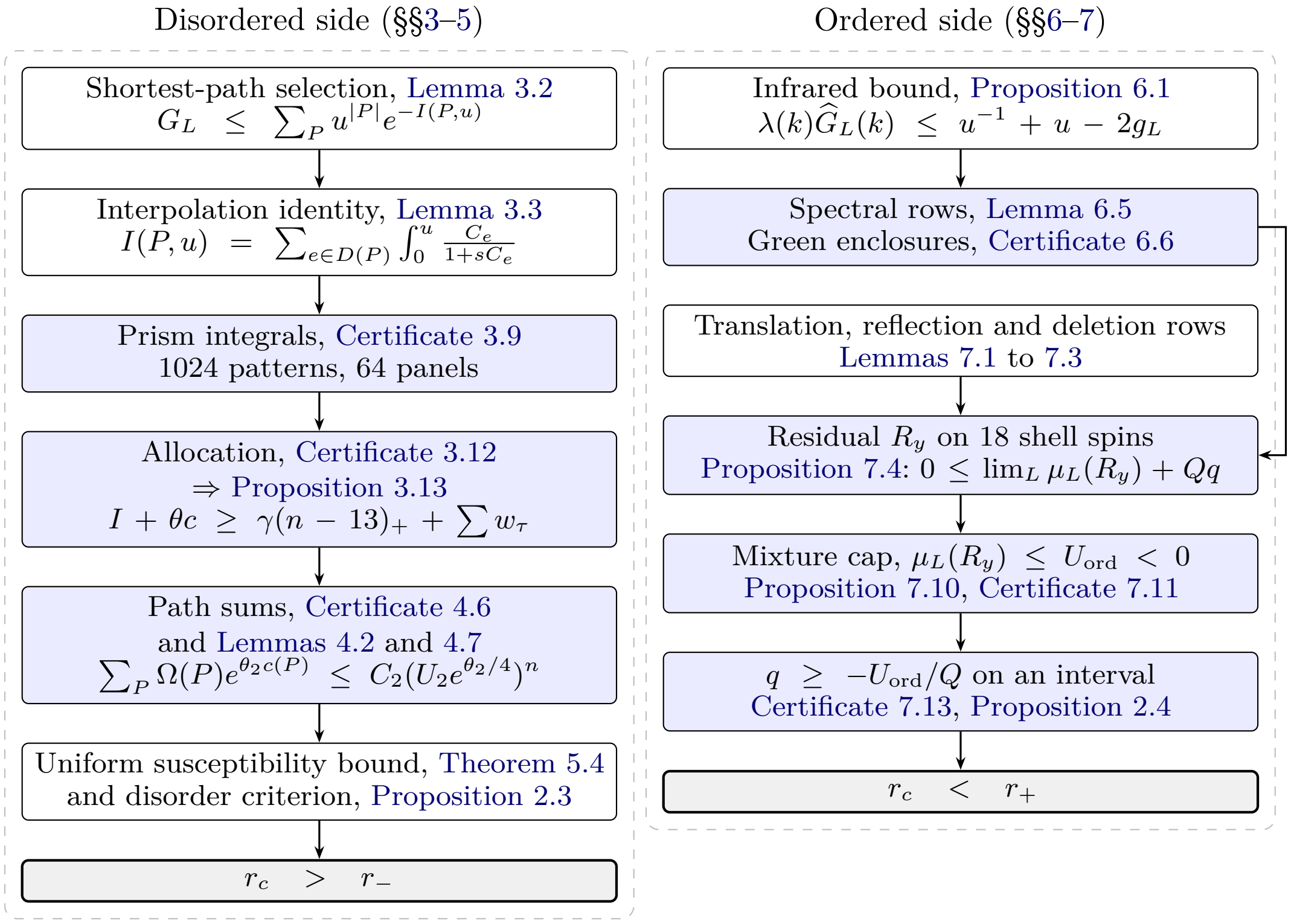

\Cref{fig:flow} summarizes the logical structure of the proof.

\subsection{The computer-assisted part}\label{sec:intro-cap}

The proof uses finitely many explicit inequalities between rational numbers. We call them \emph{certificates}. Each certificate is stated in the text in the form in which it is used, together with the mathematical argument that reduces the analytic claim to it. Their verification is described in Appendix~\ref{app:verification}. The certificates fall into three groups.
\begin{itemize}
\item \emph{Local Ising computations.} These are exact rational lower bounds for $1024$ correlation integrals on the prism (\cref{cert:prism}). On the ordered side they are the $120$ local rows and the conditional expectations on the shell (\cref{cert:multipliers,cert:fields}). All are exact finite sums over spin configurations, evaluated with directed rounding.
\item \emph{Combinatorial certificates.} These are the half-edge allocations (\cref{cert:allocation}) and the transfer bounds on the $16$-step automaton (\cref{cert:transfer}).
\item \emph{Scalar arithmetic.} This covers the Green-function enclosures (\cref{cert:green}), the endpoint inequalities (\cref{cert:endpoint-lower,cert:interval}) and the derivative bounds used to extend the ordered certificate to an interval.
\end{itemize}
The large certificates were checked by two programs with different arithmetic or different enumeration (\cref{tab:programs}). The chord-moment bound of \cref{lem:rosenbluth}, on every cubic torus, and the integer arithmetic of the allocation in \cref{cert:allocation}(a) are in addition verified in the proof assistant Lean~4. The analytic inputs, namely the high-temperature expansion, correlation inequalities, reflection positivity, Fourier analysis and the thermodynamic limits, are proved or cited in the text.

\subsection{Organization}

\Cref{sec:prelim} collects standard facts. It also proves the two thermodynamic criteria (\cref{prop:disorder,prop:order}) that turn finite-volume estimates into statements about $m_+$. \Cref{sec:paths,sec:counting,sec:lower} prove the lower bound. \Cref{sec:paths} treats the cost of induced bonds, \cref{sec:counting} the summation over paths, and \cref{sec:lower} the final arithmetic. \Cref{sec:infrared,sec:ordered} prove the upper bound. \Cref{sec:discussion} discusses possible improvements. Appendix~\ref{app:verification} describes the verification of the certificates and Appendix~\ref{app:green} the Green-function enclosures.

\section{Preliminaries}\label{sec:prelim}

This section fixes notation, recalls standard facts about the Ising model, and proves two propositions (\cref{prop:disorder,prop:order}) that convert estimates on tori into statements about the plus-state magnetization. Both halves of the proof end with one of them.

\subsection{Periodic boxes and standard inequalities}

For an even integer $L\ge4$ let $\T_L=(\Z/L\Z)^3$ be the discrete torus, $E_L$ its set of unoriented nearest-neighbour bonds, and $V=L^3$ its number of sites. Every site has six distinct neighbours and $|E_L|=3V$. We write $e_1,e_2,e_3$ for the coordinate unit vectors and $\|\cdot\|_1$ for the $\ell^1$-norm. The periodic Gibbs measure at coupling $K$ and field $h$ is
\[
 \mu_{L,K,h}(\sigma)=\frac{1}{Z_L(K,h)}\exp\Big(K\sum_{\{x,y\}\in E_L}\sigma_x\sigma_y+h\sum_{x\in\T_L}\sigma_x\Big),
 \qquad \sigma\in\{-1,1\}^{\T_L},
\]
with expectation $\avg{\cdot}_{L,K,h}$. The finite-volume pressure is $p_L(K,h)=V^{-1}\log Z_L(K,h)$. When $h=0$ we drop it from the notation. The measure $\mu_{L,K,h}$ is invariant under translations of $\T_L$ and under the $48$ signed permutations of the coordinate axes. At $h=0$ it is also invariant under the global spin flip $\sigma\mapsto-\sigma$.

We use the following classical facts.

\begin{lemma}\label{lem:standard}
Let $K>0$.
\begin{enumerate}
\item \textup{(Griffiths' inequalities~\cite{Gri67,KS68,Gin70}.)} For any finite graph with couplings $K_e\ge0$ and fields $h_x\ge0$, and any $A,B\subset V$, one has $\avg{\sigma_A}\ge0$ and $\avg{\sigma_A\sigma_B}\ge\avg{\sigma_A}\avg{\sigma_B}$, where $\sigma_A=\prod_{x\in A}\sigma_x$. Consequently $\avg{\sigma_A}$ is nondecreasing in every $K_e$ and every $h_x$.
\item \textup{(Plus states.)} For $h\ge0$, finite $\Lambda\subset\Lambda'$ and $x\in\Lambda$, one has $\avg{\sigma_x}^+_{\Lambda,K,h}\ge\avg{\sigma_x}^+_{\Lambda',K,h}$. In particular the limit $m_+(K)$ in~\eqref{eq:Kc} exists. It equals $\lim_{\Lambda\uparrow\Z^3}\avg{\sigma_x}^+_{\Lambda,K,0}$ for every $x$, and it is nondecreasing in $K$.
\item \textup{(GHS inequality~\cite{GHS70}.)} On $\T_L$ the magnetization $M_L(h)=\avg{\sigma_0}_{L,K,h}$ is concave on $[0,\infty)$.
\item \textup{(Pressure.)} For every $h\in\R$ the limits
\[
 \lim_{L\to\infty,\ L\text{ even}}p_L(K,h)\qquad\text{and}\qquad\lim_{n\to\infty}|B_n|^{-1}\log Z^+_{B_n}(K,h)
\]
exist and are equal. We denote their common value by $p(K,h)$.
\end{enumerate}
\end{lemma}

Parts~(i)--(iii) are standard; see e.g.~\cite[Chap.~3]{FV17}. For~(iv) one compares $\log Z$ for periodic, free and plus boundary conditions. Removing or inserting the $O(L^2)$ boundary bonds changes $\log Z$ by $O(KL^2)$. The existence of the free-boundary limit follows from subadditivity~\cite[Sect.~3.2]{FV17}.

\subsection{The high-temperature expansion}\label{sec:HT}

Let $G=(\mathcal V,\mathcal B)$ be a finite graph and let $a=(a_e)_{e\in\mathcal B}\in[0,1)^{\mathcal B}$ be bond weights. We consider the Ising model on $G$ with couplings $K_e=\operatorname{artanh}a_e$. For $A\subset\mathcal B$ we write $\partial A$ for the set of vertices of odd degree in $(\mathcal V,A)$. Put
\begin{equation}\label{eq:evenpoly}
 \cE_G(a)=\sum_{\substack{A\subset\mathcal B\\ \partial A=\varnothing}}\prod_{e\in A}a_e,\qquad
 \cO_G(x,y;a)=\sum_{\substack{A\subset\mathcal B\\ \partial A=\{x,y\}}}\prod_{e\in A}a_e\qquad(x\ne y).
\end{equation}
Expanding $e^{K_e\sigma_x\sigma_y}=\cosh K_e\,(1+a_e\sigma_x\sigma_y)$ and summing over spins gives
\begin{equation}\label{eq:HT}
 \avg{\sigma_x\sigma_y}_{G,a}=\frac{\cO_G(x,y;a)}{\cE_G(a)}.
\end{equation}
We call $A$ with $\partial A=\varnothing$ an \emph{even} bond set and $A$ with $\partial A=\{x,y\}$ a \emph{two-source} bond set. Both polynomials are multilinear in $a$ with nonnegative coefficients. For uniform weights $a\equiv u$ on $\T_L$ we abbreviate $\cE(u)=\cE_{\T_L}(u\one)$. For $F\subset E_L$ we write $\cE(u\one_F)$ for the even polynomial in which the bonds of $F$ have weight $u$ and the others weight $0$, i.e.\ for the even polynomial of the graph $(\T_L,F)$.

\begin{lemma}[Bond factorization]\label{lem:factor}
Let $e=\{v,w\}\in\mathcal B$ and let $G-e$ denote $G$ with $e$ removed. Then
\[
 \cE_G(a)=\cE_{G-e}(a)\bigl(1+a_e\avg{\sigma_v\sigma_w}_{G-e,a}\bigr),
 \qquad
 \frac{\partial}{\partial a_e}\log\cE_G(a)=\frac{C_e(a)}{1+a_eC_e(a)},
\]
where $C_e(a)=\avg{\sigma_v\sigma_w}_{G-e,a}$ does not depend on $a_e$.
\end{lemma}

\begin{proof}
Split the even bond sets according to whether they contain $e$. Those that do not are the even bond sets of $G-e$. If $e\in A$, then $A\setminus\{e\}$ is a two-source bond set of $G-e$ with sources $v,w$. Hence $\cE_G=\cE_{G-e}+a_e\cO_{G-e}(v,w)$. Divide by $\cE_{G-e}$ and use~\eqref{eq:HT}. The derivative formula follows because $\cE_{G-e}$ and $C_e$ do not involve $a_e$.
\end{proof}

By \cref{lem:standard}(i), $C_e(a)\ge0$ is nondecreasing in every $a_f$, $f\ne e$, and deleting bonds of $G-e$ can only decrease it. Since $x\mapsto x/(1+sx)$ is increasing on $[0,\infty)$ for $s\ge0$, the logarithmic derivative in \cref{lem:factor} inherits this monotonicity.

\subsection{Two thermodynamic criteria}\label{sec:criteria}

The finite-volume estimates of this paper are proved on tori. The following two propositions transfer them to the plus-state magnetization. Put
\[
 \chi_L(K)=\sum_{y\in\T_L}\avg{\sigma_0\sigma_y}_{L,K},\qquad
 M_L=\frac1V\sum_{x\in\T_L}\sigma_x,\qquad
 q_L(K)=\avg{M_L^2}_{L,K}=\frac{\chi_L(K)}{V}.
\]

\begin{proposition}[Disorder criterion]\label{prop:disorder}
If $X:=\sup\{\chi_L(K):L\ge L_0\text{ even}\}<\infty$ for some $L_0$, then $m_+(K)=0$.
\end{proposition}

\begin{proof}
Fix $h>0$. By translation invariance $\partial_hp_L(K,h)=M_L(h)$, $M_L(0)=0$ and $M_L'(0)=\chi_L(K)$. By \cref{lem:standard}(iii), $M_L(h')\le h'\chi_L(K)\le h'X$ for $h'\ge0$. Integrating,
\[
 p_L(K,h)-p_L(K,0)\le\tfrac12h^2X\qquad(L\ge L_0).
\]
Letting $L\to\infty$ gives $p(K,h)-p(K,0)\le\frac12h^2X$.

On the other hand, put $\overline M_n(h)=|B_n|^{-1}\sum_{x\in B_n}\avg{\sigma_x}^+_{B_n,K,h}$. This is the $h$-derivative of $|B_n|^{-1}\log Z^+_{B_n}(K,h)$. It is nondecreasing in $h$ by \cref{lem:standard}(i). By \cref{lem:standard}(ii) every term satisfies $\avg{\sigma_x}^+_{B_n,K,0}\ge m_+(K)$. Hence
\[
 |B_n|^{-1}\bigl(\log Z^+_{B_n}(K,h)-\log Z^+_{B_n}(K,0)\bigr)\ge h\,\overline M_n(0)\ge h\,m_+(K).
\]
By \cref{lem:standard}(iv), the left-hand side converges to $p(K,h)-p(K,0)$. Thus $hm_+(K)\le\frac12h^2X$ for every $h>0$, and $m_+(K)=0$.
\end{proof}

\begin{proposition}[Order criterion]\label{prop:order}
If $L_k\to\infty$ is a sequence of even integers with $q_{L_k}(K)\to q$, then $m_+(K)^2\ge q$.
\end{proposition}

\begin{proof}
We may assume $q>0$. Fix $0<a<\sqrt q$ and write $V_k=L_k^3$, $\Prob_k=\mu_{L_k,K,0}$. By spin-flip symmetry, $\pi_k:=\Prob_k(M_{L_k}\ge a)=\Prob_k(M_{L_k}\le-a)$. Since $|M_L|\le1$,
\[
 q_{L_k}\le a^2(1-2\pi_k)+2\pi_k,\qquad\text{so}\qquad \pi_k\ge\frac{q_{L_k}-a^2}{2(1-a^2)},
\]
which is bounded away from $0$ for large $k$. For $h>0$,
\[
 \frac{Z_{L_k}(K,h)}{Z_{L_k}(K,0)}=\avg{e^{hV_kM_{L_k}}}_{L_k,K}\ge\pi_ke^{hV_ka},
\]
hence $p_{L_k}(K,h)-p_{L_k}(K,0)\ge ha+V_k^{-1}\log\pi_k$. Letting $k\to\infty$ and using \cref{lem:standard}(iv) gives $p(K,h)-p(K,0)\ge ha$.

For the reverse inequality fix $R\ge1$. Convexity of $h\mapsto\log Z^+_{B_n}(K,h)$ gives
\[
 |B_n|^{-1}\bigl(\log Z^+_{B_n}(K,h)-\log Z^+_{B_n}(K,0)\bigr)\le h\,\overline M_n(h).
\]
If $x+B_R\subset B_n$, then \cref{lem:standard}(ii) and translation invariance of $\Z^3$ give $\avg{\sigma_x}^+_{B_n,K,h}\le\avg{\sigma_0}^+_{B_R,K,h}$. The remaining sites form a fraction $O(R/n)$ of $B_n$, and each contributes at most $1$. Letting $n\to\infty$ yields $ha\le p(K,h)-p(K,0)\le h\avg{\sigma_0}^+_{B_R,K,h}$. Divide by $h$ and let $h\downarrow0$. The right-hand side is a finite-volume expectation and is continuous in $h$. This gives $a\le\avg{\sigma_0}^+_{B_R,K,0}$. Letting $R\to\infty$ and then $a\uparrow\sqrt q$ proves $\sqrt q\le m_+(K)$.
\end{proof}

\begin{corollary}\label{cor:monotone}
\begin{enumerate}
\item If $m_+(K)=0$ for all $K$ in a right neighbourhood $[K_0,K_0+\delta)$, then $K_c\ge K_0+\delta>K_0$.
\item If $m_+(K_1)>0$, then $K_c\le K_1$.
\end{enumerate}
\end{corollary}

\begin{proof}
Both statements follow from the definition~\eqref{eq:Kc} and the monotonicity of $m_+$ in $K$ (\cref{lem:standard}(ii)).
\end{proof}

\subsection{Certificates}\label{sec:conventions}

A \emph{certificate} is a finite list of assertions about explicitly given rational numbers or integers. Examples are a table of rational lower bounds for finitely many integrals, an integer vector $v$ satisfying finitely many row inequalities $(Av)_i\le\Lambda v_i$, or a single strict inequality between two rationals. All certificates are formulated so that their truth can be decided by exact integer arithmetic. Where a real quantity such as an integral, a power $w^j$ or a square root must be bounded, the bound is rounded outward: down for lower bounds and up for upper bounds. The rounding rule is part of the statement of the certificate. Decimal numbers in certificates are exact decimal rationals, and integers written in units of $10^{-12}$ are exact numerators. The statement of each certificate in the text records what is asserted. Appendix~\ref{app:verification} records how it was checked.

\section{Shortest paths and the cost of induced bonds}\label{sec:paths}

This section shows that the weight of a shortest path decays exponentially in its length, up to a price paid for each chord (\cref{prop:cost}). The argument has four steps. A shortest path in a high-temperature graph forbids all bonds spanned by its vertices (\cref{lem:selection}). The cost of removing these bonds is a sum of correlation integrals (\cref{lem:interp}). Each integral is bounded below on a fixed $28$-site graph (\cref{lem:prismbound,cert:prism}). Finally the integrals are redistributed to the vertices of the path (\cref{def:allocation,cert:allocation}).

Throughout \cref{sec:paths,sec:counting,sec:lower}, $L\ge8$ is even and $G_L(x,y)=\avg{\sigma_x\sigma_y}_{L,K}$ denotes the zero-field periodic two-point function.

\subsection{Selected paths}

A \emph{path} $P=(v_0,v_1,\dots,v_n)$ of length $|P|=n\ge1$ on $\T_L$ is a sequence of distinct sites with $v_{i-1}\sim v_i$ for all $i$. Its vertex set is $V(P)$ and its bond set is $E(P)=\{\{v_{i-1},v_i\}:1\le i\le n\}$. We write $\cS_{xy}$ for the set of paths from $x$ to $y$.

\begin{definition}[Induced bonds and chords]\label{def:induced}
The \emph{induced bond set} of $P$ is $D(P)=\{\{x,y\}\in E_L:\ x,y\in V(P)\}$. Its elements outside $E(P)$ are the \emph{chords} of $P$, and $c(P)=|D(P)|-|P|$ is the \emph{chord number}. A chord joins two nonconsecutive vertices of $P$. Chords across the periodic identification are included.
\end{definition}

\begin{lemma}[Selection of a shortest path]\label{lem:selection}
For $x\ne y$ in $\T_L$,
\begin{equation}\label{eq:selection}
 G_L(x,y)\le\sum_{P\in\cS_{xy}}u^{|P|}\,e^{-I(P,u)},\qquad
 I(P,u)=\log\frac{\cE(u)}{\cE(u\one_{E_L\setminus D(P)})}\ \ge0 .
\end{equation}
\end{lemma}

\begin{proof}
By~\eqref{eq:HT}, $G_L(x,y)=\cE(u)^{-1}\sum_Au^{|A|}$, where $A$ ranges over two-source bond sets with $\partial A=\{x,y\}$. Every such $A$ contains a path from $x$ to $y$. Fix a total order on $\cS_{xy}$ and let $P(A)$ be the first among the shortest such paths contained in $A$. If a chord $\{v_i,v_j\}$, $j>i+1$, of $P(A)$ belonged to $A$, then $(v_0,\dots,v_i,v_j,\dots,v_n)$ would be a strictly shorter path in $A$. Hence $A\cap D(P(A))=E(P(A))$. The set $B=A\setminus E(P(A))$ is even, because deleting a path changes the parity of degrees only at its endpoints, and $B\subset E_L\setminus D(P(A))$. The map $A\mapsto(P(A),B)$ is injective since $A=E(P(A))\cup B$. Therefore
\[
 \sum_Au^{|A|}\le\sum_{P\in\cS_{xy}}u^{|P|}\sum_{\substack{B\subset E_L\setminus D(P)\\ \partial B=\varnothing}}u^{|B|}
 =\sum_{P\in\cS_{xy}}u^{|P|}\,\cE(u\one_{E_L\setminus D(P)}).
\]
Dividing by $\cE(u)$ gives~\eqref{eq:selection}. Finally $I(P,u)\ge0$ because $\cE$ has nonnegative coefficients.
\end{proof}

Fisher's bound~\cite{Fis67} is the special case in which the factor $e^{-I(P,u)}$ is dropped. The rest of this section bounds $I(P,u)$ from below.

\subsection{Interpolation}

Fix $P$ and write $D=D(P)$. For $s\in[0,u]$ let $a^{(s)}\in[0,1)^{E_L}$ be the bond weights
\[
 a^{(s)}_f=s\quad(f\in D),\qquad a^{(s)}_f=u\quad(f\notin D).
\]
For $e=\{v,w\}\in D$ let $C_e(s)=\avg{\sigma_v\sigma_w}_{\T_L-e,\,a^{(s)}}$ be the correlation across $e$ in the interpolated model with $e$ removed.

\begin{lemma}[Interpolation identity]\label{lem:interp}
\begin{equation}\label{eq:interp}
 I(P,u)=\sum_{e\in D(P)}\int_0^u\frac{C_e(s)}{1+sC_e(s)}\,\dd s .
\end{equation}
\end{lemma}

\begin{proof}
Let $\phi(s)=\log\cE_{\T_L}(a^{(s)})$. Then $\phi(u)=\log\cE(u)$ and $\phi(0)=\log\cE(u\one_{E_L\setminus D})$, so $I(P,u)=\phi(u)-\phi(0)$. By the chain rule, $\phi'(s)=\sum_{e\in D}\partial_{a_e}\log\cE(a)|_{a=a^{(s)}}$, and \cref{lem:factor} identifies each term with $C_e(s)/(1+sC_e(s))$.
\end{proof}

All terms in~\eqref{eq:interp} refer to the \emph{same} interpolated system. They can therefore be estimated independently, and the auxiliary graphs used for different bonds may overlap. By \cref{lem:factor} and the remark after it we obtain the following.

\begin{corollary}[Local comparison]\label{cor:local}
Let $e=\{v,w\}\in D(P)$, let $\Gamma$ be a subgraph of $\T_L-e$ containing $v,w$, and let $b$ be bond weights on $\Gamma$ with $0\le b_f\le a^{(s)}_f$ for every bond $f$ of $\Gamma$. Then $C_e(s)\ge\avg{\sigma_v\sigma_w}_{\Gamma,b}$, and the integrand in~\eqref{eq:interp} is at least $\avg{\sigma_v\sigma_w}_{\Gamma,b}/(1+s\avg{\sigma_v\sigma_w}_{\Gamma,b})$.
\end{corollary}

\subsection{The prism}

Let $e=\{v,w\}$ be a bond of $\T_L$ with $w=v+\delta$, $\delta$ a unit vector, which we call the \emph{axial} direction of $e$. Let $\pm\epsilon_1,\pm\epsilon_2$ be the four unit vectors orthogonal to $\delta$, the \emph{transverse} directions. The two endpoints $v,w$ are the \emph{hubs} of $e$.

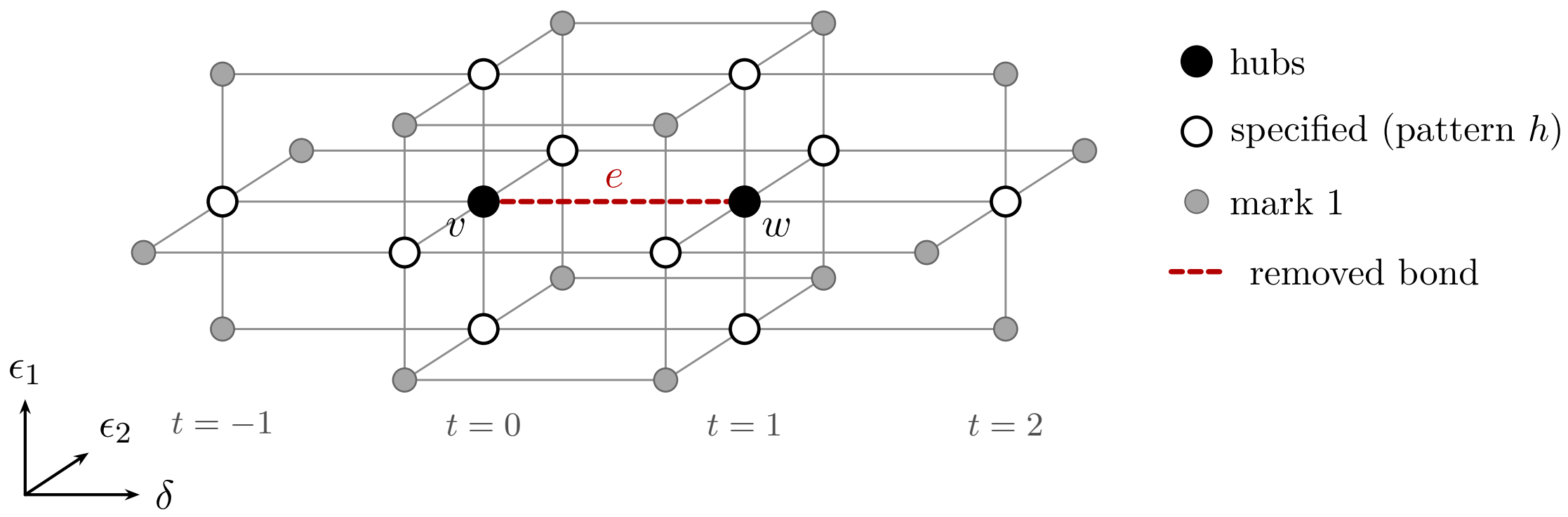
\begin{figure}[t]
\centering
\begin{tikzpicture}[scale=1.2,line cap=round]
\draw[black!45,line width=0.55pt] (0.620,-0.600) -- (0.620,0.400);
\draw[black!45,line width=0.55pt] (0.620,-0.600) -- (2.670,-0.600);
\draw[black!45,line width=0.55pt] (0.620,0.400) -- (0.620,1.400);
\draw[black!45,line width=0.55pt] (0.620,0.400) -- (2.670,0.400);
\draw[black!45,line width=0.55pt] (0.620,0.400) -- (-1.430,0.400);
\draw[black!45,line width=0.55pt] (0.620,1.400) -- (2.670,1.400);
\draw[black!45,line width=0.55pt] (2.670,-0.600) -- (2.670,0.400);
\draw[black!45,line width=0.55pt] (2.670,0.400) -- (2.670,1.400);
\draw[black!45,line width=0.55pt] (2.670,0.400) -- (4.720,0.400);
\draw[black!45,line width=0.55pt] (0.000,-1.000) -- (0.620,-0.600);
\draw[black!45,line width=0.55pt] (0.000,0.000) -- (0.620,0.400);
\draw[black!45,line width=0.55pt] (0.000,1.000) -- (0.620,1.400);
\draw[black!45,line width=0.55pt] (2.050,-1.000) -- (2.670,-0.600);
\draw[black!45,line width=0.55pt] (2.050,0.000) -- (2.670,0.400);
\draw[black!45,line width=0.55pt] (2.050,1.000) -- (2.670,1.400);
\draw[black!45,line width=0.55pt] (-2.050,0.000) -- (-1.430,0.400);
\draw[black!45,line width=0.55pt] (4.100,0.000) -- (4.720,0.400);
\draw[black!45,line width=0.55pt] (0.000,-1.000) -- (0.000,0.000);
\draw[black!45,line width=0.55pt] (0.000,-1.000) -- (2.050,-1.000);
\draw[black!45,line width=0.55pt] (0.000,-1.000) -- (-2.050,-1.000);
\draw[black!45,line width=0.55pt] (0.000,0.000) -- (0.000,1.000);
\draw[black!45,line width=0.55pt] (0.000,0.000) -- (-2.050,0.000);
\draw[black!45,line width=0.55pt] (0.000,1.000) -- (2.050,1.000);
\draw[black!45,line width=0.55pt] (0.000,1.000) -- (-2.050,1.000);
\draw[black!45,line width=0.55pt] (2.050,-1.000) -- (2.050,0.000);
\draw[black!45,line width=0.55pt] (2.050,-1.000) -- (4.100,-1.000);
\draw[black!45,line width=0.55pt] (2.050,0.000) -- (2.050,1.000);
\draw[black!45,line width=0.55pt] (2.050,0.000) -- (4.100,0.000);
\draw[black!45,line width=0.55pt] (2.050,1.000) -- (4.100,1.000);
\draw[black!45,line width=0.55pt] (-2.050,0.000) -- (-2.050,1.000);
\draw[black!45,line width=0.55pt] (-2.050,0.000) -- (-2.050,-1.000);
\draw[black!45,line width=0.55pt] (4.100,0.000) -- (4.100,1.000);
\draw[black!45,line width=0.55pt] (4.100,0.000) -- (4.100,-1.000);
\draw[black!45,line width=0.55pt] (-0.620,-1.400) -- (0.000,-1.000);
\draw[black!45,line width=0.55pt] (-0.620,-0.400) -- (0.000,0.000);
\draw[black!45,line width=0.55pt] (-0.620,0.600) -- (0.000,1.000);
\draw[black!45,line width=0.55pt] (1.430,-1.400) -- (2.050,-1.000);
\draw[black!45,line width=0.55pt] (1.430,-0.400) -- (2.050,0.000);
\draw[black!45,line width=0.55pt] (1.430,0.600) -- (2.050,1.000);
\draw[black!45,line width=0.55pt] (-2.050,0.000) -- (-2.670,-0.400);
\draw[black!45,line width=0.55pt] (4.100,0.000) -- (3.480,-0.400);
\draw[black!45,line width=0.55pt] (-0.620,-1.400) -- (-0.620,-0.400);
\draw[black!45,line width=0.55pt] (-0.620,-1.400) -- (1.430,-1.400);
\draw[black!45,line width=0.55pt] (-0.620,-0.400) -- (-0.620,0.600);
\draw[black!45,line width=0.55pt] (-0.620,-0.400) -- (1.430,-0.400);
\draw[black!45,line width=0.55pt] (-0.620,-0.400) -- (-2.670,-0.400);
\draw[black!45,line width=0.55pt] (-0.620,0.600) -- (1.430,0.600);
\draw[black!45,line width=0.55pt] (1.430,-1.400) -- (1.430,-0.400);
\draw[black!45,line width=0.55pt] (1.430,-0.400) -- (1.430,0.600);
\draw[black!45,line width=0.55pt] (1.430,-0.400) -- (3.480,-0.400);
\draw[red!70!black,line width=1.4pt,dash pattern=on 3pt off 2pt] (0.000,0.000) -- (2.050,0.000);
\filldraw[fill=black!35,draw=black!60,line width=0.5pt] (0.620,-0.600) circle (2.6pt);
\filldraw[fill=black!35,draw=black!60,line width=0.5pt] (2.670,-0.600) circle (2.6pt);
\filldraw[fill=white,draw=black,line width=0.9pt] (0.620,0.400) circle (3.2pt);
\filldraw[fill=white,draw=black,line width=0.9pt] (2.670,0.400) circle (3.2pt);
\filldraw[fill=black!35,draw=black!60,line width=0.5pt] (4.720,0.400) circle (2.6pt);
\filldraw[fill=black!35,draw=black!60,line width=0.5pt] (-1.430,0.400) circle (2.6pt);
\filldraw[fill=black!35,draw=black!60,line width=0.5pt] (0.620,1.400) circle (2.6pt);
\filldraw[fill=black!35,draw=black!60,line width=0.5pt] (2.670,1.400) circle (2.6pt);
\filldraw[fill=white,draw=black,line width=0.9pt] (0.000,1.000) circle (3.2pt);
\filldraw[fill=black!35,draw=black!60,line width=0.5pt] (4.100,1.000) circle (2.6pt);
\filldraw[fill=white,draw=black,line width=0.9pt] (2.050,1.000) circle (3.2pt);
\filldraw[fill=white,draw=black,line width=0.9pt] (0.000,-1.000) circle (3.2pt);
\filldraw[fill=black!35,draw=black!60,line width=0.5pt] (-2.050,1.000) circle (2.6pt);
\filldraw[fill=white,draw=black,line width=0.9pt] (2.050,-1.000) circle (3.2pt);
\filldraw[fill=black!35,draw=black!60,line width=0.5pt] (4.100,-1.000) circle (2.6pt);
\fill[black] (0.000,0.000) circle (3.6pt);
\fill[black] (2.050,0.000) circle (3.6pt);
\filldraw[fill=white,draw=black,line width=0.9pt] (4.100,0.000) circle (3.2pt);
\filldraw[fill=black!35,draw=black!60,line width=0.5pt] (-2.050,-1.000) circle (2.6pt);
\filldraw[fill=white,draw=black,line width=0.9pt] (-2.050,0.000) circle (3.2pt);
\filldraw[fill=black!35,draw=black!60,line width=0.5pt] (-0.620,-1.400) circle (2.6pt);
\filldraw[fill=black!35,draw=black!60,line width=0.5pt] (1.430,-1.400) circle (2.6pt);
\filldraw[fill=white,draw=black,line width=0.9pt] (-0.620,-0.400) circle (3.2pt);
\filldraw[fill=white,draw=black,line width=0.9pt] (1.430,-0.400) circle (3.2pt);
\filldraw[fill=black!35,draw=black!60,line width=0.5pt] (3.480,-0.400) circle (2.6pt);
\filldraw[fill=black!35,draw=black!60,line width=0.5pt] (-2.670,-0.400) circle (2.6pt);
\filldraw[fill=black!35,draw=black!60,line width=0.5pt] (-0.620,0.600) circle (2.6pt);
\filldraw[fill=black!35,draw=black!60,line width=0.5pt] (1.430,0.600) circle (2.6pt);
\node[below left=1pt] at (0.000,0.000) {$v$};
\node[below right=1pt] at (2.050,0.000) {$w$};
\node[red!70!black,above] at (1.025,0.000) {$e$};
\draw[-{Stealth[length=4pt]},thick] (-3.600,-2.300) -- ++(0.9,0) node[right] {$\delta$};
\draw[-{Stealth[length=4pt]},thick] (-3.600,-2.300) -- ++(0,0.75) node[above] {$\epsilon_1$};
\draw[-{Stealth[length=4pt]},thick] (-3.600,-2.300) -- ++(0.50,0.33) node[above right=-2pt] {$\epsilon_2$};
\node[font=\footnotesize,black!70] at (-2.050,-1.75) {$t=-1$};
\node[font=\footnotesize,black!70] at (0.000,-1.75) {$t=0$};
\node[font=\footnotesize,black!70] at (2.050,-1.75) {$t=1$};
\node[font=\footnotesize,black!70] at (4.100,-1.75) {$t=2$};
\fill[black] (5.6,1.1) circle (3.6pt) node[right=5pt,font=\small] {hubs};
\filldraw[fill=white,draw=black,line width=0.9pt] (5.6,0.55) circle (3.2pt) node[right=5pt,font=\small] {specified (pattern $h$)};
\filldraw[fill=black!35,draw=black!60] (5.6,0.0) circle (2.6pt) node[right=5pt,font=\small] {mark $1$};
\draw[red!70!black,line width=1.4pt,dash pattern=on 3pt off 2pt] (5.3999999999999995,-0.5499999999999998) -- (5.8,-0.5499999999999998) node[right=3pt,font=\small,black] {removed bond};
\end{tikzpicture}
\caption{The prism $\Pi(e)$ of \cref{def:prism}: two $3\times3$ squares at axial coordinates $t=0,1$ and two crosses at $t=-1,2$, with all $51$ lattice bonds among the $28$ sites except $e$. The pattern $h(e)$ records which of the ten specified sites belong to the path. The remaining $16$ sites are treated as occupied. Bonds between two sites of mark $1$ carry the interpolation weight $s$, all other bonds the reference weight $u_0$.}
\label{fig:prism}
\end{figure}

\begin{definition}[Prism]\label{def:prism}
The \emph{prism} of $e$ is the induced subgraph of $\T_L$ on the $28$ sites
\[
 \Pi(e)=\{v+t\delta+y\epsilon_1+z\epsilon_2:\ t\in\{0,1\},\ |y|,|z|\le1\}\ \cup\
 \{v+t\delta+y\epsilon_1+z\epsilon_2:\ t\in\{-1,2\},\ |y|+|z|\le1\},
\]
with the bond $e$ removed. It has $51-1=50$ bonds (\cref{fig:prism}). Since $L\ge8$, the sites are distinct and the bonds are those of $\Z^3$.
\end{definition}

Ten sites of $\Pi(e)$ are adjacent to a hub: the four transverse neighbours $v\pm\epsilon_i$ and the opposite axial neighbour $v-\delta$ of $v$, and likewise $w\pm\epsilon_i$ and $w+\delta$. These are the \emph{specified sites}. The other $16$ sites are the eight corners of the two $3\times3$ squares and the eight transverse arms of the two end crosses.

\begin{definition}[Pattern]\label{def:pattern}
Let $P$ be a path and $e\in D(P)$ with hubs $v,w$. The \emph{pattern} $h(e)$ consists of the ten indicators $\one\{x\in V(P)\}$ of the specified sites. We write it as $h=(\mathbf a,o;\mathbf a',o')$, where $\mathbf a\in\{0,1\}^4$ records the transverse neighbours of $v$, $o$ the opposite axial neighbour of $v$, and $\mathbf a',o'$ the same at $w$. There are $2^{10}=1024$ patterns.
\end{definition}

Fix a \emph{reference weight} $u_0\in(0,1)$. For a pattern $h$ assign a \emph{mark} $m(x)\in\{0,1\}$ to each site of $\Pi(e)$. The hubs get mark $1$. A specified site gets the indicator recorded in $h$. The $16$ other sites get mark $1$. For $s\in[0,u_0]$ give the bond $f=\{x,x'\}$ of $\Pi(e)$ the weight
\begin{equation}\label{eq:prismweights}
 b^{h,s}_f=s\ \text{ if } m(x)=m(x')=1,\qquad b^{h,s}_f=u_0\ \text{ otherwise}.
\end{equation}
Let $N_h(s)$ and $E_h(s)$ be the two-source polynomial $\cO_{\Pi(e)}(v,w;b^{h,s})$ and the even polynomial $\cE_{\Pi(e)}(b^{h,s})$. They are polynomials in $s$ with nonnegative rational coefficients when $u_0\in\mathbb Q$. They depend only on $h$ and $u_0$, not on $e$ or $L$. Define
\begin{equation}\label{eq:ellh}
 \ell_h(u_0)=\int_0^{u_0}\frac{N_h(s)}{E_h(s)+sN_h(s)}\,\dd s .
\end{equation}

\begin{lemma}[Local lower bound]\label{lem:prismbound}
Let $u\ge u_0$. For every path $P$ on $\T_L$,
\[
 I(P,u)\ \ge\ \sum_{e\in D(P)}\ell_{h(e)}(u_0).
\]
\end{lemma}

\begin{proof}
Fix $e\in D(P)$ and $s\in[0,u_0]$. We claim $b^{h(e),s}_f\le a^{(s)}_f$ for every bond $f=\{x,x'\}$ of $\Pi(e)$. If $f\in D(P)$, then $x,x'\in V(P)$. The marks dominate the true indicators of $V(P)$: the marks of hubs and specified sites are the true indicators, and the remaining sites have mark $1$. Hence $m(x)=m(x')=1$ and $b_f=s=a^{(s)}_f$. If $f\notin D(P)$, then $a^{(s)}_f=u\ge u_0\ge b_f$. By \cref{cor:local} with $\Gamma=\Pi(e)$ and~\eqref{eq:HT}, the integrand in~\eqref{eq:interp} is at least $N_h(s)/(E_h(s)+sN_h(s))$ for $s\le u_0$. The integrand is nonnegative on $[u_0,u]$, so the integral over $[0,u]$ is at least $\ell_{h(e)}(u_0)$. Summing over $e\in D(P)$ and using \cref{lem:interp} gives the claim.
\end{proof}

The rule ``unknown sites are occupied'' covers every possible continuation of the path outside the ten specified sites, including returns of the path into the prism, without enumerating them. The price is a pessimistic prism integral. Because $\ell_h(u_0)$ does not depend on $u\ge u_0$, a certificate computed at $u_0$ applies to every stronger coupling.

\begin{lemma}[Panel quadrature]\label{lem:panel}
For $0\le\alpha<\beta$,
\[
 \int_\alpha^\beta\frac{N_h(s)}{E_h(s)+sN_h(s)}\,\dd s\ \ge\ (\beta-\alpha)\,\frac{N_h\bigl(\tfrac{\alpha+\beta}2\bigr)}{E_h(\beta)+\beta N_h(\beta)} .
\]
\end{lemma}

\begin{proof}
The denominator has nonnegative coefficients, so it is nondecreasing on $[0,\infty)$ and bounded above by its value at $\beta$. The numerator has nonnegative coefficients and is therefore convex on $[0,\infty)$. Its mean over $[\alpha,\beta]$ is at least its value at the midpoint (Jensen).
\end{proof}

\begin{certificate}[Prism integrals]\label{cert:prism}
Let $u_0$ be one of the two reference weights
\[
 u_*=\frac{10829}{50829}\ \ (r_*=\tfrac{30829}{20000}),\qquad u_2=\frac{677}{3177}\ \ (r_2=\tfrac{1927}{1250}).
\]
Split $[0,u_0]$ into $64$ equal panels. For every pattern $h$ and every panel, the right-hand side of \cref{lem:panel} is evaluated exactly, multiplied by $10^{12}$ and rounded down to an integer. The sum of these $64$ integers, divided by $10^{12}$, is a rational number $\underline\ell_h(u_0)\le\ell_h(u_0)$. The $1024$ patterns fall into $123$ orbits under the symmetries of the transverse square and the exchange of the two hubs. The tables $(\underline\ell_h(u_*))_h$ and $(\underline\ell_h(u_2))_h$ are the input to \cref{cert:allocation}.
\end{certificate}

To give an idea of the size of these numbers, at $u_0=u_*$ the pattern of a straight bond with no transverse neighbours and both axial neighbours occupied ($\mathbf a=\mathbf a'=0$, $o=o'=1$) has $\underline\ell_h=0.010880485789$. The fully occupied pattern has $\underline\ell_h=0.002731285336$. The corresponding upper sums, with midpoint and right endpoint replaced by the trapezoid and the left endpoint, differ by less than $10^{-5}$. The local cost of an induced bond thus decreases by a factor of about four as its neighbourhood fills up. This is the effect that the allocation below has to control.

\subsection{Half-edge charges}

We now redistribute the integrals $\underline\ell_{h(e)}$ to the vertices of $P$. The scheme is a discharging argument. Each induced bond splits its prism integral between its two endpoints according to a fixed table, and the table is chosen so that every interior vertex of the path collects at least $\gamma$, whatever its occupied neighbourhood. A chord may hand out slightly more than its integral, and the excess is charged to the chord price $\theta$. Path bonds may in addition keep a shape charge $w_\tau$, which is summed separately in \cref{sec:counting}. For $e\in D(P)$ with endpoint $v$, let $\delta$ be the direction from $v$ to the other endpoint. The \emph{type} of the half-edge $(e,v)$ is
\begin{equation}\label{eq:type}
 t_v(e)=j+5o\in\{0,\dots,9\},\qquad j=\#\{\epsilon\perp\delta:\ v+\epsilon\in V(P)\},\quad o=\one\{v-\delta\in V(P)\}.
\end{equation}
The \emph{role} of $e$ is \emph{path} if $e\in E(P)$ and \emph{chord} otherwise. The type of each half-edge of $e$ is a function of the pattern $h(e)$.

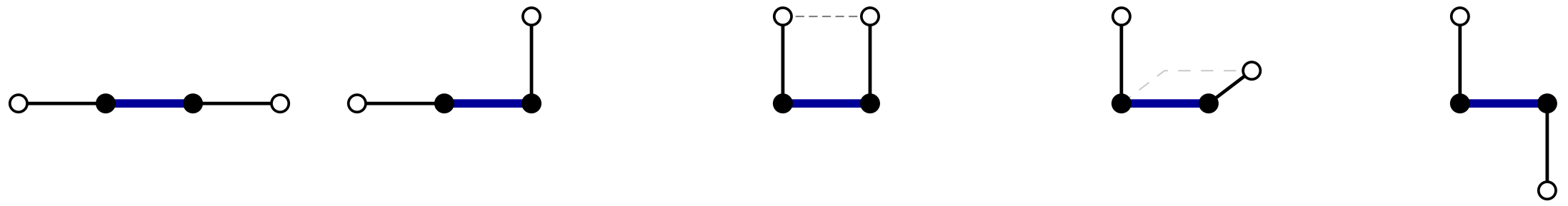
\begin{figure}[t]
\centering
\begin{tikzpicture}[scale=0.95,line cap=round,line join=round]
\draw[black,line width=0.9pt] (-0.850,0.000) -- (0.000,0.000);
\draw[black,line width=0.9pt] (0.850,0.000) -- (1.700,0.000);
\draw[blue!60!black,line width=2.2pt] (0.000,0.000) -- (0.850,0.000);
\filldraw[fill=white,draw=black,line width=0.7pt] (-0.850,0.000) circle (2.4pt);
\filldraw[fill=black,draw=black,line width=0.7pt] (0.000,0.000) circle (2.4pt);
\filldraw[fill=black,draw=black,line width=0.7pt] (0.850,0.000) circle (2.4pt);
\filldraw[fill=white,draw=black,line width=0.7pt] (1.700,0.000) circle (2.4pt);
\node[font=\small] at (0.420,-1.25) {$\mathrm{S}$ (straight)};
\node[font=\footnotesize] at (0.420,-1.7) {$1127558275$};
\draw[black,line width=0.9pt] (2.450,0.000) -- (3.300,0.000);
\draw[black,line width=0.9pt] (4.150,0.000) -- (4.150,0.850);
\draw[blue!60!black,line width=2.2pt] (3.300,0.000) -- (4.150,0.000);
\filldraw[fill=white,draw=black,line width=0.7pt] (2.450,0.000) circle (2.4pt);
\filldraw[fill=black,draw=black,line width=0.7pt] (3.300,0.000) circle (2.4pt);
\filldraw[fill=black,draw=black,line width=0.7pt] (4.150,0.000) circle (2.4pt);
\filldraw[fill=white,draw=black,line width=0.7pt] (4.150,0.850) circle (2.4pt);
\node[font=\small] at (3.720,-1.25) {$\mathrm{A}$ (one axial)};
\node[font=\footnotesize] at (3.720,-1.7) {$544046375$};
\draw[black!50,dash pattern=on 2pt off 1.5pt] (6.600,0.850) -- (7.450,0.850);
\draw[black,line width=0.9pt] (6.600,0.850) -- (6.600,0.000);
\draw[black,line width=0.9pt] (7.450,0.000) -- (7.450,0.850);
\draw[blue!60!black,line width=2.2pt] (6.600,0.000) -- (7.450,0.000);
\filldraw[fill=white,draw=black,line width=0.7pt] (6.600,0.850) circle (2.4pt);
\filldraw[fill=black,draw=black,line width=0.7pt] (6.600,0.000) circle (2.4pt);
\filldraw[fill=black,draw=black,line width=0.7pt] (7.450,0.000) circle (2.4pt);
\filldraw[fill=white,draw=black,line width=0.7pt] (7.450,0.850) circle (2.4pt);
\node[font=\small] at (7.020,-1.25) {$\mathrm{U}$ (equal)};
\node[font=\footnotesize] at (7.020,-1.7) {$544046375$};
\draw[black!20,dashed] (9.900,0.000) -- (10.320,0.320) -- (11.170,0.320);
\draw[black,line width=0.9pt] (9.900,0.850) -- (9.900,0.000);
\draw[black,line width=0.9pt] (10.750,0.000) -- (11.170,0.320);
\draw[blue!60!black,line width=2.2pt] (9.900,0.000) -- (10.750,0.000);
\filldraw[fill=white,draw=black,line width=0.7pt] (9.900,0.850) circle (2.4pt);
\filldraw[fill=black,draw=black,line width=0.7pt] (9.900,0.000) circle (2.4pt);
\filldraw[fill=black,draw=black,line width=0.7pt] (10.750,0.000) circle (2.4pt);
\filldraw[fill=white,draw=black,line width=0.7pt] (11.170,0.320) circle (2.4pt);
\node[font=\small] at (10.320,-1.25) {$\mathrm{T}$ (perpendicular)};
\node[font=\footnotesize] at (10.320,-1.7) {$94806903$};
\draw[black,line width=0.9pt] (13.200,0.850) -- (13.200,0.000);
\draw[black,line width=0.9pt] (14.050,0.000) -- (14.050,-0.850);
\draw[blue!60!black,line width=2.2pt] (13.200,0.000) -- (14.050,0.000);
\filldraw[fill=white,draw=black,line width=0.7pt] (13.200,0.850) circle (2.4pt);
\filldraw[fill=black,draw=black,line width=0.7pt] (13.200,0.000) circle (2.4pt);
\filldraw[fill=black,draw=black,line width=0.7pt] (14.050,0.000) circle (2.4pt);
\filldraw[fill=white,draw=black,line width=0.7pt] (14.050,-0.850) circle (2.4pt);
\node[font=\small] at (13.620,-1.25) {$\mathrm{Z}$ (opposite)};
\node[font=\footnotesize] at (13.620,-1.7) {$0$};
\end{tikzpicture}
\caption{The five shapes of a triple of consecutive path bonds (\cref{def:shape}). The middle bond, whose endpoints are the two hubs, is drawn thick. The numbers are the shape charges $10^{12}w_\tau$ of \cref{tab:shapes}. In shape $\mathrm U$ the two outer vertices span a chord (dashed).}
\label{fig:shapes}
\end{figure}

\begin{definition}[Shapes]\label{def:shape}
Let $P=(v_0,\dots,v_n)$ and $1\le i\le n-2$. Put $\delta=v_{i+1}-v_i$, $d_-=v_{i-1}-v_i$ and $d_+=v_{i+2}-v_{i+1}$. Then $d_-\ne\delta$ and $d_+\ne-\delta$. The \emph{shape} $\tau_i$ of the triple $(v_{i-1}v_i,v_iv_{i+1},v_{i+1}v_{i+2})$ is
\begin{center}
\begin{tabular}{ll}
 $\mathrm S$ (straight) & $d_-=-\delta$ and $d_+=\delta$;\\
 $\mathrm A$ (one axial) & exactly one of $d_-=-\delta$, $d_+=\delta$ holds;\\
 $\mathrm U$ (equal transverse) & $d_\pm\perp\delta$ and $d_-=d_+$;\\
 $\mathrm T$ (perpendicular transverse) & $d_\pm\perp\delta$ and $d_-\perp d_+$;\\
 $\mathrm Z$ (opposite transverse) & $d_\pm\perp\delta$ and $d_-=-d_+$.
\end{tabular}
\end{center}
See \cref{fig:shapes}. In shape $\mathrm U$ the outer vertices $v_{i-1},v_{i+2}$ are adjacent and span a chord.
\end{definition}

\begin{definition}[Admissible allocation]\label{def:allocation}
Let $\ell=(\ell_h)_h$ be a table of $1024$ nonnegative numbers. An \emph{allocation} consists of path charges $\alpha^{\rm P}_t\ge0$ and chord charges $\alpha^{\rm C}_t\ge0$ for $t=0,\dots,9$, shape charges $w_\tau\ge0$ for $\tau\in\{\mathrm S,\mathrm A,\mathrm U,\mathrm T,\mathrm Z\}$, a chord price $\theta\ge0$ and a vertex gain $\gamma$. It is \emph{admissible for $\ell$} if the following hold.
\begin{enumerate}[label=\textup{(A\arabic*)}]
\item \emph{Path inequalities.} For every pattern $h$, every transverse or opposite-axial direction $d_-$ at $v$ whose site is occupied in $h$, and every such direction $d_+$ at $w$, one has
$\alpha^{\rm P}_{t_v}+\alpha^{\rm P}_{t_w}+w_\tau\le\ell_h$. Here $t_v,t_w$ are the types determined by $h$ and $\tau$ is the shape determined by $(d_-,d_+)$.
\item \emph{Chord inequalities.} For every pattern $h$ in which each hub has at least two occupied specified neighbours, $\alpha^{\rm C}_{t_v}+\alpha^{\rm C}_{t_w}\le\ell_h+\theta$.
\item \emph{Vertex inequalities.} For every $S\subset\{\pm e_1,\pm e_2,\pm e_3\}$ and every two-element subset $\{d_1,d_2\}\subset S$,
\[
 \sum_{d\in\{d_1,d_2\}}\alpha^{\rm P}_{t(d,S)}+\sum_{d\in S\setminus\{d_1,d_2\}}\alpha^{\rm C}_{t(d,S)}\ \ge\ \gamma,
\]
where $t(d,S)=\#\{d'\in S:\ d'\perp d\}+5\cdot\one\{-d\in S\}$.
\end{enumerate}
\end{definition}

Condition (A3) has $\binom62\cdot2^4=240$ instances, and they depend only on a few counts. If $|S|=d$ and $S$ contains $a$ opposite pairs, then $2a$ directions in $S$ have type $d+3$ and $d-2a$ have type $d-1$. The sum in (A3) depends only on $(d,a)$ and on the number $k\in\{0,1,2\}$ of path directions among the $2a$ paired ones. There are $12$ admissible triples $(d,a,k)$.

The two allocations used in this paper are listed in \cref{tab:charges1,tab:charges2,tab:shapes}.

\begin{table}[t]
\centering\small
\caption{Allocation at $u_*$: half-edge charges in units of $10^{-12}$, indexed by $j$ and $o$ with $t=j+5o$. Shape charges vanish, $\theta_1=9/10000$ and $\gamma_1=8362020128\cdot10^{-12}$.}\label{tab:charges1}
\begin{tabular}{rrrrr}
\toprule
$j$ & $\alpha^{\rm P}$, $o=0$ & $\alpha^{\rm C}$, $o=0$ & $\alpha^{\rm P}$, $o=1$ & $\alpha^{\rm C}$, $o=1$\\
\midrule
0 & 0 & 0 & 4181010088 & 0 \\
1 & 4181010088 & 0 & 2796622230 & 2768775703 \\
2 & 2796622230 & 2768775703 & 1696196670 & 2772376814 \\
3 & 1408633261 & 2484813406 & 1287882389 & 2428165575 \\
4 & 0 & 929924211 & 549724752 & 1815642656 \\
\bottomrule
\end{tabular}
\end{table}

\begin{table}[t]
\centering\small
\caption{Allocation at $u_2$: half-edge charges in units of $10^{-12}$, with $j,o$ as in \cref{tab:charges1}. The chord price is $\theta_2=1457355254\cdot10^{-12}$ and the vertex gain $\gamma_2=8382656215\cdot10^{-12}$. Shape charges are in \cref{tab:shapes}.}\label{tab:charges2}
\begin{tabular}{rrrrr}
\toprule
$j$ & $\alpha^{\rm P}$, $o=0$ & $\alpha^{\rm C}$, $o=0$ & $\alpha^{\rm P}$, $o=1$ & $\alpha^{\rm C}$, $o=1$\\
\midrule
0 & 0 & 0 & 4191328333 & 0 \\
1 & 4191328333 & 0 & 3406623940 & 2276558740 \\
2 & 2699473836 & 2983708843 & 1425837850 & 2882262869 \\
3 & 1309065313 & 2765490332 & 738905393 & 2547916991 \\
4 & 0 & 1809011447 & 0 & 2095664091 \\
\bottomrule
\end{tabular}
\end{table}

\begin{table}[t]
\centering\small
\caption{Shape charges $w_\tau$ at $u_2$ and the rounded weights $\overline\Omega_\tau=10^{-6}\ceil{10^6(1-w_\tau+w_\tau^2/2)}\ge e^{-w_\tau}$ used in \cref{sec:counting}.}\label{tab:shapes}
\begin{tabular}{llrr}
\toprule
$\tau$ & outer neighbours relative to the middle bond & $10^{12}w_\tau$ & $10^6\,\overline\Omega_\tau$\\
\midrule
$\mathrm S$ & both axial & 1127558275 & 998874 \\
$\mathrm A$ & one axial, one transverse & 544046375 & 999457 \\
$\mathrm U$ & equal transverse directions & 544046375 & 999457 \\
$\mathrm T$ & perpendicular transverse directions & 94806903 & 999906 \\
$\mathrm Z$ & opposite transverse directions & 0 & 1000000 \\
\bottomrule
\end{tabular}
\end{table}

\begin{certificate}[Allocations]\label{cert:allocation}
\begin{enumerate}[label=\textup{(\alph*)}]
\item The charges of \cref{tab:charges1}, with $w\equiv0$, $\theta_1=9/10000$ and $\gamma_1=8362020128\cdot10^{-12}$, form an allocation admissible for $\underline\ell(u_*)$. Conditions (A1)--(A2) comprise $1637$ inequalities on the $1024$ patterns, with minimal slack $24\cdot10^{-12}$. Condition (A3) holds with equality for at least one configuration.
\item The charges of \cref{tab:charges2,tab:shapes}, with $\theta_2=1457355254\cdot10^{-12}$ and $\gamma_2=8382656215\cdot10^{-12}$, form an allocation admissible for $\underline\ell(u_2)$. After grouping patterns with identical constraints, (A1) consists of $392$ inequalities with minimal slack $301\cdot10^{-12}$ and (A2) of $87$ inequalities with minimal slack $1301\cdot10^{-12}$. Condition (A3) holds, again with equality in at least one of the $12$ reduced cases.
\end{enumerate}
\end{certificate}

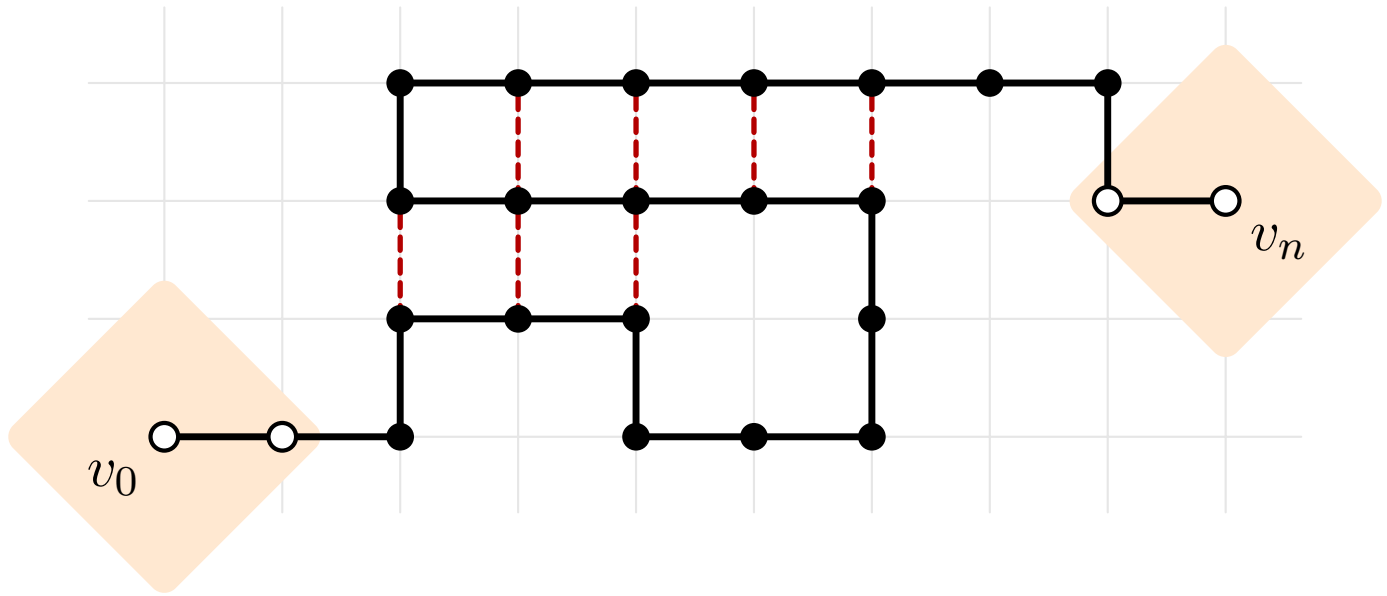
\begin{figure}[t]
\centering
\begin{tikzpicture}[line cap=round]
\draw[black!10,step=0.78] (-0.5,-0.5) grid (7.52,2.84);
\fill[orange!18,rounded corners=3pt] (5.944,1.560) -- (7.020,2.636) -- (8.096,1.560) -- (7.020,0.484) -- cycle;
\fill[orange!18,rounded corners=3pt] (-1.076,0.000) -- (0.000,1.076) -- (1.076,0.000) -- (0.000,-1.076) -- cycle;
\draw[red!70!black,line width=1.0pt,dash pattern=on 2.5pt off 1.8pt] (1.560,0.780) -- (1.560,1.560);
\draw[red!70!black,line width=1.0pt,dash pattern=on 2.5pt off 1.8pt] (2.340,0.780) -- (2.340,1.560);
\draw[red!70!black,line width=1.0pt,dash pattern=on 2.5pt off 1.8pt] (3.120,0.780) -- (3.120,1.560);
\draw[red!70!black,line width=1.0pt,dash pattern=on 2.5pt off 1.8pt] (4.680,1.560) -- (4.680,2.340);
\draw[red!70!black,line width=1.0pt,dash pattern=on 2.5pt off 1.8pt] (3.900,1.560) -- (3.900,2.340);
\draw[red!70!black,line width=1.0pt,dash pattern=on 2.5pt off 1.8pt] (3.120,1.560) -- (3.120,2.340);
\draw[red!70!black,line width=1.0pt,dash pattern=on 2.5pt off 1.8pt] (2.340,1.560) -- (2.340,2.340);
\draw[black,line width=1.3pt] (0.000,0.000) -- (0.780,0.000);
\draw[black,line width=1.3pt] (0.780,0.000) -- (1.560,0.000);
\draw[black,line width=1.3pt] (1.560,0.000) -- (1.560,0.780);
\draw[black,line width=1.3pt] (1.560,0.780) -- (2.340,0.780);
\draw[black,line width=1.3pt] (2.340,0.780) -- (3.120,0.780);
\draw[black,line width=1.3pt] (3.120,0.780) -- (3.120,0.000);
\draw[black,line width=1.3pt] (3.120,0.000) -- (3.900,0.000);
\draw[black,line width=1.3pt] (3.900,0.000) -- (4.680,0.000);
\draw[black,line width=1.3pt] (4.680,0.000) -- (4.680,0.780);
\draw[black,line width=1.3pt] (4.680,0.780) -- (4.680,1.560);
\draw[black,line width=1.3pt] (4.680,1.560) -- (3.900,1.560);
\draw[black,line width=1.3pt] (3.900,1.560) -- (3.120,1.560);
\draw[black,line width=1.3pt] (3.120,1.560) -- (2.340,1.560);
\draw[black,line width=1.3pt] (2.340,1.560) -- (1.560,1.560);
\draw[black,line width=1.3pt] (1.560,1.560) -- (1.560,2.340);
\draw[black,line width=1.3pt] (1.560,2.340) -- (2.340,2.340);
\draw[black,line width=1.3pt] (2.340,2.340) -- (3.120,2.340);
\draw[black,line width=1.3pt] (3.120,2.340) -- (3.900,2.340);
\draw[black,line width=1.3pt] (3.900,2.340) -- (4.680,2.340);
\draw[black,line width=1.3pt] (4.680,2.340) -- (5.460,2.340);
\draw[black,line width=1.3pt] (5.460,2.340) -- (6.240,2.340);
\draw[black,line width=1.3pt] (6.240,2.340) -- (6.240,1.560);
\draw[black,line width=1.3pt] (6.240,1.560) -- (7.020,1.560);
\filldraw[fill=white,draw=black,line width=0.8pt] (0.000,0.000) circle (2.6pt);
\filldraw[fill=white,draw=black,line width=0.8pt] (0.780,0.000) circle (2.6pt);
\fill[black] (1.560,0.000) circle (2.6pt);
\fill[black] (1.560,0.780) circle (2.6pt);
\fill[black] (2.340,0.780) circle (2.6pt);
\fill[black] (3.120,0.780) circle (2.6pt);
\fill[black] (3.120,0.000) circle (2.6pt);
\fill[black] (3.900,0.000) circle (2.6pt);
\fill[black] (4.680,0.000) circle (2.6pt);
\fill[black] (4.680,0.780) circle (2.6pt);
\fill[black] (4.680,1.560) circle (2.6pt);
\fill[black] (3.900,1.560) circle (2.6pt);
\fill[black] (3.120,1.560) circle (2.6pt);
\fill[black] (2.340,1.560) circle (2.6pt);
\fill[black] (1.560,1.560) circle (2.6pt);
\fill[black] (1.560,2.340) circle (2.6pt);
\fill[black] (2.340,2.340) circle (2.6pt);
\fill[black] (3.120,2.340) circle (2.6pt);
\fill[black] (3.900,2.340) circle (2.6pt);
\fill[black] (4.680,2.340) circle (2.6pt);
\fill[black] (5.460,2.340) circle (2.6pt);
\fill[black] (6.240,2.340) circle (2.6pt);
\filldraw[fill=white,draw=black,line width=0.8pt] (6.240,1.560) circle (2.6pt);
\filldraw[fill=white,draw=black,line width=0.8pt] (7.020,1.560) circle (2.6pt);
\node[below left=1pt] at (0.000,0.000) {$v_0$};
\node[below right=1pt] at (7.020,1.560) {$v_n$};
\end{tikzpicture}
\caption{A path (solid) with its 7 chords (dashed), drawn in a plane for illustration; on $\T_L$ chords may also point out of the plane or across the periodic identification. Open circles are the uncharged vertices: the endpoints and their lattice neighbours on the path (shaded). Every induced bond at a charged vertex (black) has two interior endpoints, so the local inequalities (A1)--(A2) apply to it.}
\label{fig:path}
\end{figure}

\begin{proposition}[Cost of induced bonds]\label{prop:cost}
Let $u\ge u_0$ and let $(\alpha^{\rm P},\alpha^{\rm C},w,\theta,\gamma)$ be an allocation admissible for a table $\ell$ with $\ell_h\le\ell_h(u_0)$ for all $h$. Then for every even $L\ge8$ and every path $P=(v_0,\dots,v_n)$ on $\T_L$,
\begin{equation}\label{eq:cost}
 I(P,u)+\theta\,c(P)\ \ge\ \gamma\,(n-13)_+ +\sum_{i=1}^{n-2}w_{\tau_i}.
\end{equation}
\end{proposition}

\begin{proof}
Call a vertex of $P$ \emph{charged} if it is neither an endpoint $v_0,v_n$ nor a lattice neighbour of an endpoint lying in $V(P)$. There are at most $2+2\cdot6=14$ uncharged vertices, so at least $(n+1-14)_+=(n-13)_+$ charged ones. If $v$ is charged and $\{v,x\}\in D(P)$, then $x$ is not an endpoint, since otherwise $v$ would be a neighbour of an endpoint. Thus both ends of every induced bond touching a charged vertex are interior vertices of $P$.

Let $e=\{v,w\}\in D(P)$ have two interior endpoints. If $e=v_iv_{i+1}$ is a path bond, then $1\le i\le n-2$. The directions $d_-$ and $d_+$ of \cref{def:shape} point to occupied specified sites of the prism of $e$, so (A1) gives
\begin{equation}\label{eq:pathrow}
 \alpha^{\rm P}_{t_v(e)}+\alpha^{\rm P}_{t_w(e)}+w_{\tau_i}\le\ell_{h(e)}.
\end{equation}
If $e$ is a chord, each of $v,w$ is met by two path bonds different from $e$. Each hub thus has at least two occupied specified neighbours, and (A2) gives $\alpha^{\rm C}_{t_v(e)}+\alpha^{\rm C}_{t_w(e)}\le\ell_{h(e)}+\theta$.

Let $D^\circ$ be the set of induced bonds with two interior endpoints. By \cref{lem:prismbound}, the nonnegativity of $\ell$, and these two inequalities, each $e\in D^\circ$ satisfies $\ell_{h(e)}+\theta\one\{e\text{ chord}\}\ge w_e+\sum_{z}\alpha_{t_z(e)}^{\rm role(e)}$. Here $w_e=w_{\tau_i}$ if $e=v_iv_{i+1}$ is a path bond, $w_e=0$ for chords, and the sum runs over the \emph{charged} endpoints $z$ of $e$. Dropping the charge of an uncharged endpoint is allowed since all charges are nonnegative. Summing over $e\in D^\circ$,
\[
 I(P,u)+\theta\,c(P)\ \ge\ \sum_{e\in D^\circ}\Bigl(\ell_{h(e)}+\theta\one\{e\text{ chord}\}\Bigr)
 \ \ge\ \sum_{i=1}^{n-2}w_{\tau_i}+\sum_{z\text{ charged}}\ \sum_{e\in D(P),\,e\ni z}\alpha^{\mathrm{role}(e)}_{t_z(e)} .
\]
Every path bond $v_iv_{i+1}$ with $1\le i\le n-2$ belongs to $D^\circ$. For a charged vertex $z$ let $S\subset\{\pm e_1,\pm e_2,\pm e_3\}$ be the set of directions from $z$ to its neighbours in $V(P)$, and let $d_1,d_2$ be the directions of its two path bonds. The induced bonds at $z$ are exactly the bonds in the directions of $S$, and the half-edge in direction $d$ has type $t(d,S)$. By (A3) the inner sum is at least $\gamma$, which completes the proof.
\end{proof}

\begin{remark}
\Cref{prop:cost} involves no lift of $P$ to $\Z^3$ and no choice of orientation. It holds on every even torus with $L\ge8$, including paths that wind around the torus and chords across the periodic identification. The constant $13$ comes only from the endpoint neighbourhoods.
\end{remark}

\begin{remark}[Why a chord price is needed]\label{rem:hamilton}
A bound of the form $I(P,u)\ge\gamma'|P|-C$, valid for all paths with $C$ independent of $L$, forces $\gamma'\le f(K):=p(K,0)-\log2-3\log\cosh K$. Indeed, an even torus admits a Hamiltonian path $P$ (a ``snake'' visiting every site). For it $D(P)=E_L$, $|P|=V-1$ and $I(P,u)=\log\cE(u)=\log Z_L-V\log2-3V\log\cosh K$. Dividing by $V$ and letting $L\to\infty$ gives the claim. The quantity $f(K)$, the free energy of the high-temperature expansion per site, is small near $K_c$. In~\eqref{eq:cost} the snake has $c(P)\approx2V$, and most of its cost is paid by the chord price $\theta$. The chord price is in turn controlled by the chord moments of \cref{sec:counting}.
\end{remark}

\section{Summing over paths}\label{sec:counting}

By \cref{lem:selection,prop:cost}, the susceptibility is bounded by sums, over paths $P$ from a fixed root, of $u^{|P|}e^{\theta c(P)}$ times shape factors. This section bounds such sums uniformly in $L$. The two ingredients are an exact moment of the chord number (\cref{lem:rosenbluth}) and transfer bounds on a finite automaton that remembers the last $16$ steps (\cref{lem:transfer,cert:transfer}). H\"older's inequality combines them (\cref{lem:holder}). Fix an even $L\ge8$ and a root $x\in\T_L$. All paths below start at $x$.

\subsection{Contacts}

For a path $P=(v_0,\dots,v_n)$ and $1\le i\le n$ let
\[
 a_i(P)=\#\{j\le i-2:\ v_j\sim v_i\}
\]
be the number of \emph{contacts} of the tip $v_i$ with earlier vertices other than its predecessor. For $0\le i\le n$ let $B_i(P)$ be the set of directions $d\in\{\pm e_1,\pm e_2,\pm e_3\}$ such that $v_i+d\notin\{v_0,\dots,v_i\}$.

\begin{lemma}\label{lem:contacts}
$c(P)=\sum_{i=1}^na_i(P)$, and $|B_i(P)|=5-a_i(P)$ for $1\le i\le n$. Moreover $|B_0(P)|=6$, and $v_{i+1}-v_i\in B_i(P)$ for $i<n$.
\end{lemma}

\begin{proof}
A chord $\{v_j,v_i\}$ with $j\le i-2$ is counted exactly once, in $a_i$. The six neighbours of $v_i$ are distinct, since $L\ge3$. One of them is $v_{i-1}$, $a_i$ of them are earlier vertices of $P$, and the others are unvisited at time $i$.
\end{proof}

\subsection{An exact chord moment}

\begin{lemma}[Chord moment]\label{lem:rosenbluth}
For every $n\ge1$,
\[
 \sum_{|P|=n}\Bigl(\frac54\Bigr)^{c(P)}\ \le\ 6\Bigl(\frac54\Bigr)^5 5^{n-1}.
\]
\end{lemma}

\begin{proof}
Put $b=5/4$. The five integer inequalities $(5-a)b^a\le5$, $a=0,\dots,5$, are checked directly. The cases $a=0,1$ are equalities, and the values for $a=2,3,4,5$ are $4.6875$, $3.90625$, $2.44140625$ and $0$. For a path $p$ of length $\ell\ge1$ and $k\ge0$ let
\[
 F_k(p)=\sum_{q}\ \prod_{i=\ell}^{\ell+k-1}b^{a_i(q)},
\]
the sum running over all paths $q$ of length $\ell+k$ extending $p$. Then $F_0=1$, and by \cref{lem:contacts}
\[
 F_{k+1}(p)=b^{a_\ell(p)}\sum_{d\in B_\ell(p)}F_k(p+d)\le b^{a_\ell(p)}(5-a_\ell(p))\max_dF_k(p+d).
\]
By induction $F_k(p)\le5^k$. The root has six continuations, and the final tip contributes $b^{a_n}\le b^5$. Hence $\sum_{|P|=n}b^{c(P)}\le b^5\cdot6\cdot5^{n-1}$.
\end{proof}

The argument is the Rosenbluth growth construction~\cite{RR55} read as a counting inequality. The weight $\prod_i(5-a_i)^{-1}$ is the probability that a walk choosing uniformly among unvisited neighbours produces $P$. The lemma has been verified in Lean on $(\Z/L\Z)^3$ for every $L\ge3$ (Appendix~\ref{app:lean}).

\subsection{A finite-memory automaton}

Self-avoidance is a global constraint. We replace it by a local one: a step is forbidden only if it returns to one of the recently visited vertices that are still close to the tip. Paths on $\T_L$ are first lifted to $\Z^3$. The \emph{lift} of $P$ is the lattice walk $\omega=(x_0,\dots,x_n)$ with $x_0=0$ and the same step directions $d_i=v_i-v_{i-1}$. Distinct sites of $\T_L$ have distinct lifts, so $\omega$ is self-avoiding.

\begin{definition}[Suffix automaton]\label{def:automaton}
Fix the memory $M=16$.
\begin{enumerate}
\item A \emph{retained suffix} is a sequence $(y_0,\dots,y_m)$ of distinct sites of $\Z^3$ with $y_k\sim y_{k-1}$ for all $k$.
\item For a direction $d$ with $y_m+d\notin\{y_0,\dots,y_m\}$, the \emph{update} by $d$ appends $y_{m+1}=y_m+d$. It then deletes $y_0,\dots,y_{j-1}$, where $j$ is the smallest index with $(m+1-j)+\|y_{m+1}-y_j\|_1\le M$.
\item A \emph{state} is a retained suffix modulo translations and the $48$ signed permutations of the axes. The \emph{initial state} $s_0$ is the one-point suffix at the root.
\item For a state $s$, the \emph{available labels} $A(s)$ are the directions $d$ with $y_m+d$ outside the suffix, and $s_d$ is the state after the update by $d$. Labels are counted with multiplicity, also when different labels lead to the same state.
\end{enumerate}
The \emph{automaton} is the finite set of states reachable from $s_0$, together with the labelled transitions $s\to s_d$.
\end{definition}

A vertex $y_j$ is kept as long as a self-avoiding continuation could return to it within $M$ steps. Only two properties of the rule are used below. First, the retained suffix consists of actual vertices of the walk. Second, after at least two steps it contains the last three vertices, since $2+\|y_m-y_{m-2}\|_1\le4\le M$. Consequently the state determines the last two step directions, and with them the shape (\cref{def:shape}) of the triple completed by the next step.

\begin{lemma}[Domination by the automaton]\label{lem:automaton}
Let $P$ be a path on $\T_L$ and let $s_0,s_1,\dots,s_n$ be the states obtained from the initial state by feeding the step directions of $P$. Then $s_{i+1}=(s_i)_{d_{i+1}}$ is defined for $i<n$. Moreover $B_i(P)\subseteq A(s_i)$ for $0\le i\le n$, where the labels of $s_i$ are identified with physical directions through the canonicalizing symmetry.
\end{lemma}

\begin{proof}
By induction the retained suffix of $s_i$, transported back to the physical frame, is a final segment of the lift $(x_0,\dots,x_i)$. If $d\notin A(s_i)$, then $x_i+d$ is a retained vertex $x_j$, $j<i$. Its projection $v_j$ is visited, so $d\notin B_i(P)$. In particular $d_{i+1}\in B_i(P)\subseteq A(s_i)$, so the update is defined.
\end{proof}

Periodic identifications, and returns to vertices older than the retained suffix, can remove further directions from $B_i(P)$. The bounds below are designed so that removing directions only decreases them. They are therefore uniform in $L$.

\subsection{Transfer bounds}

Let $\varphi$ assign a weight $\varphi(s,d)\ge0$ to every labelled transition and let $\kappa>0$. For a vector $v$ indexed by states put
\begin{equation}\label{eq:transfer}
 (\cT v)(s)=\max_{B\subseteq A(s)}\kappa^{5-|B|}\sum_{d\in B}\varphi(s,d)\,v(s_d)\quad(s\ne s_0),\qquad
 (\cT v)(s_0)=\sum_{d\in A(s_0)}\varphi(s_0,d)\,v((s_0)_d).
\end{equation}
For $v\ge0$ and fixed $|B|$, the maximum is attained by the $|B|$ largest summands, counted with multiplicity. If $\kappa\le1$, it is attained at $B=A(s)$, and $\cT$ is linear.

\begin{lemma}[Transfer bound]\label{lem:transfer}
Suppose $v>0$ and $\cT v\le\Lambda v$ componentwise. For a path $P$ write $\Phi(P)=\prod_{i=0}^{n-1}\varphi(s_i,d_{i+1})$. Then for every even $L\ge8$ and $n\ge1$,
\[
 \sum_{|P|=n}\kappa^{c(P)}\,\Phi(P)\ \le\ \max(1,\kappa^5)\,\frac{v(s_0)}{\min_sv(s)}\,\Lambda^n .
\]
\end{lemma}

\begin{proof}
For a path $p$ of length $\ell$ with state $s(p)$, and $k\ge0$, let
\[
 F_k(p)=\sum_q\ \prod_{i=\ell}^{\ell+k-1}\kappa^{a_i(q)\one\{i\ge1\}}\varphi(s_i(q),d_{i+1}(q)),
\]
summed over all extensions $q$ of $p$ of length $\ell+k$. We show $F_k(p)\le\Lambda^kv(s(p))/\min v$. This is clear for $k=0$. Write $s=s(p)$ and $B=B_\ell(p)$, which satisfies $\kappa^{a_\ell}=\kappa^{5-|B|}$ for $\ell\ge1$ by \cref{lem:contacts}. Then
\[
\begin{aligned}
 F_{k+1}(p)&=\kappa^{(5-|B|)\one\{\ell\ge1\}}\sum_{d\in B}\varphi(s,d)F_k(p+d)\\
 &\le\frac{\Lambda^k}{\min v}\,\kappa^{(5-|B|)\one\{\ell\ge1\}}\sum_{d\in B}\varphi(s,d)v(s_d)
 \le\frac{\Lambda^k(\cT v)(s)}{\min v}.
\end{aligned}
\]
The first inequality uses the induction hypothesis and $s(p+d)=s_d$. The second uses $B\subseteq A(s)$ from \cref{lem:automaton} and the definition of $\cT$. Finally $(\cT v)(s)\le\Lambda v(s)$. Since $c(P)=\sum_{i=1}^na_i$, we get $\sum_{|P|=n}\kappa^{c(P)}\Phi(P)\le\max_Pw^{a_n(P)}F_n(\text{root})$, and $a_n\le5$.
\end{proof}

We use two instances of \cref{lem:transfer}. The \emph{contact operator} has $\varphi\equiv1$ and $\kappa=\kappa_1:=10000/9991=(1-\theta_1)^{-1}$. Since $e^{\theta_1}<\kappa_1$, it bounds $\sum_Pe^{\theta_1c(P)}$. The \emph{shape operator} has $\kappa=1$, $\varphi(s,d)=\overline\Omega_\tau$ if $s$ has at least two steps and $\tau$ is the shape of the triple completed by $d$, and $\varphi(s,d)=1$ otherwise. For it
\[
 \Phi(P)=\prod_{i=1}^{n-2}\overline\Omega_{\tau_i}\ \ge\ \Omega(P):=\exp\Bigl(-\sum_{i=1}^{n-2}w_{\tau_i}\Bigr),
\]
because $\overline\Omega_\tau\ge1-w_\tau+w_\tau^2/2\ge e^{-w_\tau}$ for $w_\tau\ge0$.

\begin{certificate}[Transfer bounds]\label{cert:transfer}
The automaton of \cref{def:automaton} has $12\,994\,574$ states, including the initial state, and $53\,221\,763$ labelled transitions.
\begin{enumerate}[label=\textup{(\alph*)}]
\item For the contact operator, with each power $\kappa_1^j$, $0\le j\le5$, rounded up to the grid $10^{-8}\Z$, there is a positive integer vector $v$ with $\cT v\le\Lambda_1v$ and
\[
 \Lambda_1=\frac{473271745}{10^8}.
\]
The largest row quotient is $7337944694907297/1550471750000000<\Lambda_1$.
\item For the shape operator, with the weights $\overline\Omega_\tau$ of \cref{tab:shapes}, there is a positive integer vector $v$ with $\cT v\le U_2v$ and
\[
 U_2=\frac{1182583}{250000}=4.730332 .
\]
\item For the \emph{unweighted operator}, with $\varphi\equiv1$ and $\kappa=1$, there is a positive integer vector $v$ with $\cT v\le U_0v$ and
\[
 U_0=\frac{473195}{100000}=4.73195 .
\]
The largest row quotient is $151907537380697/32102539197381<U_0$. The root counts of the automaton agree with the numbers of self-avoiding walks on $\Z^3$ of length $n\le17$~\cite{SBB11}.
\end{enumerate}
\end{certificate}

The vectors were obtained by iterating upward-rounded versions of $\cT$ from a constant vector, renormalizing by a common power of two, and adding one. How they were found plays no role. The certificate is the row-by-row inequality for the final vector. That inequality is checked with exact integer cross-multiplication.

\subsection{Combining shapes and chords}

The shape operator ignores contacts. The chord price is restored by interpolating between it and \cref{lem:rosenbluth}.

\begin{lemma}[H\"older combination]\label{lem:holder}
Let $t=\theta_2/\log(5/4)$. Then $0<t<1$, $(5/U_2)^4<5/4$, and for every even $L\ge8$ and $n\ge1$,
\[
 \sum_{|P|=n}\Omega(P)\,e^{\theta_2c(P)}\ \le\ C_2\bigl(U_2e^{\theta_2/4}\bigr)^n,
 \qquad C_2=\Bigl(\frac{v(s_0)}{\min v}\Bigr)^{1-t}\Bigl(\frac65\Bigl(\frac54\Bigr)^5\Bigr)^t,
\]
where $v$ is the vector of \cref{cert:transfer}\textup{(b)}.
\end{lemma}

\begin{proof}
The two numerical claims are elementary; see \cref{cert:endpoint-lower}. Write $e^{\theta_2c}=(5/4)^{tc}$ and $\Omega=\Omega^{1-t}\Omega^t$. H\"older's inequality with exponents $1/(1-t)$ and $1/t$ gives
\[
 \sum_P\Omega(P)e^{\theta_2c(P)}\le\Bigl(\sum_P\Omega(P)\Bigr)^{1-t}\Bigl(\sum_P\Omega(P)(5/4)^{c(P)}\Bigr)^t .
\]
By \cref{lem:transfer} for the shape operator, the first sum is at most $(v(s_0)/\min v)\,U_2^n$. Since $0<\Omega\le1$, \cref{lem:rosenbluth} bounds the second sum by $6(5/4)^55^{n-1}$. Hence the left-hand side is at most $C_2U_2^n(5/U_2)^{tn}$. Finally $(5/U_2)^t\le e^{\theta_2/4}$ is equivalent to $t\log(5/U_2)\le\frac t4\log\frac54$, i.e.\ to $(5/U_2)^4\le5/4$.
\end{proof}

\begin{remark}
The exponent $1/4$ in $e^{\theta_2/4}$ reflects how close $5$ is to $U_2$. The chord moment of \cref{lem:rosenbluth} grows like $5^n$ rather than $U_2^n$. The H\"older combination therefore costs a factor $(5/U_2)^t$ per step, and the inequality $(5/U_2)^4<5/4$ converts this into $e^{\theta_2/4}$. Its margin is small: $(5/U_2)^4=1.24828\ldots$.
\end{remark}

\section{Proof of the lower bound}\label{sec:lower}

\subsection{Fisher's bound with the automaton}

\Cref{sec:paths,sec:counting} provide the two ingredients of the lower bound: a cost for every path, and bounds for sums over paths. We now combine them. We begin with what the automaton gives on its own, without the induced bonds.

\begin{proposition}\label{prop:fisher}
The connective constant of $\Z^3$ satisfies $\mu\le U_0=4.73195$. Consequently
\[
 e^{2K_c}\ \ge\ \frac{U_0+1}{U_0-1}\ >\ 1.53591,\qquad K_c>0.21456 .
\]
\end{proposition}

\begin{proof}
The domination argument of \cref{lem:automaton} applies verbatim to self-avoiding walks on $\Z^3$, which are their own lifts. \Cref{lem:transfer} with $\varphi\equiv1$, $\kappa=1$ and \cref{cert:transfer}(c) therefore bounds the number of $n$-step self-avoiding walks on $\Z^3$, and also the number of $n$-step paths from a fixed root on every even torus with $L\ge8$, by $(v(s_0)/\min v)\,U_0^n$. The first bound gives $\mu\le U_0$. For the second, drop the factor $e^{-I(P,u)}\le1$ in \cref{lem:selection}. This gives $\chi_L(K)\le1+(v(s_0)/\min v)\sum_{n\ge1}(uU_0)^n$, which is bounded uniformly in $L$ when $uU_0<1$. By \cref{prop:disorder}, $m_+(K)=0$ whenever $\tanh K<1/U_0$, and \cref{cor:monotone}(i) gives $\tanh K_c\ge1/U_0$, which is the claim.
\end{proof}

\Cref{prop:fisher} improves the bound $r_c>1.5349$ obtained from~\cite{PT00}. The remainder of this section improves it further by means of the factor $e^{-I(P,u)}$.

\subsection{A susceptibility criterion}

Write $T_3(x)=1+x+x^2/2+x^3/6$. For $x>0$ one has $T_3(x)<e^x$.

\begin{proposition}[Geometric susceptibility bound]\label{prop:geometric}
Let $u\ge u_0$ and let $(\alpha^{\rm P},\alpha^{\rm C},w,\theta,\gamma)$, with $\gamma\ge0$, be an allocation admissible for a table $\ell\le\ell(u_0)$. Put $\Omega(P)=\exp(-\sum_{i=1}^{|P|-2}w_{\tau_i})$. Suppose that for some $C,\rho>0$ and all even $L\ge8$ and $n\ge1$
\begin{equation}\label{eq:pathsum}
 \sum_{|P|=n}\Omega(P)e^{\theta c(P)}\le C\rho^n,
\end{equation}
the sum running over paths from a fixed root. Then for all even $L\ge8$
\[
 \chi_L(K)\ \le\ 1+Ce^{13\gamma}\sum_{n\ge1}\bigl(u\rho e^{-\gamma}\bigr)^n .
\]
\end{proposition}

\begin{proof}
By \cref{prop:cost}, $e^{-I(P,u)}\le e^{\theta c(P)}\Omega(P)e^{-\gamma(n-13)_+}\le e^{13\gamma}e^{-\gamma n}\Omega(P)e^{\theta c(P)}$ for $|P|=n$. Insert this in \cref{lem:selection}. Summing over $y\ne0$ amounts to summing over all paths from the root $0$, and the term $y=0$ contributes $1$. Grouping paths by length and applying~\eqref{eq:pathsum} gives the claim.
\end{proof}

\begin{certificate}[Endpoint arithmetic, disordered side]\label{cert:endpoint-lower}
Let $u_1=6769/31769$, $u_2=677/3177$, and
\[
 \bar\gamma_2=\gamma_2-\frac{\theta_2}4=\frac{16036634803}{2\cdot10^{12}} .
\]
The following hold in exact rational arithmetic.
\begin{enumerate}[label=\textup{(\alph*)}]
\item One has
\[
 \Bigl\lfloor10^{12}\Bigl(\frac{T_3(\bar\gamma_2)}{U_2u_2}-1\Bigr)\Bigr\rfloor=44289252,\qquad
 \frac{U_2+T_3(\bar\gamma_2)}{U_2-T_3(\bar\gamma_2)}=r_- .
\]
\item $(5/U_2)^4<5/4$, and $\theta_2<\frac15<\log\frac54$.
\item One has $u_*<u_1$ and
\[
 \Bigl\lfloor10^{12}\Bigl(\frac{T_3(\gamma_1)}{\Lambda_1u_1}-1\Bigr)\Bigr\rfloor=74673 .
\]
\end{enumerate}
\end{certificate}

\subsection{The lower bound}

\begin{theorem}\label{thm:lower}
There is $\delta>0$ such that for all $r\in[r_-,r_-+\delta)$ the susceptibilities $\chi_L(K)$, $L\ge8$ even, are bounded uniformly in $L$. Consequently $m_+(K)=0$ on this interval, and $e^{2K_c}>r_-$.
\end{theorem}

\begin{proof}
Put $u'=T_3(\bar\gamma_2)/U_2$. By \cref{cert:endpoint-lower}(a), $u'>u_2$ and $(1+u')/(1-u')=r_-$. Since $\bar\gamma_2>0$, $u'U_2e^{-\bar\gamma_2}=T_3(\bar\gamma_2)e^{-\bar\gamma_2}<1$. Choose $\delta'>0$ with $(u'+\delta')U_2e^{-\bar\gamma_2}<1$.

Let $u\in[u',u'+\delta')$. Then $u\ge u_2$. By \cref{cert:prism,cert:allocation}(b) the allocation of \cref{tab:charges2,tab:shapes} is admissible for $\underline\ell(u_2)\le\ell(u_2)$. By \cref{lem:holder}, \eqref{eq:pathsum} holds with $\theta=\theta_2$, $C=C_2$ and $\rho=U_2e^{\theta_2/4}$. \Cref{prop:geometric} gives
\[
 \chi_L(K)\le1+C_2e^{13\gamma_2}\sum_{n\ge1}\bigl(uU_2e^{-\bar\gamma_2}\bigr)^n<\infty,
\]
uniformly in $L$, because $u\rho e^{-\gamma_2}=uU_2e^{-\bar\gamma_2}<1$. By \cref{prop:disorder}, $m_+(K)=0$ for $\tanh K\in[u',u'+\delta')$. Since $r=(1+u)/(1-u)$ is increasing in $u$, \cref{cor:monotone}(i) gives $e^{2K_c}>r_-$.
\end{proof}

\begin{remark}[Exponential decay]
The proof gives more. In the bound of \cref{lem:selection} only paths of length at least the graph distance $|y|$ contribute. On the same interval, therefore, $G_L(0,y)\le C'\varrho^{|y|}$ with $\varrho=uU_2e^{-\bar\gamma_2}<1$ and $C'$ independent of $L$.
\end{remark}

\subsection{A variant without shape charges}

The allocation of \cref{cert:allocation}(a) has no shape charges. Combined with the contact operator it gives a slightly weaker bound, which we record because it uses the chord price in a different way. Here the contacts are paid for inside the transfer operator, and no H\"older step is needed.

\begin{proposition}\label{prop:r1}
For $r$ in a right neighbourhood of $r_1=19269/12500=1.54152$, the susceptibilities $\chi_L$ are uniformly bounded. In particular $e^{2K_c}>r_1$.
\end{proposition}

\begin{proof}
Let $u\ge u_1>u_*$. The allocation of \cref{tab:charges1} is admissible for $\underline\ell(u_*)$ with $\Omega\equiv1$. Since $e^{\theta_1}<\kappa_1$, \cref{lem:transfer,cert:transfer}(a) give~\eqref{eq:pathsum} with $\theta=\theta_1$ and $\rho=\Lambda_1$. By \cref{cert:endpoint-lower}(c), $u_1\Lambda_1e^{-\gamma_1}<u_1\Lambda_1/T_3(\gamma_1)<1$, and the strict inequality persists on a right neighbourhood. Conclude with \cref{prop:geometric,prop:disorder}.
\end{proof}

\section{An infrared bound that retains the zero mode}\label{sec:infrared}

We now turn to the ordered side. Whereas the lower bound controls the two-point function from above by sums over paths, the upper bound requires a lower bound on the magnetization, and we obtain it from the Fourier transform of the two-point function. This section proves an infrared bound with a coefficient that involves the nearest-neighbour correlation (\cref{prop:infrared}). Integrated against five test polynomials, it yields linear inequalities between local correlations and the zero Fourier mode, which is kept as an unknown (\cref{lem:spectral}).

In \cref{sec:infrared,sec:ordered}, $L\ge4$ is even and all expectations are taken in the zero-field periodic measure $\avg\cdot_L=\avg\cdot_{L,K,0}$. Put
\[
 G_L(x)=\avg{\sigma_0\sigma_x}_L,\qquad \widehat G_L(k)=\sum_{x\in\T_L}e^{ik\cdot x}G_L(x),\qquad
 g_L=G_L(e_1),\qquad \lambda(k)=2\sum_{j=1}^3(1-\cos k_j),
\]
for $k\in\T_L^*=(2\pi/L)\{0,\dots,L-1\}^3$. Positive definiteness of $G_L$ gives $\widehat G_L(k)\ge0$, and $\widehat G_L(0)=Vq_L$. Inverting the Fourier transform,
\begin{equation}\label{eq:sumrule}
 1=G_L(0)=q_L+\frac1V\sum_{k\ne0}\widehat G_L(k).
\end{equation}
We also write $c_0=u^{-1}+u=2(r^2+1)/(r^2-1)$.

\subsection{The bound for nonzero modes}

\begin{proposition}[Infrared bound]\label{prop:infrared}
For every real function $h$ on $\T_L$,
\begin{equation}\label{eq:varbound}
 \Bigl\langle\Bigl(\sum_x\sigma_x(-\Delta h)_x\Bigr)^2\Bigr\rangle_L\le(c_0-2g_L)\sum_{\{x,y\}\in E_L}(h_x-h_y)^2,
\end{equation}
where $(-\Delta h)_x=\sum_{y\sim x}(h_x-h_y)$. Consequently
\begin{equation}\label{eq:IR}
 \lambda(k)\,\widehat G_L(k)\le c_0-2g_L\qquad(k\in\T^*_L\setminus\{0\}).
\end{equation}
\end{proposition}

Since $g_L\ge u$ by \cref{lem:factor} and \cref{lem:standard}(i), the coefficient satisfies $c_0-2g_L\le u^{-1}-u=2/\sinh(2K)<1/K$. Thus~\eqref{eq:IR} is a strengthening of the Gaussian-domination bound $\lambda\widehat G_L\le1/K$ of~\cite{FSS76}. The improvement comes from a representation of the Ising weight by inner products of planar unit vectors, for which a rotation of the vectors is a symmetry.

Let $0<\alpha<\pi/4$ satisfy $\cos(2\alpha)=e^{-2K}$, and for $\sigma=\pm1$ put $n_\sigma=(\cos\alpha,\sigma\sin\alpha)\in\R^2$. Then $n_\sigma\cdot n_{\sigma'}=1$ if $\sigma=\sigma'$ and $\cos2\alpha=e^{-2K}$ otherwise, i.e.\
\[
 e^{K\sigma\sigma'}=e^K\,n_\sigma\cdot n_{\sigma'} .
\]
Let $R(\phi)$ denote rotation by the angle $\phi$. For an angle field $a\in\R^{\T_L}$ define
\[
 Z(a)=\sum_\sigma\prod_{\{x,y\}\in E_L}R(a_x)n_{\sigma_x}\cdot R(a_y)n_{\sigma_y}
 =\sum_\sigma\prod_{\{x,y\}\in E_L}\cos\bigl(\alpha(\sigma_x-\sigma_y)+a_x-a_y\bigr).
\]
Then $Z(a)$ is real, $Z(0)=e^{-3VK}Z_L(K,0)>0$, and $Z(a)=Z(0)$ whenever $a$ is constant.

\begin{lemma}\label{lem:rotation}
$|Z(a)|\le Z(0)$ for every $a\in\R^{\T_L}$.
\end{lemma}

\begin{proof}
Fix a finite set $\mathcal A\subset\R$ containing the values of $a$, and consider all fields with values in $\mathcal A$. Let $\mathfrak M$ be the maximum of $|Z|$ over these fields, so $\mathfrak M\ge Z(0)>0$. Among the maximizers choose one, $a$, with the smallest number $N(a)$ of \emph{disagreeing} bonds, i.e.\ bonds with $a_x\ne a_y$.

Suppose $N(a)>0$, and let $b$ be a disagreeing bond, say in direction $e_1$. Consider the reflection $\vartheta$ of $\T_L$ in the pair of planes orthogonal to $e_1$ that bisect $b$ and the bond $L/2$ steps away. Since $L$ is even, these planes cut $\T_L$ into two halves $\T^\ell,\T^r$ exchanged by $\vartheta$. Write each crossing bond factor as $\sum_{\iota=1}^2(R(a_x)n_{\sigma_x})_\iota(R(a_y)n_{\sigma_y})_\iota$. Summing over the spins of each half gives
\[
 Z(a)=\sum_{\boldsymbol\iota}\Xi_{\boldsymbol\iota}(a|_{\T^\ell})\,\Xi_{\boldsymbol\iota}(\vartheta^*a|_{\T^r}),
\]
where $\boldsymbol\iota$ ranges over the component indices of the $2L^2$ crossing bonds and $\Xi_{\boldsymbol\iota}$ is a real tensor depending only on the field on one half. By Cauchy--Schwarz,
\[
 Z(a)^2\le Z(a^{\ell\ell})\,Z(a^{rr}),
\]
where $a^{\ell\ell}$ equals $a$ on $\T^\ell$ and $a\circ\vartheta$ on $\T^r$, and $a^{rr}$ is defined symmetrically. Both factors are nonnegative and at most $\mathfrak M$. Since $Z(a)^2=\mathfrak M^2$, both $a^{\ell\ell}$ and $a^{rr}$ are maximizers. In a reflected field every crossing bond agrees, and the bonds inside each half are duplicated. Hence $N(a^{\ell\ell})+N(a^{rr})=2(N(a)-N_\times(a))\le2N(a)-2$, where $N_\times(a)\ge1$ counts the disagreeing crossing bonds. This contradicts the minimality of $N(a)$. Therefore $N(a)=0$, $a$ is constant, and $\mathfrak M=Z(0)$.
\end{proof}

\begin{proof}[Proof of \cref{prop:infrared}]
For a real function $h$ and $\epsilon\in\R$ consider $\psi(\epsilon)=\log Z(\epsilon h)$. Near $\epsilon=0$ every bond factor is positive. By \cref{lem:rotation}, $\psi$ has a maximum at $\epsilon=0$, so $\psi''(0)\le0$. Write $\delta_{xy}=h_x-h_y$ and $\beta_{xy}=\log\cos(\alpha(\sigma_x-\sigma_y)+\epsilon\delta_{xy})$. At $\epsilon=0$ the normalized weight $\prod\cos(\alpha(\sigma_x-\sigma_y))/Z(0)$ is the Ising measure, so
\[
 \psi''(0)=\mathrm{Var}_L\Bigl(\sum_{\{x,y\}}\partial_\epsilon\beta_{xy}\Bigr)+\Bigl\langle\sum_{\{x,y\}}\partial^2_\epsilon\beta_{xy}\Bigr\rangle_L .
\]
If $\sigma_x=\sigma_y$, then $\partial_\epsilon\beta_{xy}=0$ and $\partial_\epsilon^2\beta_{xy}=-\delta_{xy}^2$. If $\sigma_x\ne\sigma_y$, then $\partial_\epsilon\beta_{xy}=-\frac12\tan(2\alpha)(\sigma_x-\sigma_y)\delta_{xy}$ and $\partial_\epsilon^2\beta_{xy}=-\sec^2(2\alpha)\delta_{xy}^2$. The first formula covers both cases. Since $\Prob(\sigma_x\ne\sigma_y)=(1-g_L)/2$ for neighbours and $\tan^2(2\alpha)=\sec^2(2\alpha)-1=e^{4K}-1$, the mean of $\partial^2_\epsilon\beta_{xy}$ is $-[1+(e^{4K}-1)(1-g_L)/2]\delta_{xy}^2$. Moreover $\sum_{\{x,y\}}(\sigma_x-\sigma_y)\delta_{xy}=\sum_x\sigma_x(-\Delta h)_x$, which has mean zero. Hence $\psi''(0)\le0$ reads
\[
 \frac{e^{4K}-1}4\Bigl\langle\Bigl(\sum_x\sigma_x(-\Delta h)_x\Bigr)^2\Bigr\rangle_L\le\Bigl[1+\frac{(e^{4K}-1)(1-g_L)}2\Bigr]\sum_{\{x,y\}}\delta_{xy}^2 .
\]
Dividing by $(e^{4K}-1)/4$ gives~\eqref{eq:varbound} with coefficient $4/(r^2-1)+2(1-g_L)=c_0-2g_L$.

For $k\ne0$ apply~\eqref{eq:varbound} to $h_x=\cos(k\cdot x)$ and $h_x=\sin(k\cdot x)$. In both cases $-\Delta h=\lambda(k)h$ and $\sum_{\{x,y\}}\delta_{xy}^2=\sum_xh_x(-\Delta h)_x=\lambda(k)\sum_xh_x^2$. Adding the two inequalities and using translation invariance,
\[
 \lambda(k)^2\,V\widehat G_L(k)\le(c_0-2g_L)\,\lambda(k)\,V .
\]
Since $\lambda(k)>0$, this is~\eqref{eq:IR}.
\end{proof}

The sum rule~\eqref{eq:sumrule} and~\eqref{eq:IR} already give an ordered phase. Let $\mathsf W=\int_{[-\pi,\pi]^3}(1-\frac13\sum_j\cos k_j)^{-1}\frac{\dd^3k}{(2\pi)^3}=1.516386\ldots$ be Watson's integral~\cite{Wat39}, so that $J_0:=\int\lambda^{-1}\frac{\dd^3k}{(2\pi)^3}=\mathsf W/6$.

\begin{corollary}[Single-site bound]\label{cor:singlesite}
If $r^2>1+\mathsf W$, then $m_+(K)^2\ge1-\mathsf W/(r^2-1)>0$. Hence $e^{2K_c}\le\sqrt{1+\mathsf W}<1.58632$.
\end{corollary}

\begin{proof}
Summing~\eqref{eq:IR} over $k\ne0$ and dividing by $V$ gives
\[
 6(1-g_L)=\frac1V\sum_k\lambda(k)\widehat G_L(k)\le c_0-2g_L,
\]
where the identity is $\sum_x G_L(x)\,[(-\Delta)\one_{\{0\}}](x)=6G_L(0)-6g_L$. Thus $g_L\ge(6-c_0)/4$ and $c_0-2g_L\le\frac32(c_0-2)=6/(r^2-1)$. Inserting~\eqref{eq:IR} in~\eqref{eq:sumrule} gives $q_L\ge1-\frac6{r^2-1}V^{-1}\sum_{k\ne0}\lambda(k)^{-1}$. By \cref{lem:riemann} below with $P\equiv1$, the Riemann sum converges to $J_0=\mathsf W/6$. Hence every limit point $q$ of $q_L$ satisfies $q\ge1-\mathsf W/(r^2-1)$. Conclude with \cref{prop:order,cor:monotone}(ii).
\end{proof}

To improve on \cref{cor:singlesite} we use~\eqref{eq:IR} with test functions other than $\one$, and we keep the zero mode $q$ as an unknown, to be bounded from below using further information on local correlations.

\subsection{Spectral rows}

Write $X,Y,Z$ for $\cos k_1,\cos k_2,\cos k_3$ and put
\[
 \Sigma_1=X+Y+Z,\qquad \Sigma_2=X^2+Y^2+Z^2,\qquad \Sigma_{11}=XY+YZ+ZX .
\]
For each row $i$ of \cref{tab:spectral}, let $P_i$ be the average of $F_if_i^2$ over the six permutations of $(X,Y,Z)$, divided by the $\ell^1$-norm $\|\cdot\|$ of its coefficient vector in the basis of permutation-averaged monomials. The five norms, in the order of \cref{tab:spectral}, are $63936016$, $70605774$, $59587922$, $64763232$ and $49871844$. Each $F_i$ is nonnegative on $[-1,1]^3$, hence so is $P_i$. Each $P_i$ is a polynomial of degree at most four, symmetric under permutations of the coordinates. Write $P_i(\one)=P_i(1,1,1)$ and define
\[
 J_i=\int_{[-\pi,\pi]^3}\frac{P_i(\cos k)}{\lambda(k)}\,\frac{\dd^3k}{(2\pi)^3},\qquad
 X_{i,L}=\frac1V\sum_{k\in\T_L^*}P_i(\cos k)\,\widehat G_L(k).
\]
Expanding $P_i(\cos k)=\sum_x\widehat P_i(x)e^{ik\cdot x}$ in harmonics, with $\widehat P_i$ supported on $\|x\|_1\le4$, gives $X_{i,L}=\sum_x\widehat P_i(x)G_L(x)$ for $L\ge10$. This is the mean of the local spin function $\sum_x\widehat P_i(x)\sigma_0\sigma_x$.

\begin{table}[t]
\centering\small
\caption{Spectral test polynomials and the upper ends of the rational enclosures of the Green integrals $J_i$. The lower ends are one unit smaller in the last digit.}\label{tab:spectral}
\begin{tabular}{rlll}
\toprule
$i$ & $F_i$ & $f_i$ & $J_i^+\cdot10^{18}$ \\
\midrule
29 & $(1+X)(1+Y)$ & $2126-119(X+Y)+2110Z$ & $80969192552320727$\\
32 & $1$ & $-1653-613\Sigma_1+609\Sigma_2+1159\Sigma_{11}$ & $6860377629124312$\\
33 & $(1-X)(1-Y)$ & $1722+1417(X+Y)+1421Z$ & $6673088821306827$\\
35 & $(1+X)(1-Y)$ & $1772+1322X+1479Y+1387Z$ & $24027497160169481$\\
69 & $(1+X)(1+Y)$ & $2090-330(X+Y)+2101Z$ & $83985141451046847$\\
\bottomrule
\end{tabular}
\end{table}

The indices $i$ are those of the archived row list and carry no further meaning.

\begin{lemma}[Riemann sums]\label{lem:riemann}
Let $P$ be a trigonometric polynomial. Then
\[
 \lim_{L\to\infty}\frac1V\sum_{k\in\T_L^*\setminus\{0\}}\frac{P(k)}{\lambda(k)}=\int_{[-\pi,\pi]^3}\frac{P(k)}{\lambda(k)}\,\frac{\dd^3k}{(2\pi)^3}.
\]
\end{lemma}

\begin{proof}
Represent $k=2\pi j/L$ with $j\in\{-L/2+1,\dots,L/2\}^3$. The shell $\|j\|_\infty=m\ge1$ contains at most $(2m+1)^3-(2m-1)^3=24m^2+2$ points. On it $\lambda(k)\ge16m^2/L^2$, because $1-\cos x\ge2x^2/\pi^2$ for $|x|\le\pi$. Hence for $0<\delta<1$ the modes with $0<\|k\|_\infty<\delta$ contribute at most
\[
 \frac{\|P\|_\infty}{L^3}\sum_{1\le m\le\delta L/2\pi}(24m^2+2)\frac{L^2}{16m^2}\le C\|P\|_\infty\Bigl(\delta+\frac1L\Bigr).
\]
The integral over $\{\|k\|_\infty<\delta\}$ is $O(\delta)$, since $\lambda(k)^{-1}=O(|k|^{-2})$ is integrable in three dimensions. On $\{\|k\|_\infty\ge\delta\}$ the integrand is continuous, and the Riemann sums converge. Let $L\to\infty$ and then $\delta\downarrow0$.
\end{proof}

\begin{lemma}[Spectral rows]\label{lem:spectral}
Let $L_k\to\infty$ be even integers along which $q_{L_k}\to q$, $g_{L_k}\to g$ and $X_{i,L_k}\to X_i$ for the five indices $i$. Then
\begin{equation}\label{eq:spectralrow}
 X_i\le P_i(\one)\,q+(c_0-2g)\,J_i\le P_i(\one)\,q+(c_0-2g)\,J_i^+ .
\end{equation}
\end{lemma}

\begin{proof}
Separate the term $k=0$ in $X_{i,L}$. It equals $P_i(\one)q_L$. For $k\ne0$ use $P_i\ge0$ and~\eqref{eq:IR}:
\[
 X_{i,L}\le P_i(\one)q_L+(c_0-2g_L)\,\frac1V\sum_{k\ne0}\frac{P_i(\cos k)}{\lambda(k)} .
\]
Pass to the limit along $L_k$ with \cref{lem:riemann}. The second inequality in~\eqref{eq:spectralrow} uses $J_i\le J_i^+$ (\cref{cert:green}) and $c_0-2g\ge c_0-2>0$.
\end{proof}

\begin{certificate}[Green integrals]\label{cert:green}
For the five polynomials of \cref{tab:spectral}, $J_i^+-10^{-18}\le J_i\le J_i^+$, where $10^{18}J_i^+$ is the integer in the last column.
\end{certificate}

The enclosures are obtained from the exact evaluation of the cubic lattice Green function by Glasser and Boersma~\cite{GB00}. See Appendix~\ref{app:green}.

\section{Forcing a positive zero mode}\label{sec:ordered}

\Cref{lem:spectral} bounds five local correlations $X_i$ from above in terms of the unknown zero mode $q$ and the nearest-neighbour correlation $g$. If we could also bound a suitable combination of the $X_i$ and $g$ from below by local information, we would obtain a lower bound on $q$. This section carries out this idea. The local information consists of exact identities and inequalities valid in every torus. Its combination is controlled through the law of the spins on a small shell.

\subsection{Geometry and the shell marginal}\label{sec:geometry}

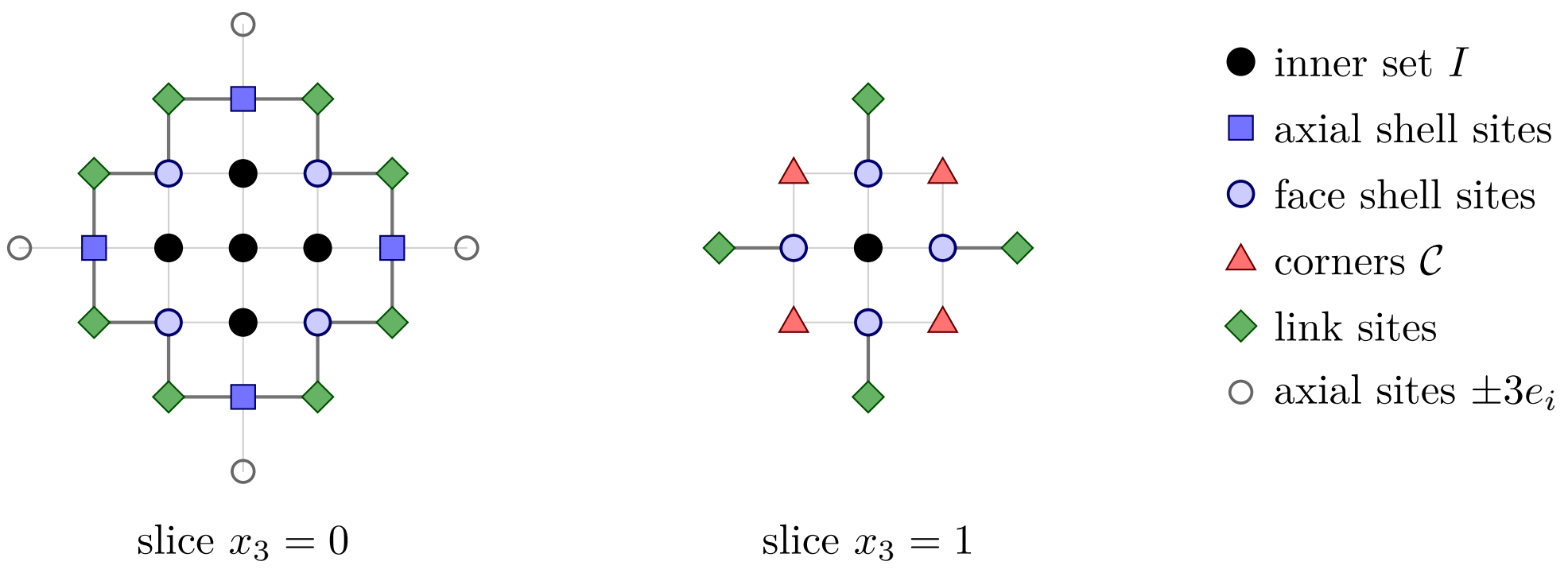
\begin{figure}[t]
\centering
\begin{tikzpicture}[line cap=round]
\draw[black!18] (-1.860,0.000) -- (-1.240,0.000);
\draw[black!55,line width=0.8pt] (-1.240,-0.620) -- (-1.240,0.000);
\draw[black!55,line width=0.8pt] (-1.240,-0.620) -- (-0.620,-0.620);
\draw[black!55,line width=0.8pt] (-1.240,0.000) -- (-1.240,0.620);
\draw[black!18] (-1.240,0.000) -- (-0.620,0.000);
\draw[black!55,line width=0.8pt] (-1.240,0.620) -- (-0.620,0.620);
\draw[black!55,line width=0.8pt] (-0.620,-1.240) -- (-0.620,-0.620);
\draw[black!55,line width=0.8pt] (-0.620,-1.240) -- (0.000,-1.240);
\draw[black!18] (-0.620,-0.620) -- (-0.620,0.000);
\draw[black!18] (-0.620,-0.620) -- (0.000,-0.620);
\draw[black!18] (-0.620,0.000) -- (-0.620,0.620);
\draw[black!18] (-0.620,0.000) -- (0.000,0.000);
\draw[black!55,line width=0.8pt] (-0.620,0.620) -- (-0.620,1.240);
\draw[black!18] (-0.620,0.620) -- (0.000,0.620);
\draw[black!55,line width=0.8pt] (-0.620,1.240) -- (0.000,1.240);
\draw[black!18] (0.000,-1.860) -- (0.000,-1.240);
\draw[black!18] (0.000,-1.240) -- (0.000,-0.620);
\draw[black!55,line width=0.8pt] (0.000,-1.240) -- (0.620,-1.240);
\draw[black!18] (0.000,-0.620) -- (0.000,0.000);
\draw[black!18] (0.000,-0.620) -- (0.620,-0.620);
\draw[black!18] (0.000,0.000) -- (0.000,0.620);
\draw[black!18] (0.000,0.000) -- (0.620,0.000);
\draw[black!18] (0.000,0.620) -- (0.000,1.240);
\draw[black!18] (0.000,0.620) -- (0.620,0.620);
\draw[black!18] (0.000,1.240) -- (0.000,1.860);
\draw[black!55,line width=0.8pt] (0.000,1.240) -- (0.620,1.240);
\draw[black!55,line width=0.8pt] (0.620,-1.240) -- (0.620,-0.620);
\draw[black!18] (0.620,-0.620) -- (0.620,0.000);
\draw[black!55,line width=0.8pt] (0.620,-0.620) -- (1.240,-0.620);
\draw[black!18] (0.620,0.000) -- (0.620,0.620);
\draw[black!18] (0.620,0.000) -- (1.240,0.000);
\draw[black!55,line width=0.8pt] (0.620,0.620) -- (0.620,1.240);
\draw[black!55,line width=0.8pt] (0.620,0.620) -- (1.240,0.620);
\draw[black!55,line width=0.8pt] (1.240,-0.620) -- (1.240,0.000);
\draw[black!55,line width=0.8pt] (1.240,0.000) -- (1.240,0.620);
\draw[black!18] (1.240,0.000) -- (1.860,0.000);
\draw[black!60,line width=0.7pt] (-1.860,0.000) circle (2.6pt);
\filldraw[fill=green!50!black!60,draw=green!30!black] (-1.240,-0.620+0.13) -- (-1.240+0.13,-0.620) -- (-1.240,-0.620-0.13) -- (-1.240-0.13,-0.620) -- cycle;
\filldraw[fill=blue!55,draw=blue!40!black] (-1.240-0.10,0.000-0.10) rectangle (-1.240+0.10,0.000+0.10);
\filldraw[fill=green!50!black!60,draw=green!30!black] (-1.240,0.620+0.13) -- (-1.240+0.13,0.620) -- (-1.240,0.620-0.13) -- (-1.240-0.13,0.620) -- cycle;
\filldraw[fill=green!50!black!60,draw=green!30!black] (-0.620,-1.240+0.13) -- (-0.620+0.13,-1.240) -- (-0.620,-1.240-0.13) -- (-0.620-0.13,-1.240) -- cycle;
\filldraw[fill=blue!20,draw=blue!40!black,line width=0.7pt] (-0.620,-0.620) circle (3.0pt);
\fill[black] (-0.620,0.000) circle (3.4pt);
\filldraw[fill=blue!20,draw=blue!40!black,line width=0.7pt] (-0.620,0.620) circle (3.0pt);
\filldraw[fill=green!50!black!60,draw=green!30!black] (-0.620,1.240+0.13) -- (-0.620+0.13,1.240) -- (-0.620,1.240-0.13) -- (-0.620-0.13,1.240) -- cycle;
\draw[black!60,line width=0.7pt] (0.000,-1.860) circle (2.6pt);
\filldraw[fill=blue!55,draw=blue!40!black] (0.000-0.10,-1.240-0.10) rectangle (0.000+0.10,-1.240+0.10);
\fill[black] (0.000,-0.620) circle (3.4pt);
\fill[black] (0.000,0.000) circle (3.4pt);
\fill[black] (0.000,0.620) circle (3.4pt);
\filldraw[fill=blue!55,draw=blue!40!black] (0.000-0.10,1.240-0.10) rectangle (0.000+0.10,1.240+0.10);
\draw[black!60,line width=0.7pt] (0.000,1.860) circle (2.6pt);
\filldraw[fill=green!50!black!60,draw=green!30!black] (0.620,-1.240+0.13) -- (0.620+0.13,-1.240) -- (0.620,-1.240-0.13) -- (0.620-0.13,-1.240) -- cycle;
\filldraw[fill=blue!20,draw=blue!40!black,line width=0.7pt] (0.620,-0.620) circle (3.0pt);
\fill[black] (0.620,0.000) circle (3.4pt);
\filldraw[fill=blue!20,draw=blue!40!black,line width=0.7pt] (0.620,0.620) circle (3.0pt);
\filldraw[fill=green!50!black!60,draw=green!30!black] (0.620,1.240+0.13) -- (0.620+0.13,1.240) -- (0.620,1.240-0.13) -- (0.620-0.13,1.240) -- cycle;
\filldraw[fill=green!50!black!60,draw=green!30!black] (1.240,-0.620+0.13) -- (1.240+0.13,-0.620) -- (1.240,-0.620-0.13) -- (1.240-0.13,-0.620) -- cycle;
\filldraw[fill=blue!55,draw=blue!40!black] (1.240-0.10,0.000-0.10) rectangle (1.240+0.10,0.000+0.10);
\filldraw[fill=green!50!black!60,draw=green!30!black] (1.240,0.620+0.13) -- (1.240+0.13,0.620) -- (1.240,0.620-0.13) -- (1.240-0.13,0.620) -- cycle;
\draw[black!60,line width=0.7pt] (1.860,0.000) circle (2.6pt);
\node[font=\small] at (0.00,-2.45) {slice $x_3=0$};
\draw[black!55,line width=0.8pt] (3.960,0.000) -- (4.580,0.000);
\draw[black!18] (4.580,-0.620) -- (4.580,0.000);
\draw[black!18] (4.580,-0.620) -- (5.200,-0.620);
\draw[black!18] (4.580,0.000) -- (4.580,0.620);
\draw[black!18] (4.580,0.000) -- (5.200,0.000);
\draw[black!18] (4.580,0.620) -- (5.200,0.620);
\draw[black!55,line width=0.8pt] (5.200,-1.240) -- (5.200,-0.620);
\draw[black!18] (5.200,-0.620) -- (5.200,0.000);
\draw[black!18] (5.200,-0.620) -- (5.820,-0.620);
\draw[black!18] (5.200,0.000) -- (5.200,0.620);
\draw[black!18] (5.200,0.000) -- (5.820,0.000);
\draw[black!55,line width=0.8pt] (5.200,0.620) -- (5.200,1.240);
\draw[black!18] (5.200,0.620) -- (5.820,0.620);
\draw[black!18] (5.820,-0.620) -- (5.820,0.000);
\draw[black!18] (5.820,0.000) -- (5.820,0.620);
\draw[black!55,line width=0.8pt] (5.820,0.000) -- (6.440,0.000);
\filldraw[fill=green!50!black!60,draw=green!30!black] (3.960,0.000+0.13) -- (3.960+0.13,0.000) -- (3.960,0.000-0.13) -- (3.960-0.13,0.000) -- cycle;
\filldraw[fill=red!55,draw=red!40!black] (4.580,-0.620+0.13) -- (4.580-0.12,-0.620-0.08) -- (4.580+0.12,-0.620-0.08) -- cycle;
\filldraw[fill=blue!20,draw=blue!40!black,line width=0.7pt] (4.580,0.000) circle (3.0pt);
\filldraw[fill=red!55,draw=red!40!black] (4.580,0.620+0.13) -- (4.580-0.12,0.620-0.08) -- (4.580+0.12,0.620-0.08) -- cycle;
\filldraw[fill=green!50!black!60,draw=green!30!black] (5.200,-1.240+0.13) -- (5.200+0.13,-1.240) -- (5.200,-1.240-0.13) -- (5.200-0.13,-1.240) -- cycle;
\filldraw[fill=blue!20,draw=blue!40!black,line width=0.7pt] (5.200,-0.620) circle (3.0pt);
\fill[black] (5.200,0.000) circle (3.4pt);
\filldraw[fill=blue!20,draw=blue!40!black,line width=0.7pt] (5.200,0.620) circle (3.0pt);
\filldraw[fill=green!50!black!60,draw=green!30!black] (5.200,1.240+0.13) -- (5.200+0.13,1.240) -- (5.200,1.240-0.13) -- (5.200-0.13,1.240) -- cycle;
\filldraw[fill=red!55,draw=red!40!black] (5.820,-0.620+0.13) -- (5.820-0.12,-0.620-0.08) -- (5.820+0.12,-0.620-0.08) -- cycle;
\filldraw[fill=blue!20,draw=blue!40!black,line width=0.7pt] (5.820,0.000) circle (3.0pt);
\filldraw[fill=red!55,draw=red!40!black] (5.820,0.620+0.13) -- (5.820-0.12,0.620-0.08) -- (5.820+0.12,0.620-0.08) -- cycle;
\filldraw[fill=green!50!black!60,draw=green!30!black] (6.440,0.000+0.13) -- (6.440+0.13,0.000) -- (6.440,0.000-0.13) -- (6.440-0.13,0.000) -- cycle;
\node[font=\small] at (5.20,-2.45) {slice $x_3=1$};
\fill[black] (8.300,1.550) circle (3.4pt);
\node[right=4pt,font=\small] at (8.30,1.55) {inner set $I$};
\filldraw[fill=blue!55,draw=blue!40!black] (8.300-0.10,1.000-0.10) rectangle (8.300+0.10,1.000+0.10);
\node[right=4pt,font=\small] at (8.30,1.00) {axial shell sites};
\filldraw[fill=blue!20,draw=blue!40!black,line width=0.7pt] (8.300,0.450) circle (3.0pt);
\node[right=4pt,font=\small] at (8.30,0.45) {face shell sites};
\filldraw[fill=red!55,draw=red!40!black] (8.300,-0.100+0.13) -- (8.300-0.12,-0.100-0.08) -- (8.300+0.12,-0.100-0.08) -- cycle;
\node[right=4pt,font=\small] at (8.30,-0.10) {corners $\mathcal C$};
\filldraw[fill=green!50!black!60,draw=green!30!black] (8.300,-0.650+0.13) -- (8.300+0.13,-0.650) -- (8.300,-0.650-0.13) -- (8.300-0.13,-0.650) -- cycle;
\node[right=4pt,font=\small] at (8.30,-0.65) {link sites};
\draw[black!60,line width=0.7pt] (8.300,-1.200) circle (2.6pt);
\node[right=4pt,font=\small] at (8.30,-1.20) {axial sites $\pm3e_i$};
\end{tikzpicture}
\caption{Two horizontal slices of the neighbourhood of the origin used in \cref{sec:ordered}. The shell $\partial I=\{\|x\|_1=2\}$ separates the inner set $I$ from the exterior. Each link site of type $(2,1,0)$ joins one axial and one face shell site (dark bonds); summing over its spin couples the two shell spins and produces the factor $\xi^{d(b)}$ in~\eqref{eq:Psi}. The corners appear in the slices $x_3=\pm1$.}
\label{fig:shells}
\end{figure}

Let
\[
 I=\{x\in\Z^3:\|x\|_1\le1\},\qquad \partial I=\{x:\|x\|_1=2\},
\]
regarded as subsets of $\T_L$ for $L\ge12$. The inner set $I$ has $7$ sites. The \emph{shell} $\partial I$ has $18$ sites: six \emph{axial} sites $\pm2e_i$ and twelve \emph{face} sites $\pm e_i\pm e_j$, $i<j$. There are no bonds within $\partial I$, since adjacent sites have $\ell^1$-norms of different parity. Every neighbour of $I$ lies in $I\cup\partial I$. The third shell $\{\|x\|_1=3\}$ consists of the eight \emph{corners} $\mathcal C=\{\pm1\}^3$, the six axial sites $\pm3e_i$, and the $24$ \emph{link sites} of type $(2,1,0)$, i.e.\ the images of $(2,1,0)$ under signed permutations. Each link site is adjacent to exactly one axial and one face site of $\partial I$. Each axial site lies on four links and each face site on two. Each corner is adjacent to three face sites. See \cref{fig:shells}.

Let $b\in\{\pm1\}^{\partial I}$ denote a \emph{shell word} and let $\mu_L(b)$ be its marginal law under $\avg\cdot_L$. Write $H_I(\sigma_I,b)$ for the sum of $\sigma_x\sigma_y$ over the $36$ bonds with at least one endpoint in $I$, and put
\[
 Z_I(b)=\sum_{\sigma_I}e^{KH_I(\sigma_I,b)},\qquad
 f^I(b)=\frac1{Z_I(b)}\sum_{\sigma_I}f(\sigma_I,b)\,e^{KH_I(\sigma_I,b)}
\]
for functions $f$ of the spins in $I\cup\partial I$. By the Markov property, $\avg f_L=\mu_L(f^I):=\sum_b\mu_L(b)f^I(b)$. We call $f^I$ the \emph{conditional representative} of $f$. It is an explicit rational function of $r$ for each $b$. The group $\mathfrak G$ generated by the $48$ signed permutations of the axes and the global spin flip acts on shell words, and $\mu_L$ is $\mathfrak G$-invariant.

\subsection{Four families of rows}\label{sec:rows}

A \emph{row} is a function $\rho:\{\pm1\}^{\partial I}\to\R$ together with an inequality for $\mu_L(\rho)$ valid for all large even $L$. All rows below are averaged over $\mathfrak G$, so they are $\mathfrak G$-invariant. There are four families.

\begin{lemma}[Translation rows]\label{lem:translation}
Let $f$ be a function of the spins in a set $S$ and let $a\in\Z^3$ be such that $S$ and $S+a$ are contained in $I\cup\partial I$. Let $f_a$ denote the translate of $f$ to $S+a$. Then $\rT=f^I-f_a^I$ satisfies $\mu_L(\rT)=0$.
\end{lemma}

\begin{proof}
$\mu_L(\rT)=\avg f_L-\avg{f_a}_L=0$ by translation invariance of the periodic measure.
\end{proof}

The certificate uses $104$ translation rows $\rT_j$.

The second family comes from reflection positivity~\cite{FILS78,Bis09}, which we use in the following form. Let $\vartheta$ be the reflection of $\T_L$ in a pair of bond-bisecting planes orthogonal to a coordinate axis, as in the proof of \cref{lem:rotation}, and let $\T^\ell,\T^r$ be the two halves. Write each crossing bond weight as $e^{K\sigma\sigma'}=e^{-K}M_{\sigma\sigma'}$ with
\[
 M=\begin{pmatrix}r&1\\1&r\end{pmatrix}=M^{1/2}M^{1/2},
\]
and insert $M_{\sigma_x\sigma_{\vartheta x}}=\sum_{\iota=1,2}(M^{1/2})_{\sigma_x\iota}(M^{1/2})_{\iota\sigma_{\vartheta x}}$ on every crossing bond. As in \cref{lem:rotation}, the torus weight becomes $\sum_{\boldsymbol\iota}\Xi_{\boldsymbol\iota}(\sigma^\ell)\Xi_{\boldsymbol\iota}(\vartheta\sigma^r)$. Consequently, for every collection of functions $F_{\boldsymbol\iota}$ of the spins of $\T^\ell$,
\begin{equation}\label{eq:RP}
 \sum_{\boldsymbol\iota}\Bigl\langle\!\Bigl\langle F_{\boldsymbol\iota}\,\vartheta F_{\boldsymbol\iota}\Bigr\rangle\!\Bigr\rangle\ge0,
\end{equation}
where $\langle\!\langle\cdot\rangle\!\rangle$ denotes the normalized sum with the split weights. Two refinements make~\eqref{eq:RP} applicable to local functions that straddle the cutting plane.
\begin{itemize}
\item \emph{Transported spins.} Since $\sigma_xM_{\sigma_x\sigma'}=(\mathrm{diag}(1,-1)M)_{\sigma_x\sigma'}$, a spin $\sigma_x$ at an endpoint of a crossing bond can be moved to the auxiliary index: replacing the column $M^{1/2}$ by $M^{-1/2}\mathrm{diag}(1,-1)M$ on the far side represents multiplication by $\sigma_x$.
\item \emph{Doubled insertions.} If both factors in~\eqref{eq:RP} carry a transported spin through the same crossing bond $\{x,\vartheta x\}$, the resulting Gram matrix is
\[
 (M^{-1/2}\mathrm{diag}(1,-1)M)^{\!\top}(M^{-1/2}\mathrm{diag}(1,-1)M)=\frac{M^3}{r^2-1}.
\]
Divided entrywise by $M$, it has the entries $c_0-\sigma_x\sigma_{\vartheta x}$, with the coefficient $c_0=u^{-1}+u$ of \cref{sec:infrared}.
\end{itemize}
Thus every choice of coefficient vectors for finitely many local functions near the cutting plane, with or without transported spins, yields an inequality $\avg{\mathcal F}_L\ge0$ for an explicit local function $\mathcal F$.

\begin{lemma}[Reflection rows]\label{lem:reflection}
For each of the four coefficient vectors of \cref{cert:multipliers}, which have $112$ components, the conditional representative $\rA_k$ of the corresponding function $\mathcal F_k$, multiplied by a positive normalizing constant, satisfies $\mu_L(\rA_k)\ge0$.
\end{lemma}

The third family compares the region outside $I$ with a finite free graph. Let $\mathrm D$ be the graph obtained from $\T_L$ by deleting $I$ and all bonds incident to it. Let $\Gamma\subset\mathrm D$ be the free induced subgraph on the second and third shells and on the fourth-shell orbits of $(0,1,3)$ and $(0,2,2)$. It has $92$ sites and $150$ bonds. For $A\subset\partial I$ write $\chi_A(b)=\prod_{x\in A}b_x$.

\begin{lemma}[Deletion rows]\label{lem:deletion}
For $A,T\subset\partial I$ the function
\[
 \rD_{A,T}(b)=\frac{\chi_{A\triangle T}(b)-\avg{\chi_T}_\Gamma\,\chi_A(b)}{Z_I(b)}
\]
satisfies $\mu_L(\rD_{A,T})\ge0$.
\end{lemma}

\begin{proof}
Summing the torus weight over the spins outside $I\cup\partial I$ at fixed $b$ gives $\mu_L(b)=Z_I(b)Z_{\mathrm D}(b)/Z_L$, where $Z_{\mathrm D}(b)$ is the partition function of the Ising model on $\mathrm D$ with the shell spins fixed to $b$. Hence $F_A:=\sum_b\mu_L(b)\chi_A(b)/Z_I(b)$ is proportional, with a positive constant independent of $A$, to $\avg{\chi_A}_{\mathrm D}$. By \cref{lem:standard}(i) on $\mathrm D$, $\avg{\chi_{A\triangle T}}_{\mathrm D}=\avg{\chi_A\chi_T}_{\mathrm D}\ge\avg{\chi_A}_{\mathrm D}\avg{\chi_T}_{\mathrm D}$ and $\avg{\chi_T}_{\mathrm D}\ge\avg{\chi_T}_\Gamma$. With $\avg{\chi_A}_{\mathrm D}\ge0$ this gives $F_{A\triangle T}-\avg{\chi_T}_\Gamma F_A\ge0$, which is the claim.
\end{proof}

The certificate uses seven deletion rows $\rD_s$, each multiplied by $32\min_bZ_I(b)$ and symmetrized. The observable $\chi_{A\triangle T}(b)/Z_I(b)$ is integrated against the \emph{true} marginal $\mu_L$. No independence between $b$ and the interior is assumed.

The fourth family consists of the spectral rows of \cref{lem:spectral}. Let $g^I$ and $X_i^I$ be conditional representatives of local functions supported in $I\cup\partial I$ whose means are $g_L$ and $X_{i,L}$. Such representatives exist by translation invariance, since every displacement $x$ with $\|x\|_1\le4$ is realized by a pair of sites in $I\cup\partial I$. Put
\[
 \rS_i(b)=J_i^+\bigl(c_0-2g^I(b)\bigr)-X_i^I(b).
\]

\subsection{The residual}\label{sec:residual}

Write $y=\log r$, and let $a_j\in\mathbb Q$, $d_k,e_s,v_i\in\mathbb Q_{\ge0}$ be the multipliers of \cref{cert:multipliers}. The \emph{residual} and its zero-mode coefficient are
\begin{equation}\label{eq:residual}
 R_y(b)=-\sum_ja_j\rT_j(b)+\sum_kd_k\rA_k(b)+\sum_se_s\rD_s(b)+\sum_iv_i\rS_i(b),\qquad
 Q=\sum_iv_iP_i(\one).
\end{equation}

\begin{proposition}[Lower row inequality]\label{prop:lowerrow}
Let $L_k\to\infty$ be even integers along which $q_{L_k}$, $g_{L_k}$, the $X_{i,L_k}$ and $\mu_{L_k}(R_y)$ converge, and let $q=\lim q_{L_k}$. Then
\[
 0\le\lim_{k\to\infty}\mu_{L_k}(R_y)+Q\,q .
\]
\end{proposition}

\begin{proof}
By \cref{lem:translation,lem:reflection,lem:deletion}, $\mu_L(\rT_j)=0$, $\mu_L(\rA_k)\ge0$ and $\mu_L(\rD_s)\ge0$. By \cref{lem:spectral}, $\lim_k\mu_{L_k}(\rS_i)=J_i^+(c_0-2g)-X_i\ge-P_i(\one)q$. Multiply by the multipliers, of which all except the $a_j$ are nonnegative, and sum.
\end{proof}

\begin{certificate}[Rows and multipliers]\label{cert:multipliers}
The certificate consists of the $104+4+7+5=120$ rows as explicit local functions, specified by their supports, coefficient vectors and orbit data, and of their rational multipliers. The multipliers have the common denominator $10^6\cdot17887576666057000$. The translation multipliers $a_j$ have either sign, and all others are nonnegative. Moreover
\[
 Q=\frac{8943787728743670589}{8943788333028500000}.
\]
The residual $R_y$ is $\mathfrak G$-invariant. It is therefore constant on each of the $3236$ orbits of $\mathfrak G$ on $\{\pm1\}^{\partial I}$.
\end{certificate}

The values $R_y(b)$ have either sign, and $\max_bR_y(b)$ is positive. An upper bound for $\mu_L(R_y)$ that is strong enough to be combined with \cref{prop:lowerrow} must therefore use the structure of $\mu_L$. The rest of this section derives such a bound.

\subsection{A mixture representation of the shell marginal}\label{sec:mixture}

The marginal $\mu_L$ depends on the unknown configuration outside the shell. We remove this dependence in two steps. First we sum over a thin layer of spins next to the shell: the eight corners, the $24$ link sites and three sites on each of the six half-axes. Each summed spin leaves a factor in the shell spins, and we bracket it between two extreme cases. Expanding every factor in these two cases writes $\mu_L$ as a mixture of explicit finite measures. The measures are indexed by auxiliary binary labels, and only the mixture weights depend on the exterior.

Put $\zeta=1/r=e^{-2K}$. With the normalization in which a bond contributes $1$ if its spins agree and $\zeta$ otherwise, we use the following constants:
\begin{equation}\label{eq:shellconstants}
 \xi=\frac{\zeta+\zeta^5}{\sqrt{(1+\zeta^6)(\zeta^2+\zeta^4)}},\qquad
 \eta=\frac{\zeta^2+\zeta^4}{1+\zeta^6},\qquad
 \rho_{\rm face}=\eta,\qquad \rho_{\rm axial}=\eta^2\phi_3,
\end{equation}
where $\phi_0=\zeta$ and $\phi_{m+1}=(\zeta+\zeta^4\phi_m)/(1+\zeta^5\phi_m)$. All of them lie in $(0,1)$ and are algebraic in $\zeta$. The quantities $\xi^2$, $\eta$, $\rho$ entering below are rational at rational $r$.

Two elementary facts are used repeatedly. The first describes how a ratio of weights propagates through a single spin. Write $F(x)=(\zeta+x)/(1+\zeta x)$; it is increasing on $[0,\infty)$ and satisfies $F(1/x)=1/F(x)$.

\begin{lemma}[Transfer of odds]\label{lem:oddstransfer}
Let $x$ be a site with neighbour $y$, and suppose that, with all other variables fixed, the remaining factors of the weight give $\sigma_x$ the two values $(\alpha_+,\alpha_-)$ with $\alpha_-/\alpha_+\in[\rho,\rho^{-1}]$ for some $\rho\in(0,1]$. Then summing over $\sigma_x$ gives $\sigma_y$ a factor $(\beta_+,\beta_-)$ with $\beta_-/\beta_+\in[F(\rho),F(\rho)^{-1}]$.
\end{lemma}

\begin{proof}
In the normalization of agreeing bonds, $\beta_+=\alpha_++\zeta\alpha_-$ and $\beta_-=\zeta\alpha_++\alpha_-$, so $\beta_-/\beta_+=F(\alpha_-/\alpha_+)$. The claim follows from the monotonicity of $F$ and $F(1/x)=1/F(x)$.
\end{proof}

The second fact is the decomposition device: if $0<\rho<1$ and $x=(x_+,x_-)$ is a positive vector with $\rho\le x_-/x_+\le\rho^{-1}$, then $x$ is a nonnegative combination of the two \emph{columns} $(1,\rho)$ and $(\rho,1)$. Applying it to a factor of the Gibbs weight introduces an auxiliary binary label, the choice of column, and summing over the label restores the weight.

\begin{lemma}[Mixture representation]\label{lem:mixture}
For $c\in\{\pm1\}^{\mathcal C}$, a \emph{corner source}, and $f\in\{\pm1\}^{\partial I}$, a \emph{field label}, define
\begin{equation}\label{eq:Psi}
 \Psi_{c,f}(b)=Z_I(b)\,\xi^{d(b)}\prod_{z\in\mathcal C}\bigl(\zeta^{k_z}+\zeta^{6-k_z}\bigr)\prod_{j\in\partial I}\rho_j^{\one\{b_j\ne f_j\}} .
\end{equation}
Here $d(b)$ is the number of links whose two shell endpoints disagree, which is even. The number $k_z$ counts the minus signs among the three face neighbours of $z$ and three copies of $c_z$. Finally $\rho_j=\rho_{\rm axial}$ or $\rho_{\rm face}$ according to the type of $j$. Let $\nu_{c,f}=\Psi_{c,f}/\sum_{b'}\Psi_{c,f}(b')$. Then for every even $L\ge12$ there are numbers $p_L(c,f)\ge0$ with $\sum_{c,f}p_L(c,f)=1$ such that
\begin{equation}\label{eq:mixture}
 \mu_L=\sum_{c,f}p_L(c,f)\,\nu_{c,f}.
\end{equation}
Moreover $p_L(gc,gf)=p_L(c,f)$ for $g\in\mathfrak G$.
\end{lemma}

\begin{proof}
Condition on the spins $\omega$ outside $I\cup\partial I\cup\mathcal C\cup\{\text{link sites}\}\cup\{\text{axial segments}\}$, where the axial segments are the sites $me_i$, $3\le|m|\le5$. Given $\omega$, the conditional weight of $b$ is $Z_I(b)$ times a product of factors, one for each corner, each link site and each axial segment. These are obtained by summing over the spin at that site or segment. We decompose each factor in turn.

\emph{Corners.} The three outward bonds of a corner $z$ contribute the vector $(\zeta^m,\zeta^{3-m})$ to the two values of $\sigma_z$, where $m$ is the number of minus signs among the outward neighbours. Its ratio lies in $[\zeta^3,\zeta^{-3}]$, so it is a nonnegative combination of $(1,\zeta^3)$ and $(\zeta^3,1)$. These columns are the weights produced by three outward neighbours of a common sign $c_z$. Summing over $\sigma_z$ then gives $\zeta^{k_z}+\zeta^{6-k_z}$.

\emph{Links.} A link site $t$ adjacent to the shell sites $i$ (axial) and $j$ (face) has four outward neighbours. If all four had the sign $a$, summing over $\sigma_t$ would give
\[
 (1+\zeta^6)\,\xi^{\one\{b_i\ne b_j\}}\,\eta^{(\one\{b_i\ne a\}+\one\{b_j\ne a\})/2},
\]
as one checks in the three cases $b_i=b_j=a$, $b_i=b_j=-a$ and $b_i\ne b_j$. For a general outward pattern the factor is a nonnegative mixture of the two cases $a=\pm1$. We call $a$ the \emph{preference} of the link.

\emph{Axial segments.} The three sites of an axial segment each have four outward neighbours. Tracing them from the outermost inwards and applying the computation of \cref{lem:oddstransfer} with the four outward bonds in the least favourable configuration gives the recursion $\phi_{m+1}=(\zeta+\zeta^4\phi_m)/(1+\zeta^5\phi_m)$, started from the ratio $\phi_0=\zeta$ of a single bond. Hence the factor at the axial shell site has ratio in $[\phi_3,\phi_3^{-1}]$. Combined with the $\eta^{1/2}$-factors of its four links, the total factor at an axial site has ratio in $[\rho_{\rm axial},\rho_{\rm axial}^{-1}]$. At a face site the two links give ratio in $[\eta,\eta^{-1}]$.

\emph{Field labels.} Decompose the total factor at each shell site $j$ into the columns $(1,\rho_j)$ and $(\rho_j,1)$, and call the chosen column $f_j$. Summing the products of all decomposition coefficients over $\omega$, against its true weight, yields nonnegative numbers $p_L(c,f)$ for which~\eqref{eq:mixture} holds after normalization. Symmetry follows from the $\mathfrak G$-invariance of $\avg\cdot_L$ and the covariance of each decomposition.
\end{proof}

The mixture weights $p_L(c,f)$ carry the whole influence of the exterior and are not known explicitly. The labels at the two ends of a link are coupled through its preference, and the corner sources are coupled through the exterior. The following two families of inequalities constrain them.

Fix a link $t$ joining the axial site $i$ and the face site $j$. For a field label $f$ write $f^{(i)}$ for $f$ with the bit at $i$ flipped.

\begin{lemma}[Bridge inequality]\label{lem:bridge}
If $f_i\ne f_j$, then for every $c$
\[
 p_L(c,f)\le\omega_i^{-1}p_L(c,f^{(i)})+\omega_j^{-1}p_L(c,f^{(j)}),\qquad
 \omega_k=\rho_k\,\frac{\rho_k/\eta-\rho_k}{1-\rho_k^2/\eta}>0 .
\]
We call $\omega_{\rm axial}$ and $\omega_{\rm face}$ the \emph{flip weights}.
\end{lemma}

\begin{proof}
Since $f_i\ne f_j$, one of the two labels disagrees with the preference $a$ of the link $t$, say $f_i\ne a$. Given $a$, the shell spins and the exterior, the local odds at $i$ in favour of $a$ are at least $\rho_i/\eta$. Expanding a vector with odds $x$ in the columns $(1,\rho_i)$ and $(\rho_i,1)$ gives coefficient odds $(x-\rho_i)/(1-\rho_ix)$. This is increasing in $x$, and is therefore at least $(\rho_i/\eta-\rho_i)/(1-\rho_i^2/\eta)$. For either value of $b_i$, the entries of the favourable and unfavourable columns have ratio at least $\rho_i$. Hence the unnormalized weight of $(c,f)$ on the event $\{a\ne f_i\}$ is at most $\omega_i^{-1}$ times the weight of $(c,f^{(i)})$ on the same event. Summing over the two values of $a$, with the roles of $i$ and $j$ exchanged on $\{a\ne f_j\}$, and then over the exterior gives the claim.
\end{proof}

For the corner sources, $\mathfrak G$ partitions $\{\pm1\}^{\mathcal C}$ into $14$ classes. Let $n_C$ be the size of the class $C$ and $p_C=\sum_{c\in C}\sum_fp_L(c,f)$ its mass. By symmetry each source in $C$ has mass $p_C/n_C$.

\begin{lemma}[Class odds]\label{lem:classodds}
Let $C,C'$ be classes such that a representative of $C'$ is obtained from one of $C$ by flipping a single corner source $c_z$. Let $m\in\{0,1,2,3\}$ be the number of the three corners adjacent to $z$ in the cube $\{\pm1\}^3$ whose source agrees with the new value of $c_z$. Then $p_{C'}/n_{C'}\ge\beta_m\,p_C/n_C$, where
\[
 \beta_m=\frac{\varpi-\zeta^3}{1-\zeta^3\varpi}\cdot\frac{\upsilon_m+\zeta^3}{1+\zeta^3\upsilon_m},\qquad
 \varpi=F(\zeta^4F(l^5))^3,\quad \upsilon_m=a_-^{3-m}a_+^m,
\]
with $F(x)=(\zeta+x)/(1+\zeta x)$, $l=F(\zeta^5)$ and $a_\pm=F(\zeta^4F(l^2\zeta^{\mp3}))$.
\end{lemma}

\begin{proof}
Fix the configuration outside the $13$-site star consisting of $z$ and its six two-step arms $z+te$, $t\in\{1,2\}$, where $e$ runs over the three inner directions (towards the face sites) and the three outer directions. Fix also the other seven corner sources. Their decomposition coefficients are then constants, so it suffices to compare the unnormalized weights $W_\pm$ of the two values of $c_z$.

\emph{Leaves.} The \emph{leaves} are the $21$ sites outside the star adjacent to the second points of the arms: five for each outer arm and two for each inner arm. Each leaf has five neighbours outside the star. By \cref{lem:oddstransfer} with $\rho=\zeta^5$, tracing such a leaf gives its arm site a factor with ratio in $[l,l^{-1}]$, $l=F(\zeta^5)$.

\emph{Outer branches.} The second point of an outer arm receives five leaves, hence a factor with ratio in $[l^5,l^{-5}]$, and \cref{lem:oddstransfer} transfers it to the first point as $F(l^5)$. The first point has four further fixed neighbours, which contribute a ratio of at least $\zeta^4$. Transferring once more, the three outer branches together give $z$ a factor whose ratio $x$ satisfies $x\in[\varpi,\varpi^{-1}]$ with $\varpi=F(\zeta^4F(l^5))^3$. The coefficient of the column associated with the sign $\pm$ in the decomposition of this factor into the columns $(1,\zeta^3)$ and $(\zeta^3,1)$ is obtained by inverting the $2\times2$ matrix with diagonal $1$ and off-diagonal $\zeta^3$. The ratio of the two coefficients equals $(x-\zeta^3)/(1-\zeta^3x)$, which is increasing in $x$ and therefore at least $(\varpi-\zeta^3)/(1-\zeta^3\varpi)>0$ in favour of either sign.

\emph{Inner branches.} The first point of an inner arm is a face site, and its second point is a corner $z'\ne z$. The outward field of $z'$ has been replaced by its fixed source, which acts like three bonds of strength $K$, that is, with ratio $\zeta^{\mp3}$ according to whether the source of $z'$ agrees with the new value of $c_z$. The two remaining neighbours of $z'$ outside the star are leaves, and the face site has four further fixed neighbours, which contribute at least $\zeta^4$. Two applications of \cref{lem:oddstransfer} show that each inner branch gives $z$ a factor with ratio at least $a_+=F(\zeta^4F(l^2\zeta^{-3}))$ if the neighbouring source agrees and $a_-=F(\zeta^4F(l^2\zeta^{3}))$ otherwise. With $m$ agreeing sources the three inner branches give a ratio of at least $\upsilon_m=a_-^{3-m}a_+^m$.

\emph{The corner spin.} Summing over $\sigma_z$ with the column of the source $c_z$ multiplies the source odds by the factor $(\upsilon+\zeta^3)/(1+\zeta^3\upsilon)$, where $\upsilon$ is the ratio contributed by the inner branches. This factor is increasing in $\upsilon$ and hence at least $(\upsilon_m+\zeta^3)/(1+\zeta^3\upsilon_m)$.

Combining the three steps gives $W_{\rm new}\ge\beta_mW_{\rm old}$ for every fixed configuration outside the star. The inequality concerns unnormalized weights, so it can be summed over the exterior configuration and over the field labels, also on events of zero mass. This gives $\sum_fp_L(c',f)\ge\beta_m\sum_fp_L(c,f)$ for the two sources $c\in C$ and $c'=c^{(z)}\in C'$. By symmetry each source of a class $C$ has mass $p_C/n_C$, which proves the claim.

The number $m$ depends only on the pair of classes. Indeed, let $E(c)$ be the number of edges of the cube $\{\pm1\}^3$ whose endpoints carry equal sources. It is invariant under $\mathfrak G$ and under the global flip, and flipping $c_z$ changes it by $m-(3-m)=2m-3$. Hence $m=(E(c')-E(c)+3)/2$ is determined by $C$ and $C'$.
\end{proof}

The program \nolinkurl{upper/STAGE29_ORDERED_corner.py} checks the geometric facts used in this proof for all eight corners: the star has $13$ sites, there are $21$ leaves, each first point has four neighbours outside the star and each leaf has five. It also recomputes each $\beta_m$ as an exact sum over the $2^{13}$ configurations of the star, and enumerates the $40$ ordered pairs of classes joined by a single flip.

\subsection{The upper cap}\label{sec:cap}

The cap is obtained in two stages. Within a class of corner sources, the field-label average of $R_y$ is a convex combination of a large cap on a few \emph{high} labels and a smaller cap on all others. The bridge inequality limits the weight of the high labels. Across classes, the class odds, combined with nonnegative multipliers, average the class caps.

Write $\cR_{c,f}=\nu_{c,f}(R_y)$, so that $\mu_L(R_y)=\sum_{c,f}p_L(c,f)\cR_{c,f}$ by \cref{lem:mixture}. Each $\cR_{c,f}$ is a ratio of two explicit finite sums over $2^{18}$ shell words.

\begin{proposition}[Mixture cap]\label{prop:cap}
For each class $C$ with representative $c_C$, suppose given $\epsilon\ge0$, numbers $R^{\rm hi}_C\ge R^{\rm lo}_C$, a set $H_C\subset\{\pm1\}^{\partial I}$ of \emph{high labels}, and for each $f\in H_C$ a link $t(f)=\{i,j\}$ with $f_i\ne f_j$ and $f^{(i)},f^{(j)}\notin H_C$, such that
\[
 \cR_{c_C,f}\le R^{\rm hi}_C+\epsilon\ \ \text{for all }f,\qquad
 \cR_{c_C,f}\le R^{\rm lo}_C+\epsilon\ \ \text{for }f\notin H_C .
\]
Let $B_C$ be the largest total of the numbers $\omega_i^{-1}$, resp.\ $\omega_j^{-1}$, received by a single label outside $H_C$ under the assignment $f\mapsto f^{(i)},f^{(j)}$. Put $h_C=(B_CR^{\rm hi}_C+R^{\rm lo}_C)/(1+B_C)$. Finally let $\nu_{CC'}\ge0$ be multipliers indexed by the pairs of \cref{lem:classodds}, and
\begin{equation}\label{eq:Sclass}
 S_C=\frac1{n_C}\Bigl(\sum_{C'}\nu_{C'C}-\sum_{C'}\nu_{CC'}\beta_{CC'}\Bigr),\qquad
 \Uord=\epsilon+\max_C\bigl(h_C+S_C\bigr),
\end{equation}
where $\beta_{CC'}$ is the constant of \cref{lem:classodds} for the pair $(C,C')$. Then $\mu_L(R_y)\le\Uord$ for every even $L\ge12$.
\end{proposition}

\begin{proof}
Fix $C$ and write $p(f)=p_L(c_C,f)$. Summing \cref{lem:bridge} over $f\in H_C$ gives $p(H_C)\le B_C\,p(H_C^c)$. Since $R^{\rm hi}_C\ge R^{\rm lo}_C$,
\[
 \sum_fp(f)\cR_{c_C,f}\le p(H_C)R^{\rm hi}_C+p(H_C^c)R^{\rm lo}_C+\epsilon\,p(\cdot)
 \le\bigl(h_C+\epsilon\bigr)\,p(\cdot),
\]
because the middle expression is a weighted mean of $R^{\rm hi}_C$ and $R^{\rm lo}_C$ in which the weight of $R^{\rm hi}_C$ is at most $B_C/(1+B_C)$. The same bound holds for every source in $C$ by symmetry. Summing over classes, $\mu_L(R_y)\le\sum_Cp_C(h_C+\epsilon)$. By \cref{lem:classodds},
\[
 \sum_Cp_CS_C=\sum_{(C,C')}\nu_{CC'}\Bigl(\frac{p_{C'}}{n_{C'}}-\beta_{CC'}\frac{p_C}{n_C}\Bigr)\ge0 .
\]
Hence $\mu_L(R_y)\le\sum_Cp_C(h_C+S_C)+\epsilon\le\Uord$, since $\sum_Cp_C=1$.
\end{proof}

\subsection{The ordered certificate}

Let
\[
 r_b=\frac{3775}{2386},\qquad y_b=\log r_b,\qquad \Delta_b=\frac{4075}{10^{14}} .
\]

\begin{certificate}[Field caps at $r_b$]\label{cert:fields}
At $r=r_b$ the following data satisfy the hypotheses of \cref{prop:cap}.
\begin{enumerate}[label=\textup{(\alph*)}]
\item The values $R_{y_b}(o)$ on the $3236$ orbits are replaced by integer-grid approximations $\widetilde R(o)$. The common radius $\epsilon$ bounds $|R_{y_b}-\widetilde R|$, summed against the absolute values of the multipliers.
\item For each of the $14$ classes and each of the $2^{18}$ field labels, the ratio $\nu_{c_C,f}(\widetilde R)$ is compared with $R^{\rm hi}_C$, and for $f\notin H_C$ with $R^{\rm lo}_C$. This amounts to $14\cdot2^{18}=3\,670\,016$ comparisons with $R^{\rm hi}$ and $3\,661\,604$ comparisons with $R^{\rm lo}$. The sets $H_C$ contain $8412$ labels in total.
\item The six odds constants are bounded below on the grid $10^{-10}\Z$. The class odds $\beta_0,\dots,\beta_3$ have numerators $502492095$, $515270524$, $528601005$, $542501939$, and the flip weights $\omega_{\rm axial},\omega_{\rm face}$ have numerators $356890721$ and $5255754299$. These enter the $40$ directed one-flip comparisons between classes. The certificate also contains the assignment $f\mapsto t(f)$ on the $24$ links and $13$ nonzero class multipliers $\nu_{CC'}$.
\item The resulting number $\Uord(r_b)$ is negative, and
\[
 \Bigl\lfloor10^{15}L_b\Bigr\rfloor=215386342,\qquad L_b:=-\frac{\Uord(r_b)}{Q},
\]
with positive remainder.
\end{enumerate}
\end{certificate}

\begin{proposition}\label{prop:rb}
$m_+(\tfrac12y_b)^2\ge L_b>215386342\cdot10^{-15}$.
\end{proposition}

\begin{proof}
Choose a subsequence as in \cref{prop:lowerrow}, which is possible by compactness. By \cref{prop:lowerrow,prop:cap,cert:fields}, $0\le\Uord(r_b)+Qq$, i.e.\ $q\ge L_b$. Conclude with \cref{prop:order}.
\end{proof}

\subsection{Extension to an interval and the upper bound}

The proof of \cref{prop:rb} works at any $r$ for which the inequalities of \cref{cert:fields} remain valid. We keep all combinatorial data fixed: the rows, the multipliers, $J_i^+$, $Q$, the sets $H_C$, the assignment $t$, the class multipliers, and the frozen lower bounds for the six odds constants. We then bound the variation of the field ratios with $y=\log r$.

\begin{certificate}[Interval]\label{cert:interval}
On $r\in[r_b-\Delta_b,r_b]$ the following hold.
\begin{enumerate}[label=\textup{(\alph*)}]
\item Splitting $1/r\in[1/r_b,1/(r_b-\Delta_b)]$ into $16$ equal intervals and evaluating with outward rounding on the grid $2^{-120}\Z$, the six odds constants stay strictly above their frozen values. The least gap is at least $1171\cdot10^{-14}$.
\item The $y$-derivative of a single row is bounded by $144$, $228$, $57792$ and $14208$ for translation, spectral, reflection and deletion rows respectively. Summed against the absolute multipliers, this gives $|\partial_yR_y(b)|<8326$ for all $b$.
\item $|R_y(b)|<71/1000$ for all $b$ and $r$ in the interval.
\item $|\partial_y\log\Psi_{c,f}(b)|\le468$ for a suitable $b$-independent normalization of $\Psi_{c,f}$.
\item With $C_*=8326+468\cdot\frac{71}{1000}=\frac{2089807}{250}$, exact rational arithmetic gives
\[
 L_b-\frac{C_*\Delta_b}{Q(r_b-\Delta_b)}>\frac{84723}{10^{15}} .
\]
\end{enumerate}
\end{certificate}

The derivative constants in~(b) come from three facts. The $y$-derivative of the conditional representative of a $\pm1$-valued inner observable is at most $72$. One also has $0<J_i^+<1$ and $|\partial_yc_0|<12$. A normalized reflection row with $112$ components satisfies $\|v\|_1^2\le112\|v\|_2^2$. This gives $2\cdot72=144$, $3\cdot72+12=228$, $112(7\cdot72+12)=57792$ and $32(2\cdot72+300)=14208$, where $300$ bounds the derivative of the $150$-bond comparison correlations.

\begin{theorem}\label{thm:upper}
For every $r\in[r_b-\Delta_b,r_b]$, $m_+(\tfrac12\log r)^2>84723\cdot10^{-15}$. Consequently
\[
 e^{2K_c}\le r_b-\Delta_b<r_b-\frac{407}{10^{13}}=r_+ .
\]
\end{theorem}

\begin{proof}
Fix $r$ in the interval. By \cref{cert:interval}(a), \cref{lem:bridge,lem:classodds} hold with the frozen constants. For each $(c,f)$,
\[
 \partial_y\,\nu_{c,f}(R_y)=\nu_{c,f}(\partial_yR_y)+\mathrm{Cov}_{\nu_{c,f}}\bigl(R_y,\partial_y\log\Psi_{c,f}\bigr).
\]
By \cref{cert:interval}(b)--(d) and the Cauchy--Schwarz inequality, $|\partial_y\nu_{c,f}(R_y)|\le C_*$. Since $|y-y_b|=\log(r_b/r)\le(r_b-r)/r\le \Delta_b/(r_b-\Delta_b)$, every field ratio at $r$ exceeds its value at $r_b$ by at most $C_*\Delta_b/(r_b-\Delta_b)$. Raising both caps $R^{\rm hi}_C,R^{\rm lo}_C$ by this amount raises each $h_C$, and hence $\Uord$, by the same amount. \Cref{prop:cap} gives $\mu_L(R_y)\le\Uord(r_b)+C_*\Delta_b/(r_b-\Delta_b)$, and $Q$ does not depend on $r$. As in the proof of \cref{prop:rb}, $m_+^2\ge L_b-C_*\Delta_b/(Q(r_b-\Delta_b))$, which exceeds $84723\cdot10^{-15}$ by \cref{cert:interval}(e). The last assertion follows from \cref{cor:monotone}(ii) at $r=r_b-\Delta_b$ and from the identity $r_b-407\cdot10^{-13}=r_+$.
\end{proof}

\begin{remark}
The multipliers were optimized so that $\Uord(r_b)$ is negative at a point $r_b$ as far to the left as possible. The resulting lower bounds on $m_+^2$, about $2.2\cdot10^{-7}$ at $r_b$ and $8.5\cdot10^{-11}$ at $r_b-\Delta_b$, are accordingly small.
\end{remark}

\begin{proof}[Proof of \cref{thm:main}]
Combine \cref{thm:lower,thm:upper}. The decimal bounds in~\eqref{eq:main} and~\eqref{eq:mainK} are exact rational comparisons; see Appendix~\ref{app:arith}.
\end{proof}

\section{Concluding remarks}\label{sec:discussion}

The two halves of the proof are independent, and each can be sharpened separately. We discuss them in turn.

\subsection{The disordered side}

Three quantities determine the lower bound: the vertex gain $\gamma$, the chord price $\theta$ and the growth rate of the shape-weighted path count. Each can be improved independently.

The gain $\gamma$ is set by the prism, which treats every site outside its ten specified sites as occupied. A larger local graph, or one that records the occupation of second neighbours, increases the certified prism integrals. Joint gains from pairs of nearby middle bonds are a further source. At $r=309/200=1.545$ even the most favourable degree-two integrals give a shape-weighted growth factor $u\rho>1.0039$ on the $16$-step automaton. Gains of this kind are therefore what is needed to move the lower bound substantially further.

The chord price enters through the H\"older step of \cref{lem:holder}, which converts the chord moment, growing like $5^n$, into the factor $e^{\theta/4}$. A transfer operator that records contacts and shapes simultaneously avoids this conversion. The contact operator of \cref{prop:r1} already records contacts. At $r=1.545$, contact operators with $\kappa=4/5$ and $\kappa=6/5$ show that paths with chord density $c(P)/|P|\le0.137$, and those with $c(P)/|P|\ge0.253$, give a uniformly summable contribution to the susceptibility.

\subsection{The ordered side}

The upper bound rests on a linear programme. Its constraints are exact identities and inequalities for local correlations, and its unknown is the zero mode $q$. The infrared bound of \cref{prop:infrared} enters through five test polynomials. The law of the shell spins enters through the mixture representation of \cref{lem:mixture}, with constraints that couple field labels across single links and corner sources through single flips. Additional valid local inequalities, a larger shell, or constraints coupling several labels at once would each enlarge the ordered region certified by the same method.

\appendix
\section{Verification of the certificates}\label{app:verification}

All programs, input data and output logs are available at \repo. Paths below are relative to the root of that repository, whose directories \nolinkurl{lower/}, \nolinkurl{upper/} and \nolinkurl{lean/} contain the material for \cref{sec:paths,sec:counting,sec:lower}, for \cref{sec:infrared,sec:ordered}, and the Lean modules. The programs require Python~3 with its standard library and a C++17 compiler. The full ordered-side field traversal also uses NumPy, SciPy and GMP.

\subsection{Programs and independent checks}

\Cref{tab:programs} lists, for each certificate, the program that produces it and the independent check. Two checks count as independent if they use different arithmetic or different enumerations. Inputs that are \emph{shared} between two checks are listed explicitly. Their derivation is part of the mathematics in the main text.

\begin{table}[ht]
\centering\footnotesize
\caption{Certificates, programs and independent checks.}\label{tab:programs}
\renewcommand{\arraystretch}{1.15}
\begin{tabular}{>{\raggedright}p{2.2cm}>{\raggedright}p{5.2cm}>{\raggedright\arraybackslash}p{5.6cm}}
\toprule
Certificate & Primary computation & Independent check\\
\midrule
\Cref{cert:prism} (prism integrals) &
\nolinkurl{lower/STAGE53_DISORDER_prism.py} ($u_*$), \nolinkurl{lower/STAGE54_WEIGHTED_shapes.py} ($u_2$). Spin sums by layer transfer over the two hub squares, integer common denominator. &
\nolinkurl{lower/STAGE53_LOWER_AUDIT_verify.py} (coordinate spin transfer) and \nolinkurl{lower/STAGE53_PRISM_CHECK.py} (enumeration of all $2^{23}$ even and $2^{23}$ two-source bond sets via the cycle space). All $136$ stored integral rows agree at both panel ends at all test couplings.\tabularnewline
\Cref{cert:allocation} (allocations) &
Same programs. They enumerate all patterns, roles and shapes. &
Separate coordinate replay of $6400$ path inequalities, $676$ chord inequalities and $240$ vertex inequalities at $u_2$. Lean: all $1637$ edge inequalities and $240$ vertex inequalities of part (a), \nolinkurl{STAGE54Prism.lean}.\tabularnewline
\Cref{cert:transfer} (transfer bounds) &
\nolinkurl{lower/STAGE54_CONTACT_robust.cpp} (a), \nolinkurl{lower/STAGE54_WEIGHTED_transfer.cpp} (b) and \nolinkurl{lower/STAGE42_DISORDER_PEER_saw.cpp} (c). The automaton is built by \nolinkurl{lower/verify_saw16_direct_straight_20261121.cpp}. &
(a) \nolinkurl{lower/STAGE54_CONTACT_AUDIT.cpp} rechecks all rows with exact unrounded powers and all possible successor counts. (b) Reconstruction of every transition and its shape mark from coordinates, and comparison with all $81\,390$ seven-step lattice paths. (c) The root counts for $n\le17$ are compared with the exact enumeration of~\cite{SBB11}. The row inequalities of (b) and (c) are checked by the primary program only.\tabularnewline
\Cref{lem:rosenbluth} &
Proof in text. &
Lean, \nolinkurl{STAGE54Torus.lean} and \nolinkurl{STAGE54Kinetic.lean}, for $(\Z/L\Z)^3$, $L\ge3$.\tabularnewline
\Cref{cert:endpoint-lower} &
\nolinkurl{lower/verify_stage54_weighted_prism.py}. &
\nolinkurl{manuscript/check_numbers.py} accompanying this manuscript.\tabularnewline
\Cref{cert:green} (Green integrals) &
\nolinkurl{upper/STAGE23_HOLDBACK_GREEN.py}. &
\nolinkurl{upper/check_physical_inputs.py}.\tabularnewline
\Cref{cert:multipliers} (rows, multipliers) &
Row generator \nolinkurl{upper/STAGE28_ORDERED_ENDPOINT_precise.py}, multipliers \nolinkurl{upper/STAGE21_ORDERED_780_493.json}. &
Hash-checked rebuild, \nolinkurl{upper/new_endpoint/prepare.py}.\tabularnewline
\Cref{cert:fields} (field caps) &
Directed integer traversal, \nolinkurl{upper/new_endpoint/replay.py}, mode \texttt{directed}. &
Unreduced integer traversal in disjoint blocks, \nolinkurl{replay.py}, mode \texttt{raw}, with \nolinkurl{blocked_raw.cpp}. It also checks the $3\,661\,604$ complementary caps. A further $56$ exact comparisons on sampled fields use the direct Hamming kernel, \nolinkurl{crosscheck.py}.\tabularnewline
\Cref{cert:interval} (interval) &
\nolinkurl{upper/new_endpoint/margin.py}. &
\nolinkurl{upper/verify_stage33_upper_margin.py} (row and weight derivative bounds) and \nolinkurl{manuscript/check_numbers.py} (final comparison).\tabularnewline
\bottomrule
\end{tabular}
\end{table}

The two ordered-side traversals share the physical row generator, the lattice and shell geometry, the Green inputs of \cref{cert:green}, the multipliers and the bridge assignment. Their agreement therefore checks the field arithmetic and enumeration, while the shared inputs are covered by the arguments of \cref{sec:infrared,sec:ordered}. On the disordered side, the contact audit uses the same labelled direction graph as the primary computation. It performs $53\,221\,758$ exact row comparisons.

\subsection{Rounding}

All rounding is directed. Lower bounds for integrals, gains and odds constants are rounded down. Upper bounds for powers $w^j$, shape weights $\overline\Omega_\tau$, transfer row quotients, Green integrals and field caps are rounded up. Row inequalities $(\cT v)(s)\le\Lambda v(s)$ for integer vectors are decided by exact cross-multiplication. The power iteration that produced $v$ is not part of the certificate.

\subsection{Lean}\label{app:lean}

The Lean~4 files listed below were compiled with warnings treated as errors and rechecked with \texttt{leanchecker}. Their axioms are the standard \texttt{propext}, \texttt{Quot.sound} and \texttt{Classical.choice}.
\begin{itemize}
\item \nolinkurl{STAGE54Torus.lean}, \nolinkurl{STAGE54Kinetic.lean}: the bound of \cref{lem:rosenbluth} on the actual periodic graph $(\Z/L\Z)^3$ for every $L\ge3$, stated with cleared denominators as $\sum_P5^{c(P)}4^{5n-c(P)}\le6\cdot5^5(5\cdot4^5)^{n-1}$. The identity $c(P)=\sum_ia_i$ of \cref{lem:contacts} is proved along the way.
\item \nolinkurl{STAGE54Prism.lean}, \nolinkurl{STAGE54PrismData.lean}: the $1637$ edge inequalities and $240$ vertex inequalities of \cref{cert:allocation}(a) from the frozen integer table of prism integrals.
\item \nolinkurl{STAGE57CertificateInterface.lean}:\\ positivity of denominators, exact cross-multiplications, finite row equalities and finite list totals used in the endpoint arithmetic.
\end{itemize}
These modules take the frozen tables as input. The completeness of the enumerations and the thermodynamic arguments are proved in the text.

\subsection{Arithmetic of the statement}\label{app:arith}

The following exact comparisons are checked by \nolinkurl{manuscript/check_numbers.py}:
\[
 r_-\in(1.541630483120638,\,1.541630483120639),\qquad
\]
\[
 r_+\in(1.582145850755611,\,1.582145850755612),
\] the bounds~\eqref{eq:mainK} on $K_c=\frac12\log r_c$ using rational enclosures of the logarithm, $r_+=r_b-407\cdot10^{-13}$ and $r_b-\Delta_b<r_+$. The script also re-derives every relation between the constants quoted in the text. Examples are $u_2=677/3177$, $\bar\gamma_2=\gamma_2-\theta_2/4$, \cref{cert:endpoint-lower}, $\overline\Omega_\tau$ from $w_\tau$, $C_*=2089807/250$, and the vertex rows (A3) for both allocations from \cref{tab:charges1,tab:charges2}.

\subsection{Reproduction}

From the repository root:
\begin{verbatim}
cd lower
python3 -B STAGE53_DISORDER_prism.py
python3 -B STAGE53_LOWER_AUDIT_verify.py --final-only
python3 -B STAGE53_PRISM_CHECK.py --cross-check
c++ -O3 -std=c++17 STAGE54_CONTACT_robust.cpp -o walk_check
./walk_check 16 40 10000 9991
c++ -O3 -std=c++17 STAGE54_CONTACT_AUDIT.cpp -o walk_audit
./walk_audit bulk
python3 -B STAGE54_CONTACT_verify.py
c++ -O3 -std=c++17 STAGE54_WEIGHTED_transfer.cpp -o shape_check
./shape_check 16 28
python3 -B STAGE54_WEIGHTED_shapes.py
python3 -B verify_stage54_weighted_prism.py
c++ -O3 -std=c++17 STAGE42_DISORDER_PEER_saw.cpp -o saw_check
./saw_check
\end{verbatim}
for the ordered side
\begin{verbatim}
cd upper/new_endpoint
python3 -B replay.py arithmetic
python3 -B replay.py directed
python3 -B replay.py raw
\end{verbatim}
and for the Lean modules and the scalar arithmetic
\begin{verbatim}
python3 -B lean/STAGE54_kinetic_replay.py
python3 -B lean/STAGE57_certificate_replay.py
python3 -B manuscript/check_numbers.py
\end{verbatim}
The \texttt{arithmetic} mode reconstructs the odds, class caps and interval margin from stored integer outputs. The other two modes regenerate all field comparisons.

\section{The Green-function enclosures}\label{app:green}

Let $\Gamma(x)=\int_{[-\pi,\pi]^3}e^{ik\cdot x}\lambda(k)^{-1}\frac{\dd^3k}{(2\pi)^3}$. Expanding $P_i(\cos k)=\sum_x\widehat P_i(x)e^{ik\cdot x}$, with rational coefficients supported on $\|x\|_1\le4$, gives the finite identity $J_i=\sum_x\widehat P_i(x)\Gamma(x)$. By cubic symmetry only eleven distinct values $\Gamma(x)$ occur. Glasser and Boersma~\cite[Table~1]{GB00} express the lattice Green function $g(x)=\int e^{ik\cdot x}(3-\sum_j\cos k_j)^{-1}\frac{\dd^3k}{(2\pi)^3}$ at these displacements as rational linear combinations of $1$, $g_0=g(0)$ and $1/(\pi^2g_0)$. Since $\lambda=2(3-\sum_j\cos k_j)$, $\Gamma=g/2$. Hence each $J_i$ is $\alpha_i+\beta_ig_0+\gamma_i/(\pi^2g_0)$ with explicit rationals $\alpha_i,\beta_i,\gamma_i$.

The constant $g_0=\mathsf W/3$ is enclosed via its expression through complete elliptic integrals~\cite{Wat39,GB00}. The square roots that appear are bounded by integer squares. The elliptic series is summed exactly up to the term of index $17$, and its positive tail is bounded by the geometric majorant $k_-^{36}/(1-k_-^2)$. The number $\pi$ is enclosed by Machin's formula with alternating arctangent bounds. Propagating these enclosures with interval arithmetic in exact rationals gives \cref{tab:spectral}.

\section*{Acknowledgements}
The author thanks Professor Peng Li (School of Physics, Sichuan University), who introduced him to statistical physics.

\section*{Statements and declarations}

\noindent\textbf{Data and code availability.} All programs, input tables and output logs used for the certificates are available at \repo; an archived snapshot (version 1.0) is deposited at Zenodo, \doi{10.5281/zenodo.24487984}. Appendix~\ref{app:verification} describes their use.

\smallskip\noindent\textbf{Competing interests.} The author declares no competing interests.

\smallskip\noindent\textbf{Use of AI tools.} The author used large language models developed by OpenAI and Anthropic to assist in writing and auditing the verification programs, in cross-checking the finite certificates against the archived outputs, and in drafting and editing the manuscript. The author checked all mathematical arguments, computations and text, and takes full responsibility for the content.

\end{document}